\documentclass[journal, twoside]{IEEEtran}

\usepackage{cite}
\usepackage{graphicx}
\usepackage{float}
\usepackage{amsmath,amssymb,amsfonts}
\usepackage{array}
\usepackage{textcomp}
\usepackage{indentfirst}
\usepackage[space]{grffile}
\usepackage[colorlinks=true, citecolor=blue, linkcolor=blue, urlcolor=black]{hyperref}
\usepackage{algorithmic}
\usepackage{comment}
\usepackage{siunitx}
\usepackage{gensymb}
\usepackage{background}
\usepackage{stackengine}

\graphicspath{ {./Figures/} } %Figure repertory

\begin{document}
\bstctlcite{IEEEexample:BSTcontrol}

%Title and authors
\title{Design of the New Wideband Vivaldi Feed for the HERA Radio-Telescope Phase II}
\author{Nicolas~Fagnoni$^{1}$,
Eloy~de~Lera~Acedo$^{1}$,
Nick~Drought$^{2}$,
David~R.~DeBoer$^{3}$, \IEEEmembership{Member, IEEE},
Daniel~Riley$^{4}$,
Nima~Razavi-Ghods$^{1}$,
Steven~Carey$^{1}$,
and Aaron~R.~Parsons$^{3}$.
\thanks{Manuscript received August 30, 2020; revised April 26, 2021; accepted May 09, 2021. Date of publication XXXX XX, 2021; date of
current version XXXX XX, 2021. This material is based upon work supported by the National Science Foundation under Grant Nos. 1636646 and 1836019 and institutional support from the HERA collaboration partners. This research is also funded in part by the Gordon and Betty Moore Foundation. (\it Corresponding author: Nicolas Fagnoni.)}%
\thanks{N. Fagnoni, E. de Lera Acedo, N. Razavi-Ghods, and S. Carey are with the Department of Physics, Cavendish Astrophysics, University of Cambridge, Cambridge, UK (e-mail: nf323@mrao.cam.ac.uk; eloy@mrao.cam.ac.uk; nima@mrao.cam.ac.uk; shcarey@mrao.cam.ac.uk).}%
\thanks{N. Drought is with Cambridge Consultants, Cambridge, UK (e-mail: nick.drought@cambridgeconsultants.com).}%
\thanks{D. R. DeBoer and A. R. Parsons are with the Department of Astronomy, University of California, Berkeley, CA, USA (e-mail: ddeboer@berkeley.edu; aparsons@berkeley.edu).}%
\thanks{D. Riley is with the Department of Physics, Massachusetts Institute of Technology, Cambridge, MA, USA (e-mail: dgriley@mit.edu).}%
\thanks{Color versions of one or more of the figures in this article are available  online at http://ieeexplore.ieee.org}
\thanks{Digital Object Identifier XXXXXXX}}%

%Paper headers
\markboth{IEEE TRANSACTIONS ON ANTENNAS AND PROPAGATION,~Vol.~XX, No.~XX, MONTH~YEAR}%
{Fagnoni \MakeLowercase{\textit{et al.}}: Design of the New Wideband Vivaldi feed for HERA Phase II}

%Make the title area
\maketitle

%copyright
\setstackEOL{\\}
\SetBgContents{\stackunder[37.5cm]{\Longstack{ This article has been accepted for publication in IEEE Transactions on Antennas and Propagation. This is the author's version which has not been fully edited and\\ 
content may change prior to final publication. Citation information: DOI 10.1109/TAP.2021.3083788}}{$\copyright$ 2021 IEEE. Personal use is permitted, but republication/redistribution requires IEEE permission. See https://www.ieee.org/publications/rights/index.html for more information.}}
\SetBgScale{0.7}
\SetBgAngle{0}
\SetBgPosition{current page.center}
\SetBgVshift{0.5cm}
\SetBgColor{black}
\SetBgOpacity{1}

%%%%%%%%%%%%%%%%% ABSTRACT %%%%%%%%%%%%%%%%%%

\begin{abstract}

This paper presents the design of a new dual-polarised Vivaldi feed for the Hydrogen Epoch of Reionization Array (HERA) radio-telescope. This wideband feed has been developed to replace the Phase I dipole feed, and is used to illuminate a 14-m diameter dish. It aims to improve the science capabilities of HERA, by allowing it to characterise the redshifted 21-cm hydrogen signal from the Cosmic Dawn as well as from the Epoch of Reionization. This is achieved by increasing the bandwidth from 100 -- 200 MHz to 50 -- 250 MHz, optimising the time response of the antenna - receiver system, and improving its sensitivity. This new Vivaldi feed is directly fed by a differential front-end module placed inside the circular cavity and connected to the back-end via cables which pass in the middle of the tapered slot. We show that this particular configuration has minimal effects on the radiation pattern and on the system response.
\end{abstract}

%Keywords
\begin{IEEEkeywords}
Telescopes, Radio astronomy, Receivers, Reflector antennas, Vivaldi antennas.
\end{IEEEkeywords}

\IEEEpeerreviewmaketitle

%%%%%%%%%%%%%%%%% INTRODUCTION %%%%%%%%%%%%%%%%%%

\section{Introduction and scientific context}
\label{sec:1.Intro}

\begin{figure}
    \centering
    (a) \includegraphics[width=0.73\linewidth]{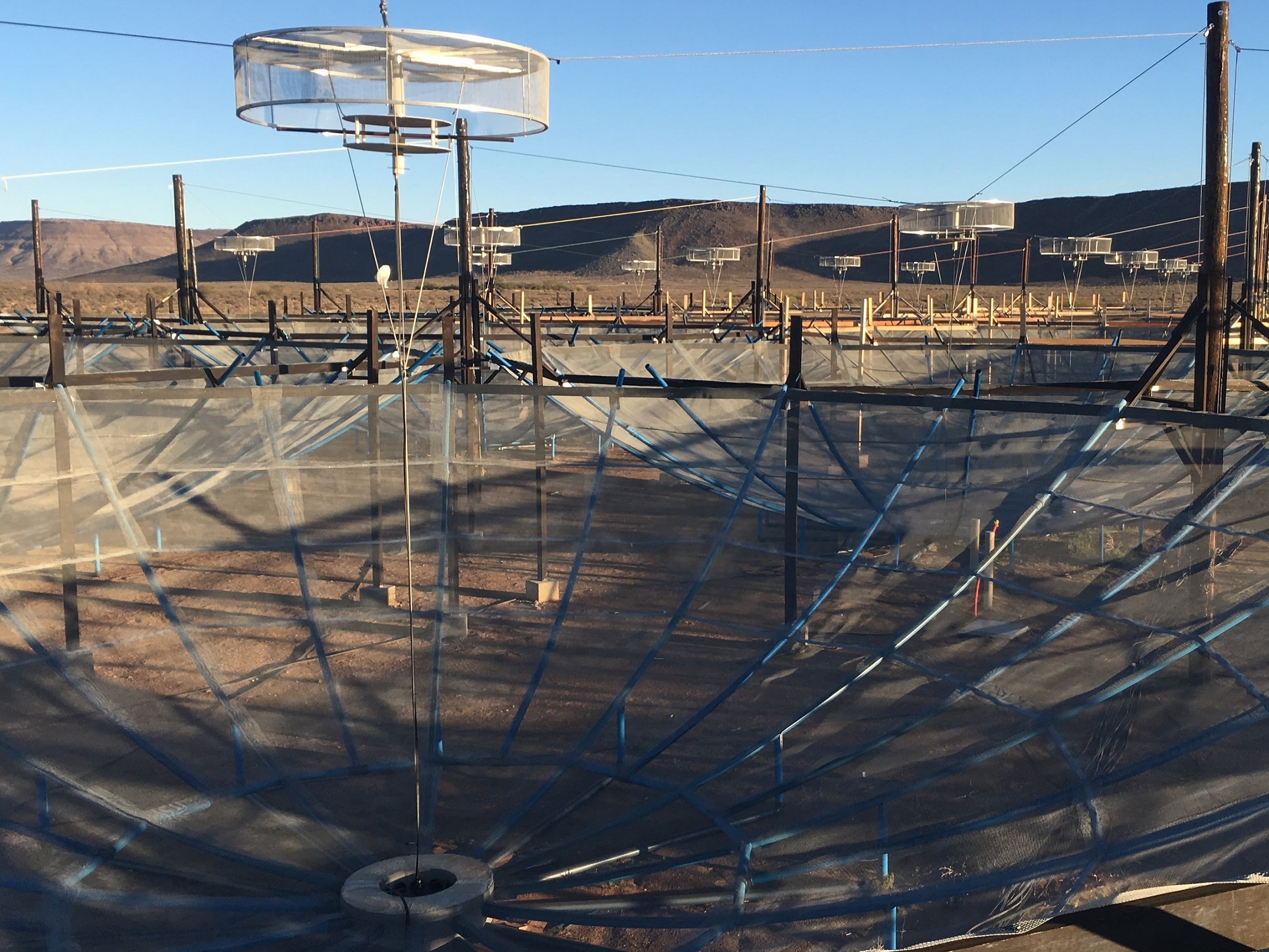}
    
    \vspace{3mm}
    
    (b) \includegraphics[width=0.73\linewidth]{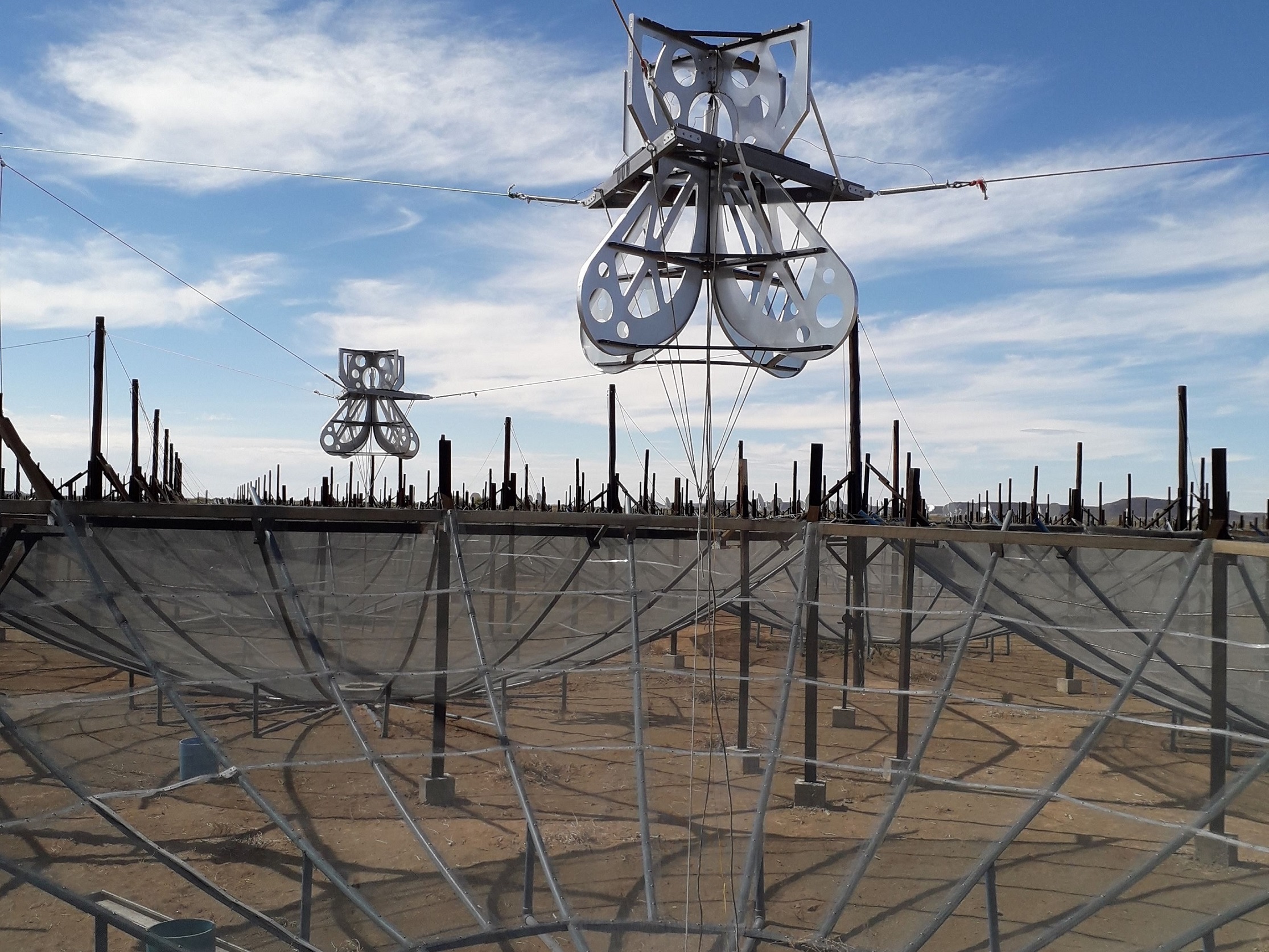}
    \caption{Pictures of the HERA array under construction in the Karoo desert (South Africa). The top picture shows the array with the former Phase I dipole feeds, and the bottom picture the new Vivaldi feeds which are being deployed.}
    \label{fig1:HERAarray}
\end{figure}

\IEEEPARstart{T}{he} Hydrogen Epoch of Reionization Array (HERA) (https://reionization.org) is a radio-telescope which has been designed to study the early universe and characterise the formation and evolution of the first stars and galaxies. It is currently being built in the Karoo desert in South Africa \cite{DeBoer2016}. In its final configuration, this radio-interferometer will comprise 320 antennas forming a dense hexagonal core, as well as 30 additional outrigger antennas \cite{Dillon2016}. Each antenna consists of a 14-m parabolic dish illuminated by a feed suspended 5 m above (cf. Fig. \ref{fig1:HERAarray}). The initial system has a nominal frequency bandwidth between 100 and 200 MHz, and it has been collecting data since 2015, while being expanded.

Before the formation of the first structures, the early universe was dark and mainly composed of clouds of neutral hydrogen. Consequently, HERA uses the 21-cm hydrogen emission line (1420.4 MHz) as a probe \cite{Furlanetto2015}. However, this signal is redshifted due to the expansion of the universe. A signal emitted 115 million years after the Big Bang, i.e. 13.7 billion years ago, reaches the Earth with a frequency of 50 MHz, and a signal received at 250 MHz was emitted 1.3 billion years after the Big Bang. Then, due to the formation of the very first stars and galaxies during the "Cosmic Dawn", the intergalactic medium started to be ionised. Thus, the quantity of neutral hydrogen progressively decreased: this is the "Epoch of Reionization" (EoR). This corresponds to emission received between about 100 and 200 MHz. Typically, the first galaxies formed "ionization bubbles" around them, and the absence of hydrogen signal in a cosmic volume reveals the presence of galaxies. Thanks to the study of this signal, it is possible to infer information about the formation and evolution of the first structures of our universe \cite{Pritchard2012}. The reionization is also expected to be an isotropic phenomenon at large spatial scales. Therefore, the dishes are fixed on the ground, point towards the zenith, and collect the signal from the drifting sky for months, in order to achieve a statistical detection \cite{Thyagarajan2013}. 

Detecting the EoR signal is extremely challenging. It is contaminated by the "foreground", which corresponds to the additional radio signals mainly coming from galactic synchrotron emission and bright compact sources. The foreground brightness temperature is about four to five orders of magnitude more intense than the expected EoR signal \cite{Bowman2009}. Therefore, it is essential to mitigate its contribution, either by “subtracting” it from the received signal, or by analysing a specific portion of a parameter space which is not affected. The first method requires a very accurate knowledge of the foreground along with the systematics induced by the instrument, which is hard to achieve in practice. That is why HERA has been initially designed to apply a "foreground avoidance" method which is based on the spectral differences between these two signals. For a cold patch of sky, the foreground frequency spectrum is rather smooth, whereas the EoR spectrum significantly varies because of the spatial and temporal fluctuations of the signal. 

Since HERA is a radio-interferometer, each pair of antennas independently measures the visibility from the sky \cite{Thomson2017-ch.03}. Then, by Fourier transforming the visibility from the frequency to the delay domain, one obtains a delay spectrum which is relatively compact for the foreground, and which spreads for the EoR signal \cite{Parsons2012b, Liu2014}. In theory, the delay spectrum of the received signal at high delays should be free from foreground contamination. By analysing this portion, it is possible to retrieve partial but essential information about the reionization associated to specific spatial scales. However, in this approach the separation between the delay spectra of the foreground and of the EoR greatly depends on the smoothness of the system response \cite{Thyagarajan2016}. Instrumental chromatic effects such as reflections can cause the foreground contribution to spread at high delays, which may jeopardise the detection of the EoR.

In HERA Phase I, the feed consists of two orthogonal 1.3-m dipoles surrounded by a metal cage to taper the beam (cf. Fig. \ref{fig1:HERAarray}). It was initially designed along with a receiver for the telescope PAPER \cite{Parsons2010}, the precursor of HERA. From this project, the dipoles were recycled, optimised, and combined with a dish \cite{DeBoer2015} in order to reach the sensitivity required for a robust detection of the EoR. By using computer simulations, measurements, and data analysis, \cite{Ewall-Wice2016, Patra2018, Fagnoni2021, Kern2020a} studied the chromatic effects in the array and their impact on the detection. The HERA Phase I system is affected by reflections occurring between the feed and the dish, inside the cage, within the coaxial cables connecting the front-end to the back-end module, and by mutual coupling between the dishes. Therefore, a new feed and a receiver have been developed to improve the performance in terms of response, bandwidth, and sensitivity. In HERA Phase II, the dipoles are replaced by wideband Vivaldi feeds \cite{Gibson1979, Tokan2013} which cover the 50 -- 250 MHz band to study the Cosmic Dawn and confirm the end of the EoR. As for the new receiver, it uses a radio-over-fibre system to mitigate the reflections caused by the coaxial cables.

HERA is not the only interferometer designed to detect the EoR. Let us mention PAPER \cite{Kolopanis2019}, LOFAR \cite{Nijboer2009, vanHaarlem2013}, MWA \cite{Tingay2013, Wayth2018}, and the SKA Low-Frequency Aperture Array (LFAA) \cite{DeLeraAcedo2015, DeLeraAcedo2018}. Table \ref{tab1:telesComp} presents their main characteristics. Before the SKA LFAA is fully operational with more than 130,000 elements after the mid-2020s, our new solution stands out with an excellent collecting area ensuring a high sensitivity to the EoR signal \cite{Pober2014}, and over a wide bandwidth. Note that LOFAR uses two different antennas to cover the low (LBA) and the high (HBA) frequencies. The low-band stations are also divided into an "inner" and an "outer" section, which cannot be simultaneously operated due to technical limitations. In this comparison, we only consider the "core stations" in the Netherlands, which are designed for the study of the EoR. 

This paper describes the development of this new Vivaldi feed, and details its performance in the framework of the EoR signal detection with the foreground avoidance method. Section \ref{sec:2.Design} presents the electromagnetic and mechanical design of the feed. Section \ref{sec:3.RecFeeding} explains how the front-end module (FEM) and the cables are integrated into the Vivaldi feed with a minimal influence on its performance. Lastly in Section \ref{sec:4.Perf}, by combining the electromagnetic model of the antenna with the parameters of the receiver, the performance of the Phase II system is assessed and compared with HERA Phase I.

\begin{table*}[t]
\caption{Main characteristics of radio-interferometers designed for the EoR detection}
\begin{tabular}{|c|c|c|c|c|c|c|c|}
\hline
                                                                                   & \multicolumn{2}{c|}{\textbf{\begin{tabular}[c]{@{}c@{}}HERA\\ Phase II\end{tabular}}}   & \textbf{\begin{tabular}[c]{@{}c@{}}HERA\\ Phase I\end{tabular}}   & \textbf{PAPER 64}                                                                                                       & \textbf{\begin{tabular}[c]{@{}c@{}}LOFAR\\ (NL "core stations")\end{tabular}}                                                                                                                                                                                           & \textbf{\begin{tabular}[c]{@{}c@{}}MWA\\ Phase II\end{tabular}}                                       & \textbf{\begin{tabular}[c]{@{}c@{}}SKA LFAA\\ Phase I\end{tabular}}                                                               \\ \hline
\textbf{Antenna}                                                                   & \multicolumn{2}{c|}{\begin{tabular}[c]{@{}c@{}}Vivaldi feed\\ + 14-m dish\end{tabular}} & \begin{tabular}[c]{@{}c@{}}Dipole feed\\ + 14-m dish\end{tabular} & \begin{tabular}[c]{@{}c@{}}Dipoles\\ +  ground screen\end{tabular}                                                      & \begin{tabular}[c]{@{}c@{}}- LBA: dipoles + ground plane     \\ - HBA: bowties + ground plane\end{tabular}                                                                                                                                                              & \begin{tabular}[c]{@{}c@{}}Bowties +\\ ground plane\end{tabular}                                      & \begin{tabular}[c]{@{}c@{}}Log-periodic\\ + ground plane\end{tabular}                                                             \\ \hline
\textbf{\begin{tabular}[c]{@{}c@{}}Bandwidth \\ (MHz)\end{tabular}}                & \multicolumn{2}{c|}{50 - 250}                                                           & 110 - 190                                                         & 110 - 190                                                                                                               & \begin{tabular}[c]{@{}c@{}}LBA: 10 - 90\\ HBA: 110 - 240\end{tabular}                                                                                                                                                                                                   & 80 - 300                                                                                              & 50 - 350                                                                                                                          \\ \hline
\textbf{\begin{tabular}[c]{@{}c@{}}Array \\ configuration\end{tabular}}            & \multicolumn{3}{c|}{\begin{tabular}[c]{@{}c@{}}- 320 el. in dense \\ hexagonal \\ core (300m x 250m)\\ - 30 outriggers\end{tabular}}                        & \begin{tabular}[c]{@{}c@{}}- 64 el. in regular \\ grid (250m x 30m)\\ - 8 x 8 array in \\ East - West dir.\end{tabular} & \begin{tabular}[c]{@{}c@{}}- Phased array\\ - 24 “core stations” (4-km diam.)\\ - LBA: 96 el. per station with \\ 48 “inner” el. (32.25-m diam.) \\ + 48 “outer” el. (81.34-m diam.)  \\ - HBA: 16 el. per tile and \\ 48 tiles per station (30.75-m diam)\end{tabular} & \begin{tabular}[c]{@{}c@{}}- Phased array\\ - 256 tiles\\ - 16 el. per tile \\ (5m x 5m)\end{tabular} & \begin{tabular}[c]{@{}c@{}}- Phased array\\ - 512 stations\\ with dense core\\ - 256 el. per station \\ (35-m diam.)\end{tabular} \\ \hline
\textbf{\begin{tabular}[c]{@{}c@{}}3-dB \\ beamwidth (°)\end{tabular}}             & \multicolumn{2}{c|}{\begin{tabular}[c]{@{}c@{}}10.1 \\ at 150 MHz\end{tabular}}         & \begin{tabular}[c]{@{}c@{}}10.5 \\ at 150 MHz\end{tabular}        & \begin{tabular}[c]{@{}c@{}}100\\ at 138 MHz\end{tabular}                                                                & \begin{tabular}[c]{@{}c@{}}- LBA “inner”: 9.8 at 60 MHz\\ - LBA "outer": 3.9 at 60 MHz\\ - HBA: 3.8 at 150 MHz\\ (for a station)\end{tabular}                                                                                                                           & \begin{tabular}[c]{@{}c@{}}28 at 150 MHz\\ (for a tile)\end{tabular}                                  & \begin{tabular}[c]{@{}c@{}}15 at 150 MHz\\ (for 16 combined el.)\end{tabular}                                                     \\ \hline
\textbf{\begin{tabular}[c]{@{}c@{}}Total\\ effective\\ aperture (m²)\end{tabular}} & \multicolumn{2}{c|}{\begin{tabular}[c]{@{}c@{}}29,645\\ at 150 MHz\end{tabular}}        & \begin{tabular}[c]{@{}c@{}}35,035\\ at 150 MHz\end{tabular}       & \begin{tabular}[c]{@{}c@{}}\textless 576 \\ (physical size)\end{tabular}                                                & \begin{tabular}[c]{@{}c@{}}- LBA “inner”: 8,844 at 60 MHz\\ - LBA "outer": 9,597 at 60 MHz\\ - HBA: 12,288 at 150 MHz\end{tabular}                                                                                                                                      & \begin{tabular}[c]{@{}c@{}}5,504\\ at 150 MHz\end{tabular}                                            & \begin{tabular}[c]{@{}c@{}}247,398\\ at 150 MHz\end{tabular}                                                                      \\ \hline
\textbf{\begin{tabular}[c]{@{}c@{}}Receiver\\ temperature (K)\end{tabular}}        & \multicolumn{2}{c|}{75}                                                                 & 160                                                               & 145                                                                                                                     & \begin{tabular}[c]{@{}c@{}}- LBA: 1,500 K\\ - HBA: 180 K\end{tabular}                                                                                                                                                                                                   & 50                                                                                                    & 40 (front-end LNA)                                                                                                                \\ \hline
\end{tabular}
\label{tab1:telesComp}
\end{table*}

%%%%%%%%%%%%%%%%% SECTION II %%%%%%%%%%%%%%%%%%

\section{Electromagnetic simulations and mechanical design}
\label{sec:2.Design}

\subsection{Scientific requirements and goals}
\label{sec:2.1.DesignReq}

The scientific context and the method used to detect the EoR signal impose stringent requirements on the design. The top priority is to extend the bandwidth to 50 -- 250 MHz while minimising the instrument chromaticity. From the delay spectrum of the received signal, the delay power spectrum of the redshifted 21-cm signal can be computed. This power spectrum is averaged over a cylindrical cosmic volume. Its transverse size depends on the baseline length and is limited by the antenna beamwidth. Shorter baselines are advantageous since they are associated with larger spatial scales. Its depth along the line-of-sight corresponds to the observation bandwidth centred on the frequency of the studied redshift. In practice, this bandwidth is limited by the cosmic evolution of the signal, and it is assumed that its characteristics do not significantly change over an 8-MHz bandwidth ($\Delta z\approx0.5)$. In the delay domain, deep cosmic volumes correspond to low delays. Therefore, if the foreground contribution in the delay spectrum spreads at high delays, then the size of the spatial scales over which the EoR signal can be characterised is reduced. However, this signal is more intense and so more easily detectable when it is measured over large volumes \cite{Pober2016}.

The instrument chromaticity can be quantified by studying the system response. It is obtained by combining the antenna impedance, the beam pattern, and the receiver Z-parameters (cf. Equation \eqref{eq5:VolSystResp}). As a goal, the system time response should be attenuated by a factor $10^5$ as quick as possible, and ideally after 150 ns \cite{Thyagarajan2016} in order to limit the foreground leakage in the delay spectrum. In the frequency domain, this implies that the antenna and receiver parameters need to be "smooth". The antenna impedance should be as flat as possible and close to 100 $\Omega$ in order to make the matching with the FEM easier in terms of power and voltage reflection coefficients, and in terms of noise. The sidelobe level must be kept as low as possible to limit the mutual coupling, and prevent bright sources from contaminating the signal received by the main lobe. In practice, the study of the frequency parameters is a first step to estimate the performance and compare the designs. However, precisely quantifying the requirements in the frequency domain is challenging, because both the level of the parameters and their spectral structures need to be characterised. Therefore, the end-to-end system time response is one of the most important metrics in this paper.

The antenna sensitivity should also be optimised in order to reduce the observation time required to obtain a significant statistical detection. This can be achieved by improving the antenna gain at zenith and by minimising the noise temperature generated by the receiver. If the receiver noise temperature is below about 80 K, then one can consider that the system temperature is dominated by the sky emission in the bandwidth of operation \cite{Rogers2008}, and therefore the contribution from the receiver is less problematic. Lastly, the feed needs to be dual-polarised to fully characterise the received radiation.

\subsection{Design constraints}
\label{sec:2.2.DesignConstr}

The feed selection and its design are also driven by some mechanical constraints. The feed weight has to be kept below 50 kg in order to be transportable by two persons and supported by the suspension system. As illustrated in Fig. \ref{fig1:HERAarray}, the array is dense. Therefore, the feed has to be compact enough so that it can be carried between the dishes, and passed through the structural lattice supporting the parabola to be positioned at its centre. The design also has to be robust in order to last several years in the desert, and withstand wind gusts which regularly exceed 50 km/h and can blow up to 100 km/h.

A horn is a wideband feed commonly used to illuminate a dish. However, under these constraints it would not be a suitable choice because it would require an aperture of about 3 m to work at 50 MHz. Moreover, the rigging system only enables the feed to be suspended 5 m above the dish maximum. Since the focal point is 4.5 m above the vertex, the feed length also has to be limited. Regarding this aspect, planar wideband antennas such as sinuous or spiral antennas \cite{Balanis2016-ch.11} may appear as a good option. But like with the dipole feed, they require a backplane or a cavity to cancel out the symmetric backlobe radiation \cite{Buck2008}, which is a significant source of chromaticity.

\subsection{Design presentation}
\label{sec:2.3.DesignPres}

Following these requirements, a Vivaldi feed was chosen as a candidate to replace the dipoles and illuminate the 14-m diameter dish. Initially developed by P. Gibson in 1979 \cite{Gibson1979}, this "travelling wave" antenna consists of a metal plate with a tapered slot terminated by a circular cavity. The antenna is fed at the beginning of the slot, and the resonant cavity acts as an open circuit which orientates the direction of radiation. Thanks to its exponential curvature which adjusts the antenna aperture to the incident or radiated wavelength, the antenna pattern is smooth and its impedance stable over a very wide frequency band. Such an antenna produces a directive radiation pattern, linearly polarised, symmetric, with low sidelobes, and a peak directivity which can easily exceed 10 dBi \cite{Balanis2016-ch.09, Bolli2016}. Mechanically, its planar configuration makes it light, cheap and easy to manufacture. Consequently, this type of design can be used as a feed \cite{Tokan2013} and presents very interesting properties in the framework of the EoR signal detection with HERA.

\begin{figure}[t]
    \centering
    \includegraphics[width=0.75\linewidth]{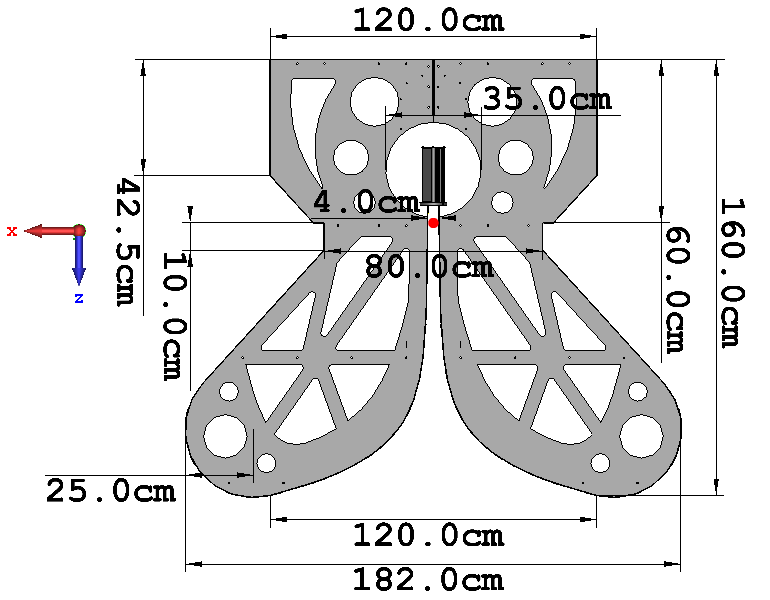}
    \caption{Dimensions of the Vivaldi blade.}
    \label{fig2:antDim}
\end{figure}

The feed is dual-polarised and made up of two blades in aluminium, orthogonally placed to generate the X and Y-polarisations. The dimensions of a blade are presented in Fig. \ref{fig2:antDim}. It can be divided into two sections: the top part which contains the circular cavity, the feed point, and the FEM, and the bottom part with the exponential slot. The feed is 160-cm high. The slot has a maximum aperture of 182 cm and a height $H_{\rm s}$ of 100 cm. The antenna is fed at the top of the slot by the FEM which is placed inside the 35-cm diameter cavity. At the feed point, the slot width $W_{\rm sMin}$ is 4 cm. In the Cartesian coordinate system defined in Fig. \ref{fig2:antDim} with the red dot as the origin, the equation of the exponential aperture $A_{\rm p}\left(z\right)$ is:
\begin{align}
    A_{\rm p}\left(z\right) = \alpha\left(\mathrm{e}^{\beta z}-1\right)+\frac{W_{\rm sMin}}{2},
	\label{eq1:aperture}
\end{align}
with:
\begin{align}
    \alpha = \frac{W_{\rm sMax}-W_{\rm sMin}}{2\left(\mathrm{e}^{\beta H_{\rm s}}-1\right)}.
	\label{eq2:alphaCoef}
\end{align}
$z$ represents the distance from the origin, and varies from 0 to $H_{\rm s}$ = 100 cm. $\beta$ controls the exponential growth rate of the slot and is equal to 0.07. In practice, the slot follows an exponential function until it reaches an aperture $W_{\rm sMax}$ of 120 cm. Then, the aperture is curved following a 25-cm radius disc, in order to reduce the diffraction from the edges.

\begin{figure}[H]
    \centering
    \includegraphics[width=0.95\linewidth]{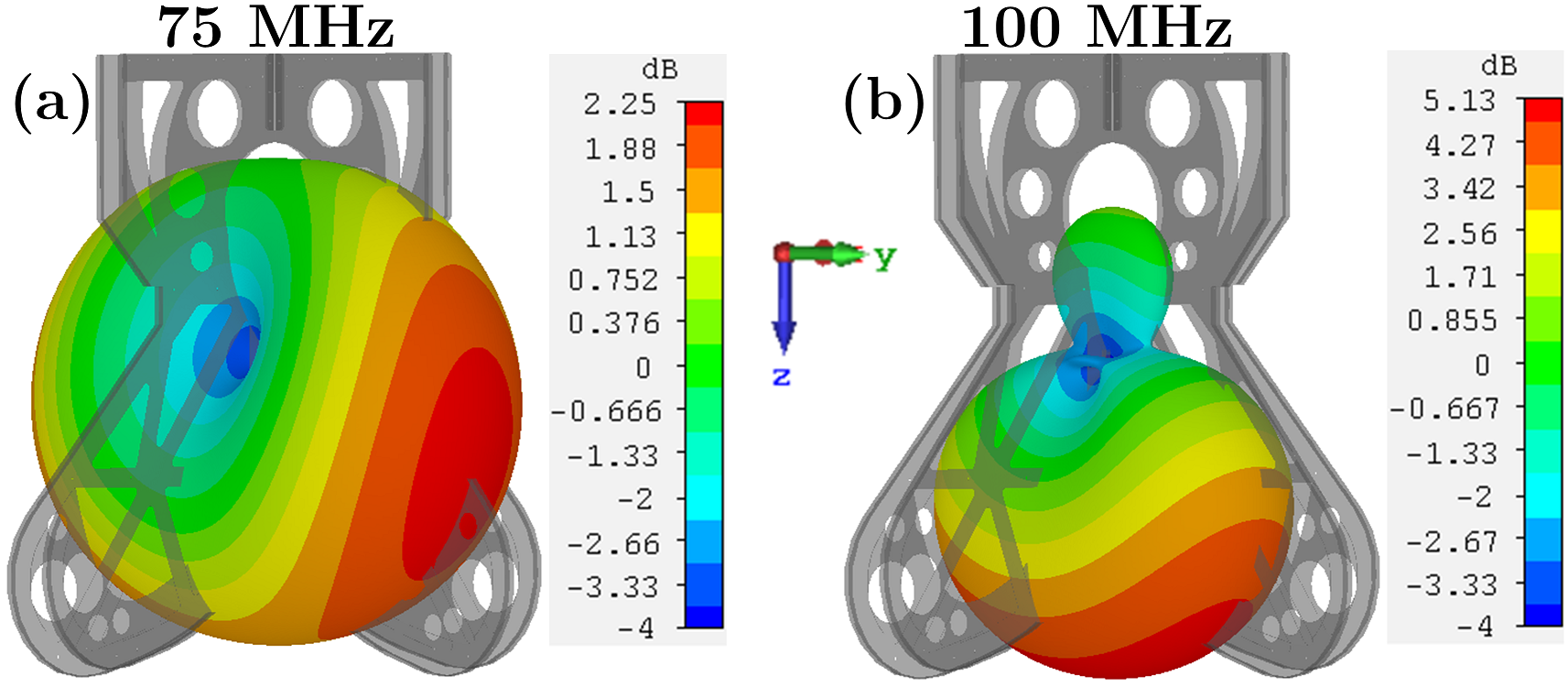}
    \caption{3D gain of the Vivaldi feed at 75 and 100 MHz, illustrating the transition between the radiation modes, for the X-polarisation.}
    \label{fig3:feedBeams}
\end{figure}

The maximum aperture is 1.82 m, which is rather small compared with the maximum wavelength at 50 MHz, $\lambda_{max} =$ 6 m. Ideally, the aperture should be about 3-m wide ($\lambda_{max}/2$) to properly guide the electromagnetic wave \cite{Balanis2016-ch.09}. A 1.82-m aperture is theoretically associated with a minimum frequency of 82 MHz. However, this cut-off frequency is not abrupt, and the antenna is still able to receive larger wavelengths. In this case, the main radiation mode changes, and the feed starts behaving like a wideband dipole. This transition smoothly occurs between 80 and 85 MHz. As illustrated in Fig. \ref{fig3:feedBeams}, this results in an increase in the beamwidth and in the backlobe. At low frequencies, the dish illumination is mediocre and the efficiency is degraded (cf. Fig. \ref{fig5:illumEfficiency}). Nevertheless, the antenna still receives a significant amount of power from the zenith and the system response remains smooth (cf. Section \ref{sec:4.Perf}).

\subsection{Optimisation process with numerical simulations}
\label{sec:2.4.ElectroSimu}

\subsubsection{Feed parametrisation}
\hfill \vspace{1.5mm}

The feed is optimised along with the dish from 50 to 250 MHz by using CST. Since the antenna has a complex geometry which includes small and large elements, is wideband, and the time response is a crucial parameter, the transient solver based on the "finite integration technique" (FIT) is used. The feed dimensions are parametrised and swept. The parameter space is explored by predefining the points to be simulated. These points are chosen in such a way that the beam pattern and the impedance slowly evolve. The range and the minimum step used to optimise each parameter are specified in Table \ref{tab2:feedDim}. 

\begin{figure}[H]
    \centering
    \includegraphics[width=0.74\linewidth]{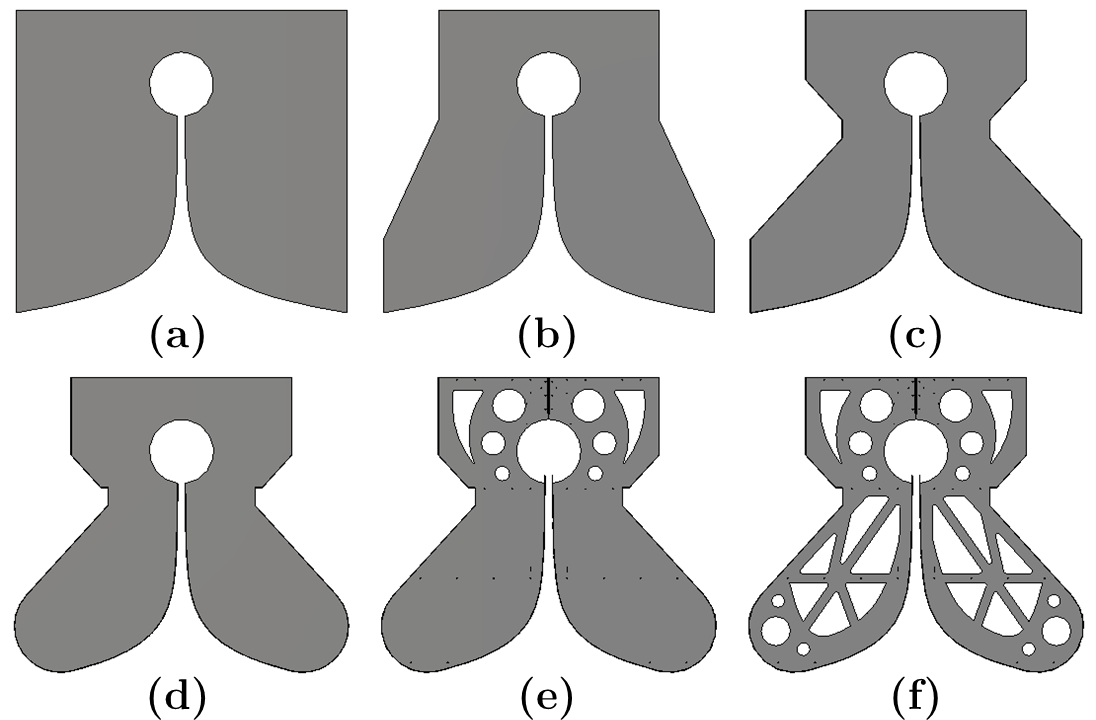}
    \caption{Evolution of the design during the optimisation process with the aim of reducing the weight and the wind load of the feed.}
    \label{fig4:designEvol}
\end{figure}

First, the main dimensions, namely the slot length, the maximum slot aperture, and the exponential growth rate along with the feed height are simultaneously optimised. These parameters control the beam pattern and the antenna impedance. In a second phase, the minimum slot width, the metal thickness of the blade, and the cavity diameter are optimised together. Moderate variations of these parameters do not significantly impact the antenna beam, but only its impedance. Thus, the FEM impedance can be easily matched by tuning this second set of parameters. A trade-off between time response obtained from the frequency parameters, which is the main objective, gain at zenith, and sidelobe level is performed to select the design after this first optimisation step (cf. Fig. \ref{fig4:designEvol} \textit{(a)}).

\begin{table}[H]
    \centering
    \caption{Main optimised parameters and swept range}
    \begin{tabular}{|p{1.0cm}<{\centering} | p{1.4cm}<{\centering} | p{0.9cm}<{\centering} | p{0.9cm}<{\centering} | p{0.9cm}<{\centering}| p{0.9cm}<{\centering}| }
        \hline
        \textbf{Param.} & \textbf{Description} & \textbf{Min. value} & \textbf{Ref. value} & \textbf{Max. value} & \textbf{Min. step} \\\hline
        $H_{\rm s}$     & Slot length               & 80 cm      & 100 cm     & 140 cm     & 5 cm      \\\hline
        $W_{\rm apMax}$ & Max. feed aperture        & 162 cm     & 182 cm     & 222 cm     & 5 cm      \\\hline
        $\beta$         & Exp. growth rate          & 0.04       & 0.07       & 0.13       & 0.01      \\\hline
        $W_{\rm sMin}$  & Min. slot width           & 2 cm       & 4 cm       & 8 cm       & 0.5 cm    \\\hline
        $T_{\rm bl}$    & Blade thickness           & 1 cm       & 3 cm       & 7 cm       & 0.5 cm    \\\hline
        $D_{\rm cav}$   & Cavity diameter           & 30 cm      & 35 cm      & 45 cm      & 2.5 cm    \\\hline
        $F_{\rm h}$     & Feed height               & 4.7 m      & 5.0 m      & 5.2 m      & 5 cm       \\
        \hline
    \end{tabular}
    \label{tab2:feedDim}
\end{table}

\subsubsection{Feed height and dish illumination}
\hfill \vspace{1.5mm}

The optimal height is reached when the distance between the vertex and the feed point (red dot in Fig. \ref{fig2:antDim}) is equal to 5 m. Simulations show that the phase centre of the feed varies as a function of the frequency and is actually about 50 $\pm$ 20 cm below this point. Thus, the phase centre is around 4.5 m above the vertex, which does correspond to the focal point.

\begin{figure}[H]
    \centering
    \includegraphics[width=1.0\linewidth]{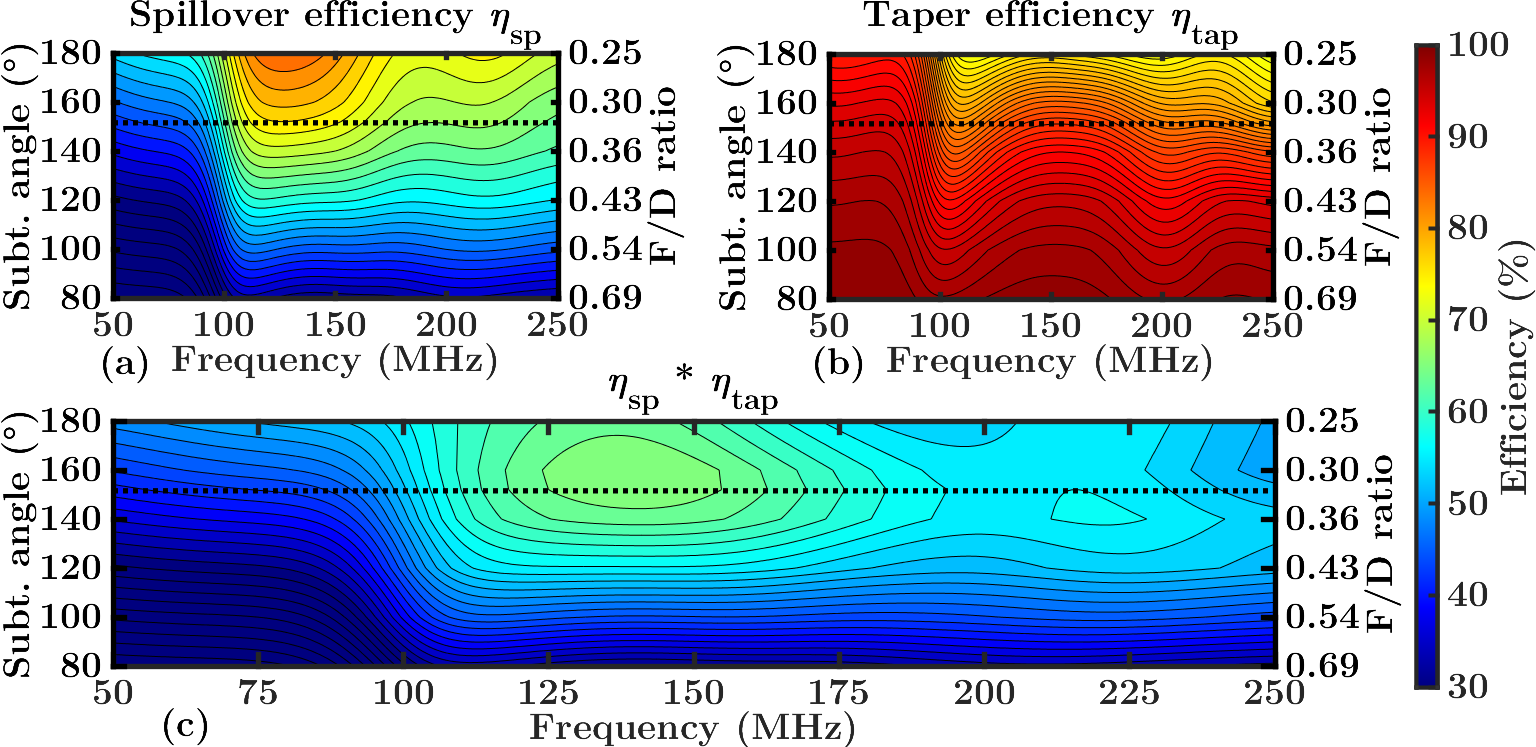}
    \caption{Spillover $\eta_{sp}$, taper $\eta_{tap}$, and $\eta_{sp}*\eta_{tap}$ efficiencies as a function of the subtended angle or F/D, for the model \textit{(f)}. The dashed line corresponds to the actual F/D of the HERA dish (0.32).}
    \label{fig5:illumEfficiency}
\end{figure}

In order to quantify the dish illumination, the spillover and taper efficiencies \cite{Collin1984, Kildal1985} are computed as a function of the subtended angle or F/D ratio, in Fig. \ref{fig5:illumEfficiency}. The dashed line represents the F/D of the HERA dish (0.32). Above 90 MHz, the spillover efficiency is between 60\% and 75\%, and most of the losses actually come from the backlobe of the feed. Below, it drops to about 50\% because the dish also becomes over-illuminated. The subtended angle at the focal point is 152$\degree$. The 3-dB beamwidth of the feed is always below 115$\degree$ in the E-plane. However, it is significantly wider in the H-plane, and well above 152$\degree$ below 90 MHz (cf. Fig. \ref{fig3:feedBeams}). Since the dishes are spaced 60 cm apart, this implies that the adjacent dishes will be also illuminated in the H-plane, which will contribute to mutual coupling. The absence of a cage around the feed limits the possibility to control the beam. The taper efficiency is high, above 80\%, which means that the E-field illuminating the dish surface is relatively uniform in amplitude and phase.

The product of these two terms provides information about the efficiency of the dish illumination, which is between 50\% and 70\% above 90 MHz and between 40\% and 50\% below. This parameter directly affects the aperture efficiency of the antenna, which is given in Section \ref{sec:4.2.efficiencies}. Lastly, the selected feed does provide the optimum performance for the F/D of the HERA dish, which is not a surprise since the feed was optimised along with the dish and the feed height. 

\subsubsection{Design evolution and fine-tuning}
\hfill \vspace{1.5mm}

\begin{figure}[t]
    \centering
    \includegraphics[width=0.87\linewidth]{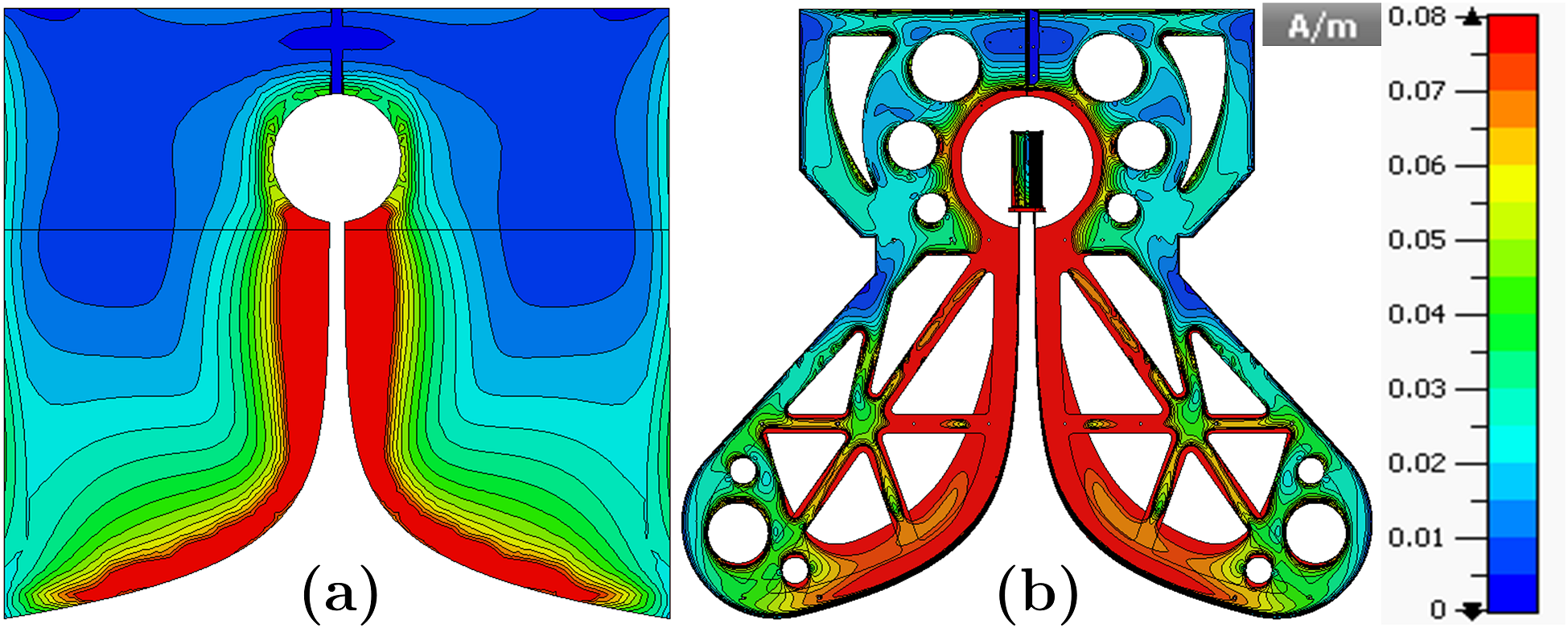}
    \caption{Simulation of the surface current distribution at 150 MHz (absolute value of the amplitude in A/m), for the models \textit{(a)} and \textit{(f)}.}
    \label{fig6:surfCur}
\end{figure}

Fig. \ref{fig4:designEvol} illustrates the evolution of the design. Because of the mechanical constraints, the weight and the wind load of the feed need to be reduced by cutting out some parts. Fig. \ref{fig6:surfCur} shows the simulated current distribution at the surface of the blade, at 150 MHz. The current mainly flows along the aperture of the slot and along the cavity edge. The contribution from the top part and from the centre of the blade to the radiation process is relatively marginal. Therefore, these areas can be trimmed without significantly affecting the performance. 

\begin{figure}[t]
    \centering
    \includegraphics[width=1.0\linewidth]{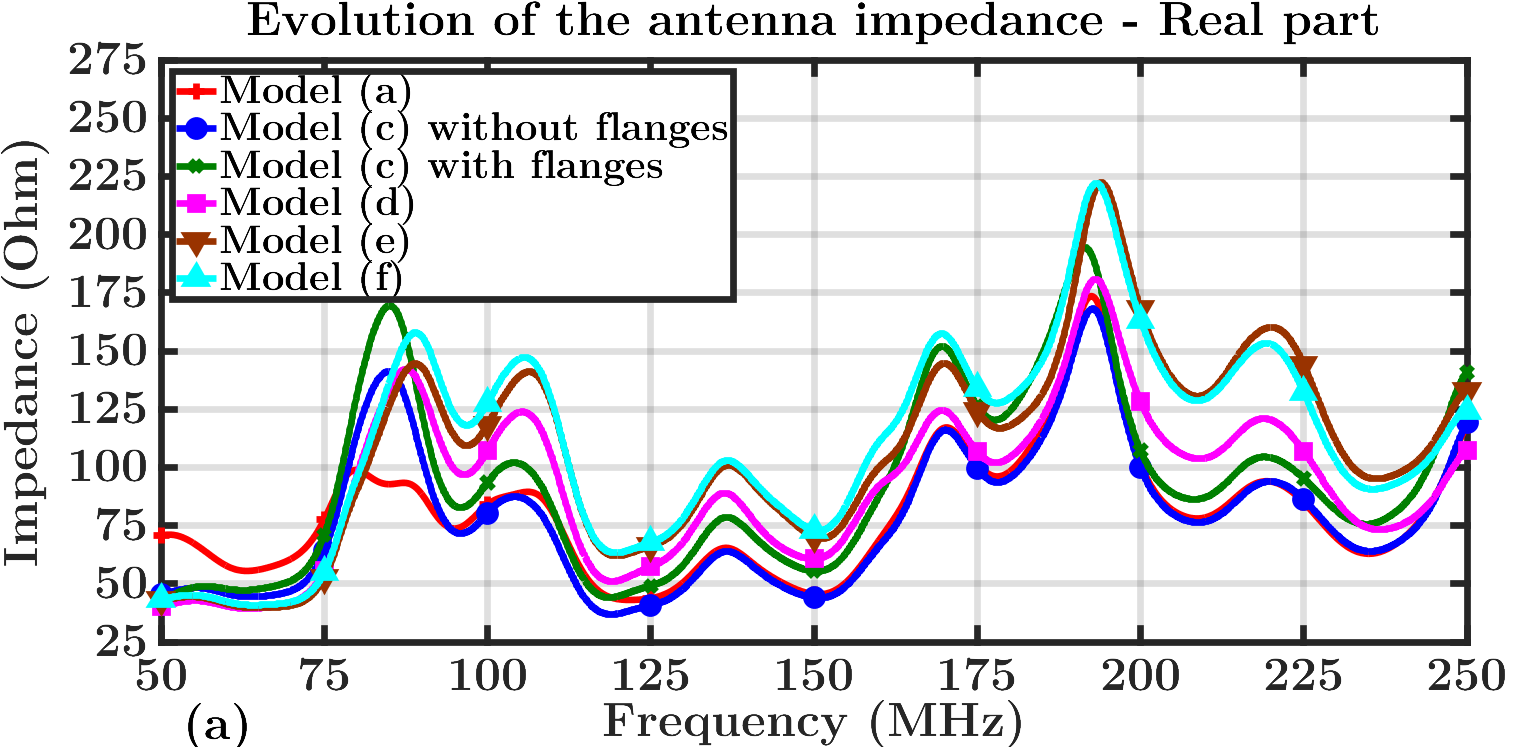}
    
    \vspace{2.7mm}
    
    \includegraphics[width=1.0\linewidth]{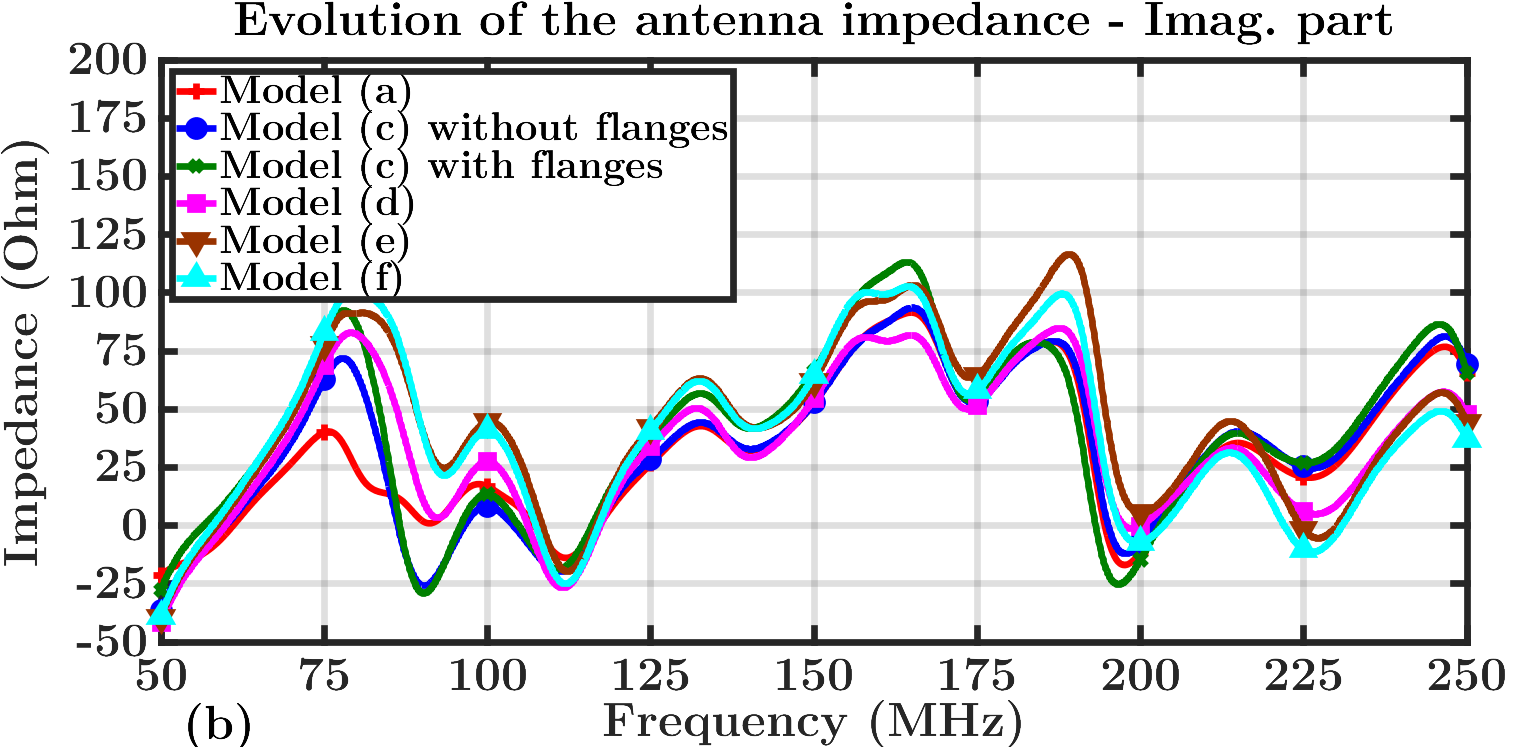}
    \caption{Evolution of the antenna impedance (with dish) during the optimisation process.}
    \label{fig7:impEvol}
\end{figure}
\begin{figure}[t!]
    \centering
    \includegraphics[width=1.0\linewidth]{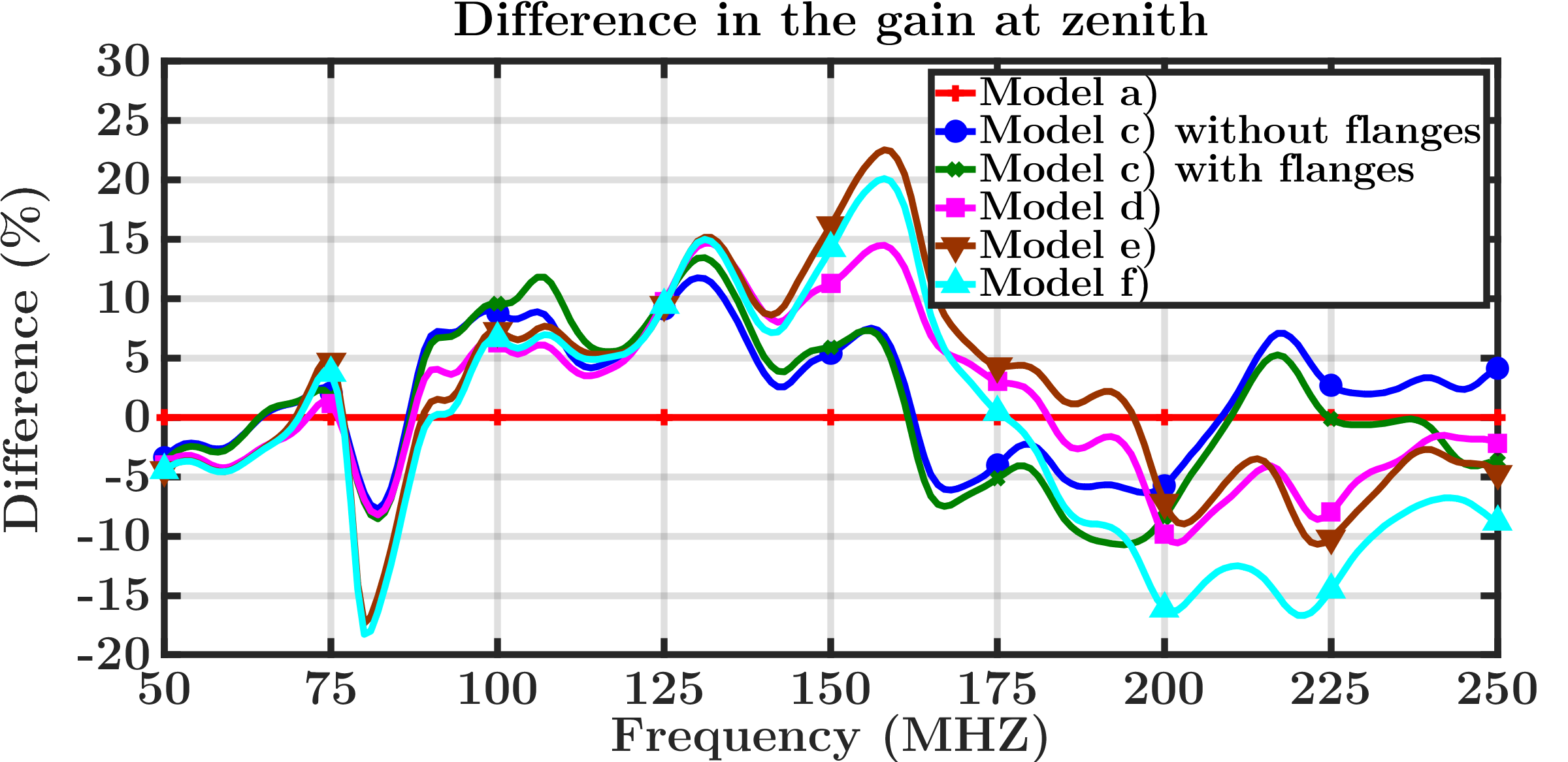}
    \caption{Evolution of the difference (in \%) in the antenna gain at zenith during the optimisation process, with respect to the model \textit{(a)}.}
    \label{fig8:gainDiffEvol}
\end{figure}

Fig. \ref{fig7:impEvol} and \ref{fig8:gainDiffEvol} show the evolution of the antenna impedance and of the gain at zenith, including the dish. After parametric optimisation, the model \textit{(b)} is obtained. The "neck" of the antenna is then shortened (model \textit{(c)}) to reduce the size of the triangular frame which supports the feed (cf. Fig. \ref{fig11:CSTmecha}). These cutouts slightly modify the gain with variations of $\pm$ 10\% with respect to the model \textit{(a)}. The blade thickness is used to control the impedance level so that it is around 100 $\Omega$. Ideally, the metal plates should be 3-cm thick, but this makes the feed extremely heavy. Therefore, 0.2-cm thick plates are used and 3-cm wide flanges are perpendicularly tack welded along the length of the slot. Since the electromagnetic wave is mainly guided by the slot, the properties are rather equivalent. On average, the impedance is increased by 15 + 10j $\Omega$ with respect to a model with 3-cm thick plates, but the profile remains the same. The addition of the flanges has no significant effects on the gain. In the model \textit{(d)}, the lower corners of the slot are rounded. This improves the performance by decreasing the amplitude between the maxima and minima of the impedance, and by increasing the gain between 125 and 200 MHz.

Lastly, holes are cut out of the blade. A large portion of the metal is removed from the top part with a minimal impact (model \textit{(e)}). The gain decreases by 0.5 dB at 80 MHz, and increases by 0.3 dB at 160 MHz. As for the bottom part, Fig. \ref{fig6:surfCur} shows that the current is maximum along the slot over a 10 -- 15-cm wide strip. Therefore, this part is preserved and the inside of the blade is trimmed. Three small holes are also cut out in the corners. They do not significantly affect the current flow which follows the curvature of the slot and naturally curls in the round corners. This last modification degrades the gain by about 0.5 dB above 175 MHz. Compared with the model \textit{(a)}, the impedance profile remains similar, but its level has been increased by about 30 + 25j $\Omega$. As for the gain, the performance is up to 20\% higher between 80 and 175 MHz.

\subsection{Mechanical design}
\label{sec:2.5.mechaDes}

The surface and the weight are significantly reduced, without compromising the performance. With respect to the model \textit{(a)}, the blade surface is decreased by 53\%, from 2.58 m$^2$ to 1.20 m$^2$. Each blade weighs 7.7 kg, and the whole feed structure 45 kg, thanks to the use of thin plates and flanges perpendicular to the slot. The flanges are only needed on the slot aperture to tune the impedance, but they are continued around the whole blade outline to improve the stiffness. 

\begin{figure}[t]
    \centering
    \includegraphics[width=0.90\linewidth]{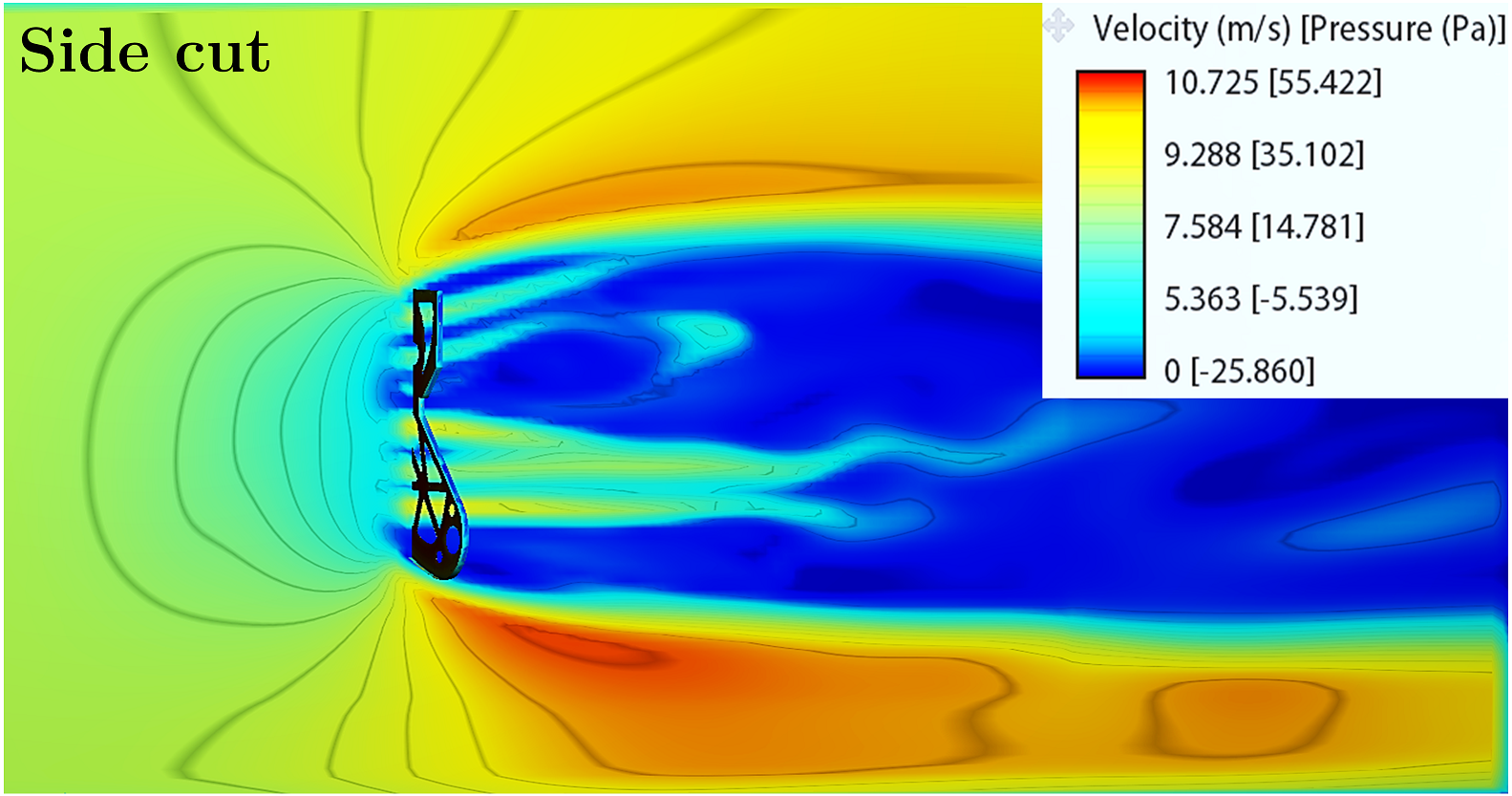}
    \caption{Flow velocity and surface pressure with the Vivaldi feed with holes, and for an incident wind perpendicularly blowing at 30 km/h, side cut.}
    \label{fig9:flowVeloCut}
\end{figure}

\begin{figure}[t]
    \centering
    \includegraphics[width=1.0\linewidth]{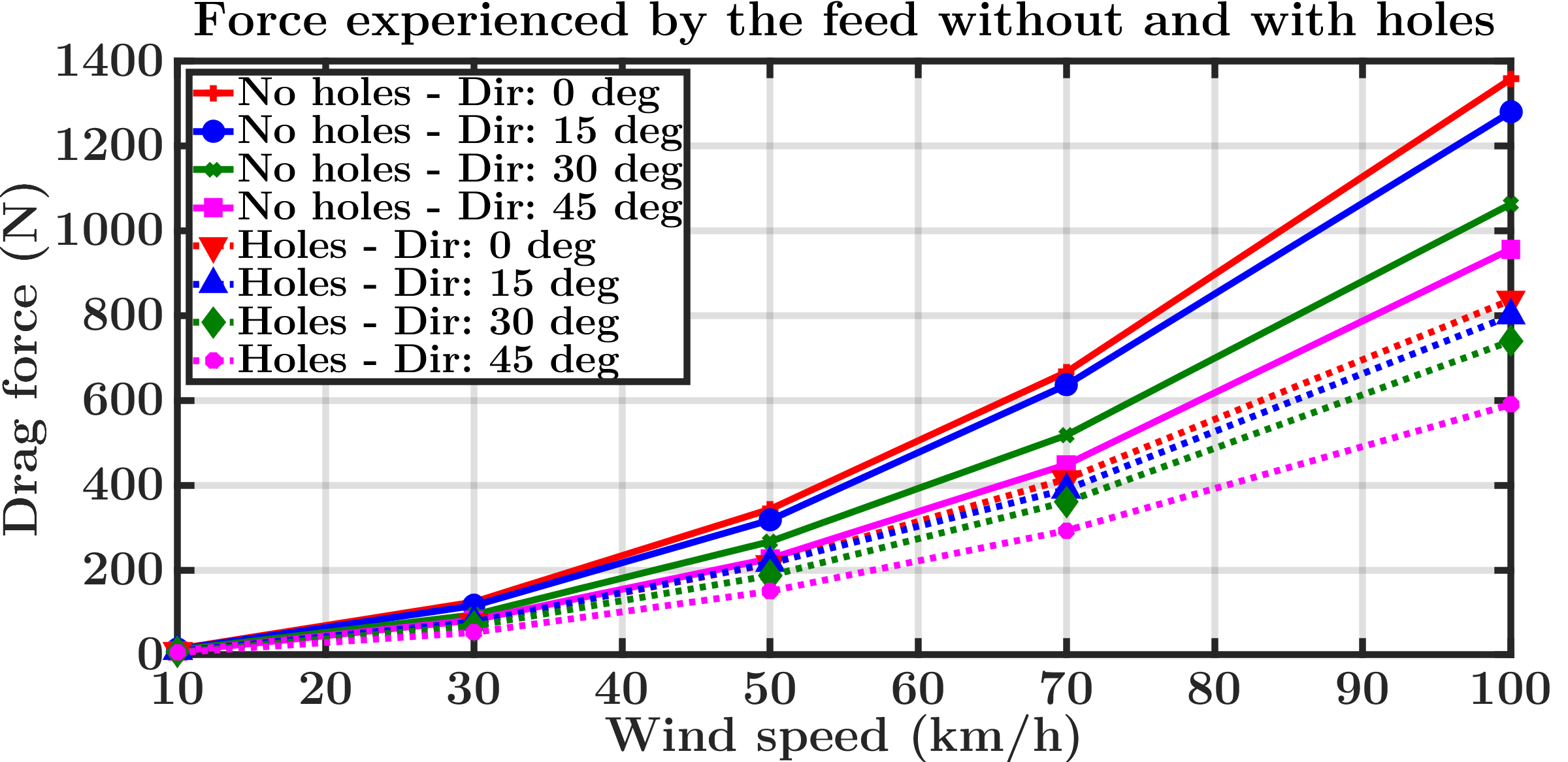}
    \caption{Drag force experienced by the feed without and with holes, as a function of the wind speed and its direction.}
    \label{fig10:dragForce}
\end{figure}

The effect of the cutouts on the wind profile has been analysed by using computational fluid dynamics with Autodesk Flow Design which simulates a wind tunnel. Thanks to this software, the wind flow around the structure (cf. Fig. \ref{fig9:flowVeloCut}), its velocity, the pressure on the blades, and the drag force which quantifies the resistance of the feed to the air flow, are computed. In the simulations, the incident wind speed varies from 10 to 100 km/s and its direction from 0$\degree$ to 45$\degree$, which is enough to extrapolate the performance up to 360$\degree$ due to the symmetry of the feed. Fig. \ref{fig10:dragForce} shows the drag force experienced by the feed without and with holes. Thanks to the reduced surface, this force is decreased by 30\% to 40\%, depending on the wind direction and speed. In the extreme case where the wind blows at 100 km/h, the feed with holes should not experienced a force higher than 850 N, unlike a design without holes where it can exceed 1350 N.


\begin{figure}[H]
    \centering
    \includegraphics[width=0.41\linewidth]{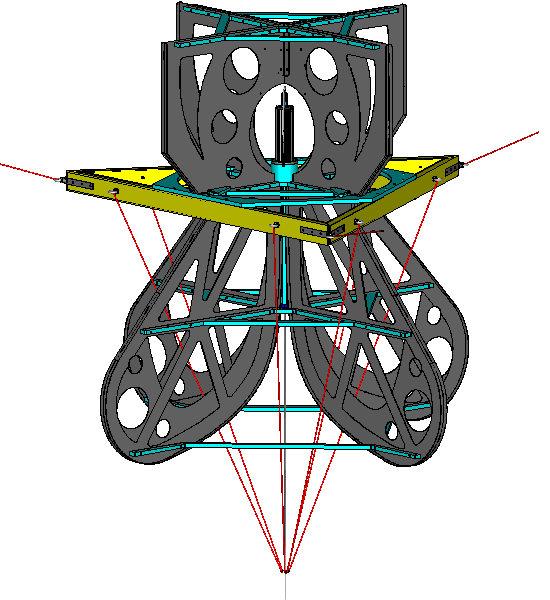}
    \caption{Mechanical model used for the CST simulations. The dielectric structure appears in light blue (UPVC) and in yellow (GRP).}
    \label{fig11:CSTmecha}
\end{figure}

The feed consists of four independent aluminium plates which are perpendicularly held together thanks to a structure made of unplasticised polyvinyl chloride (UPVC) (cf. Fig. \ref{fig11:CSTmecha}). The blades and the UPVC structure are supported by a triangular frame in glass reinforced plastic (GRP). These materials have the advantage to be sturdy, weather-resistant, in particular to hot temperatures and UV-sunlight, durable, relatively light, and cost-effective. They are also quasi-transparent to electromagnetic waves at these frequencies. The electrical properties of the materials are included in the simulations, and the radiation losses are below 0.1 dB. Compared with UPVC, GRP has superior flexural and tensile strengths. This is necessary because the triangular frame supports the weight of the feed and the tension from the three Kevlar ropes which are used to suspend it. The ropes are linked to pulleys placed at the top of the wooden poles (cf. Fig. \ref{fig1:HERAarray}), and the feed can be moved up and down thanks to crank handles. Reflective surfaces are added to the structure to position the feed by using a laser system. The feed can be positioned with a precision of 2 cm in the XY-plan, and 1 cm for the height \cite{Dynes2019}.

%

%%%%%%%%%%%%%%%%% SECTION III %%%%%%%%%%%%%%%%%%

\section{RF receiver and antenna feeding}
\label{sec:3.RecFeeding}

\subsection{RF analogue receiver}
\label{sec:3.1.RFrec}

The analogue receiver and the antenna were developed together to optimise the performance. The receiver is a radio-over-fibre system which consists of a front-end module connected to a post-amplifier module (PAM) via a 500-m optical fibre. The PAM is connected to the correlator for analogue-to-digital conversion and signal processing. When connected to the antenna, the receiver provides a total power gain which smoothly increases from 70 dB at 50 MHz to 88 dB at 220 MHz, before dropping to 70 dB again at 250 MHz, due to the bandpass filter of the system. Its roll-off is relatively steep and the gain goes below 20 dB above 280 MHz and below 30 MHz. The receiver also provides a 85-dB reverse isolation. 

Inside the FEM, a balun transforms the differential signals coming from the antenna pins into a unbalanced signal. However, the balun is not ideal and imbalance between the two input ports may cause the propagation of undesired common mode signals. The amplitude imbalance is below 0.3 dB, and the phase imbalance varies between 180$\degree$ and 178$\degree$. Thus, the common mode rejection ratio is above 30 dB. In addition, the differential to common mode conversion parameter of the measured antenna does not exceed -25 dB. Therefore, the propagation of common mode signals in the RF chain is negligible, and we assume that the FEM is only excited by differential mode signals in the rest of this analysis. 

The receiver noise temperature, taking into account the impedance noise mismatch with the antenna, is calculated by combining the measured antenna impedance with the measured noise parameters of the receiver. It is 555 K at 50 MHz, 88 K at 100 MHz, 77 K at 150 MHz, 58 K at 200 MHz, and 60 K at 250 MHz. This is low enough to consider that the system noise is dominated by the sky. As described in \cite{Hayward2012}, the frequency noise parameters (minimum noise factor, equivalent noise resistance, and optimum source impedance) are measured by using a Keysight PNA-X, thanks to its internal tuner which varies the source impedance of the connected receiver. Its noise is mainly affected by the first amplifier which is a PGA-103+ from Mini-Circuit \cite{Mini-Circuits}.

\begin{figure}[H]
    \centering
    \includegraphics[width=0.57\linewidth]{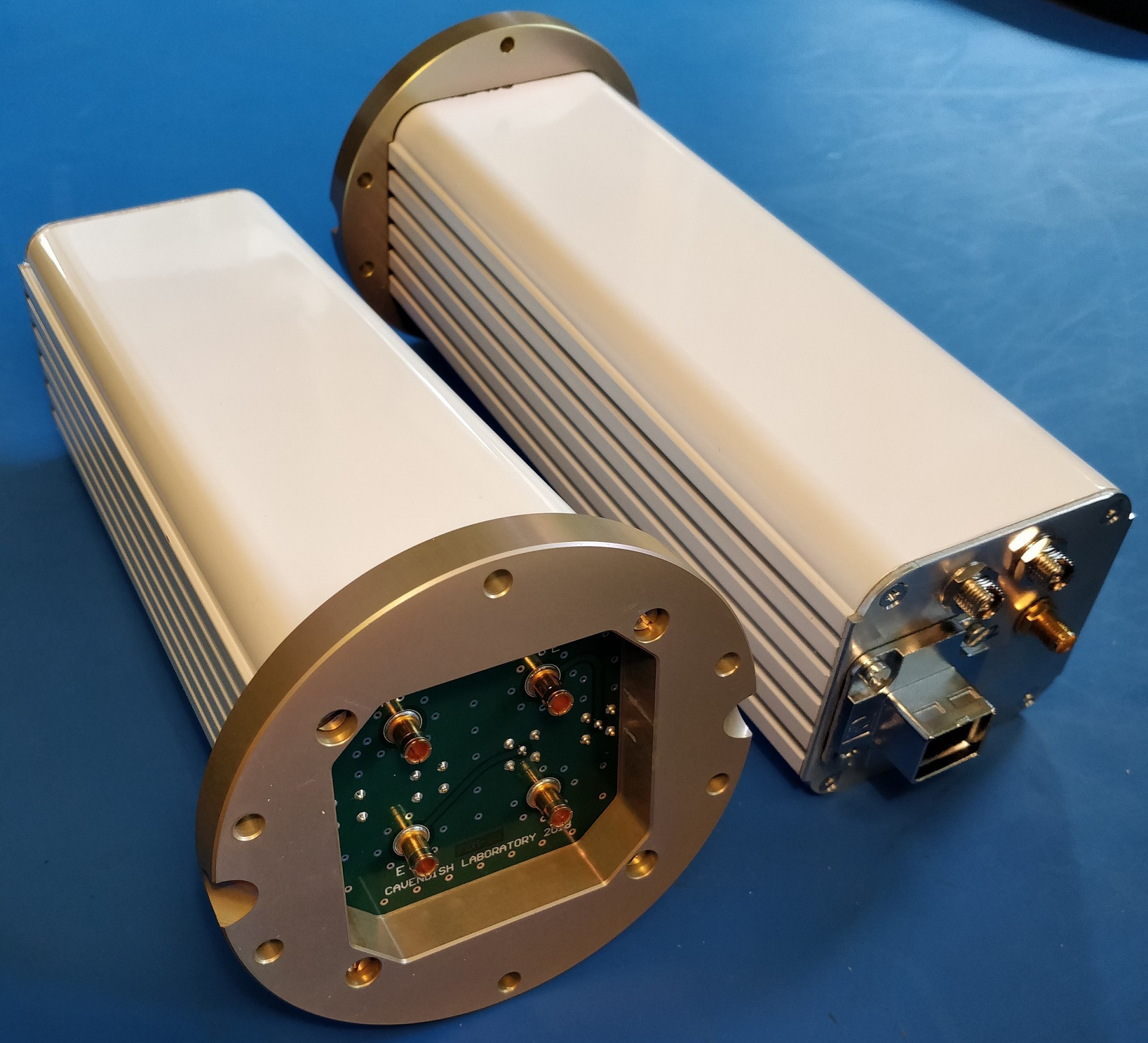}
    \caption{Picture of the front-end module (FEM).}
    \label{fig12:FEM}
\end{figure}

\subsection{Antenna feeding}
\label{sec:3.2.feeding}

\begin{figure}[H]
    \centering
    \includegraphics[width=0.57\linewidth]{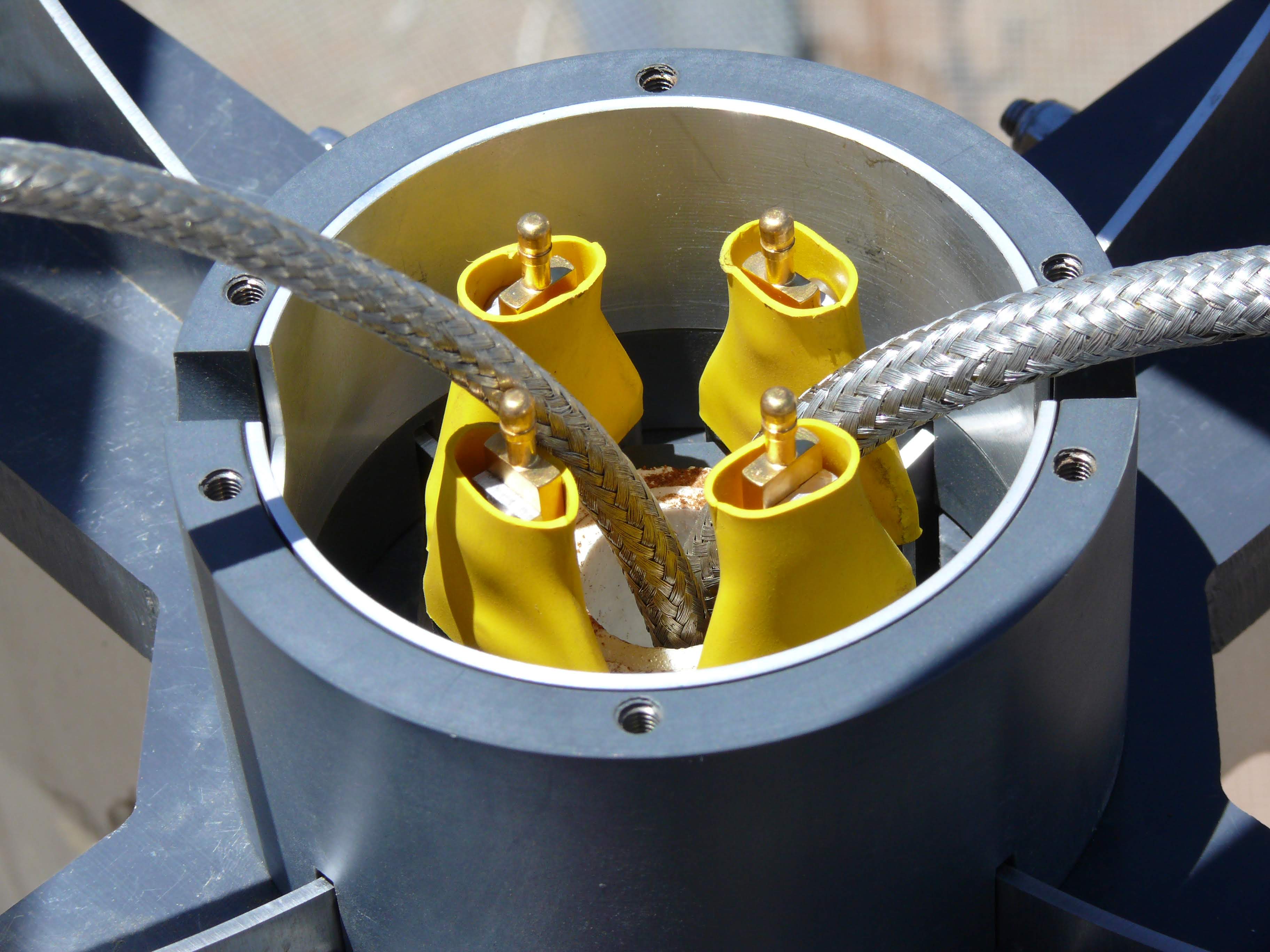}
    \caption{Picture of the balanced feed point of the antenna. The shielded cables pass between the flanges which are insulated to avoid electrical contacts.}
    \label{fig13:feedArea}
\end{figure}

The balanced feed is connected to the differential amplifiers of the FEM via four pins, two for each polarisation, as shown in Fig. \ref{fig12:FEM} and \ref{fig13:feedArea}. The FEM is positioned inside the cavity of the feed (cf. Fig. \ref{fig14:assembly}), and simulations show that the electromagnetic properties of the antenna, in particular its impedance and the radiation pattern, are not significantly affected. Thanks to this configuration, the distance between the amplifiers and the antenna feed point is extremely short. Thus, the noise caused by the ohmic losses from the feed line and entering the FEM is negligible. Moreover, the impedance matching is facilitated because the input impedance "seen" by the FEM is exactly the antenna impedance.

\subsection{Cabling}
\label{sec:3.3.cabling}

\begin{figure}[H]
    \centering
    \includegraphics[width=0.57\linewidth]{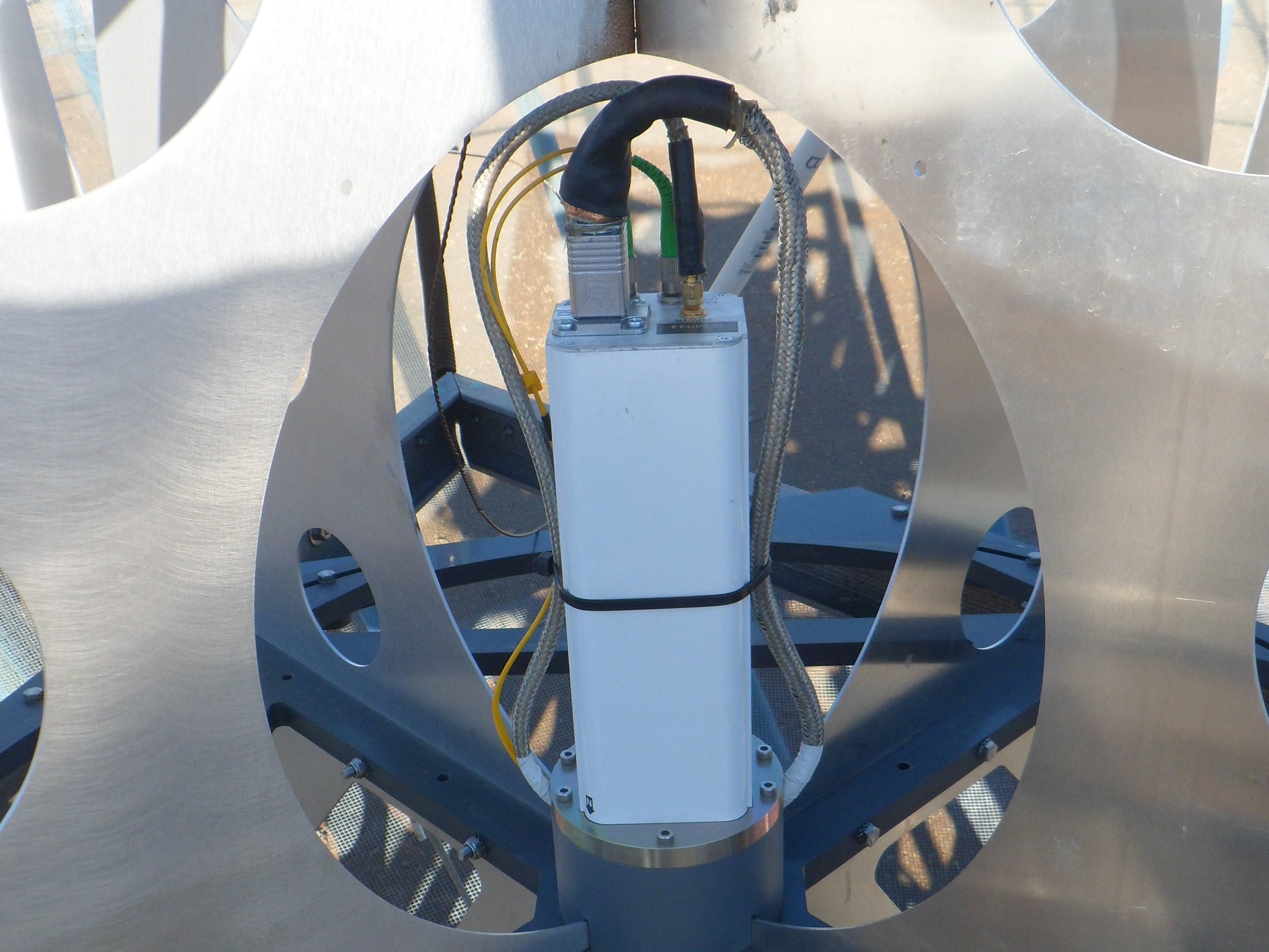}
    \caption{Picture of the antenna, FEM, and cable assembly. The SMA power cable and the CAT7 control cable are shielded with metal braided sleeves which must remain electrically in contact.}
    \label{fig14:assembly}
\end{figure}

\begin{figure}[t]
    \centering
    \includegraphics[width=0.83\linewidth]{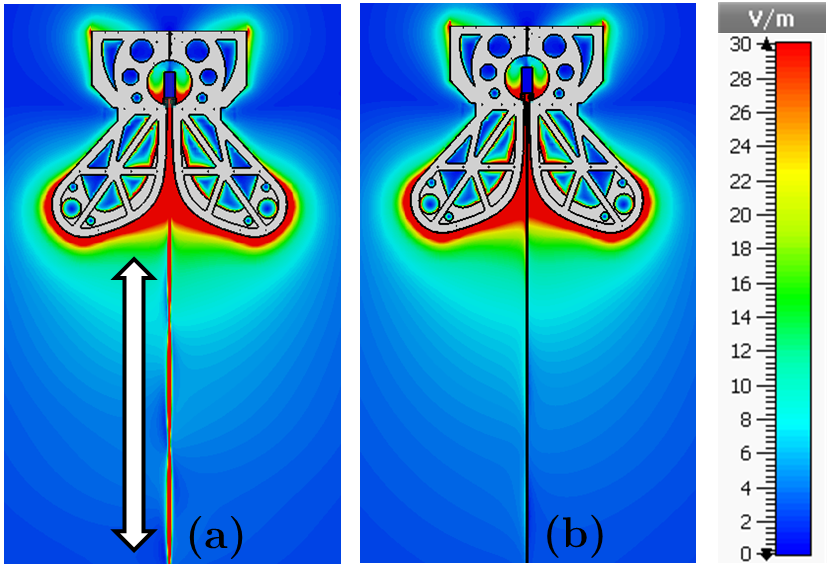}
    \caption{Simulations of the electric field (in V/m) radiated by the feed at 150 MHz. In picture \textit{(a)}, the two coaxial cables passing through the aperture are not shielded, and so the E-field can propagate along. In picture \textit{(b)}, the cables are inserted into metal braided sleeves in contact, which prevents the formation of a parallel transmission line and the presence of standing waves.}
    \label{fig15:cableSimu}
\end{figure}

Fig. \ref{fig14:assembly} shows how the FEM is connected to the antenna. Four cables exit the module: two optical fibres which transport the two polarisations, a CAT7 cable with a shielded RJ45 connector which is used to send and receive control data, and a coaxial cable with a SMA connector used to power the unit. The management of the two coaxial cables must be performed with care, since they are metallic conductors. The main challenge is to connect these cables from the FEM to the node without affecting the symmetry of the beam and the system response. The solution is to pass these cables along the symmetry axis of the feed, in other words exactly in the middle of the blades as shown in Fig. \ref{fig15:cableSimu}. Asymmetric cable configurations cause frequency and spatial structures on the beam pattern, such as ripples, which could mask the EoR signal. The cables go down as close as possible from the FEM, pass between the pins and the flanges via a notch, before being tightly guided in a dielectric conduit through the middle of the slot. After reaching the vertex, the cables pass through the central hole and then go under the dish. If the cables are correctly aligned, detailed electromagnetic simulations show that the radiated beam is not affected (cf. Fig. \ref{fig19:gainBeam}). The effects of the cable misalignment are studied in Section \ref{sec:4.6.cableMisalign}.

\begin{figure}[t]
    \centering
    \includegraphics[width=1.0\linewidth]{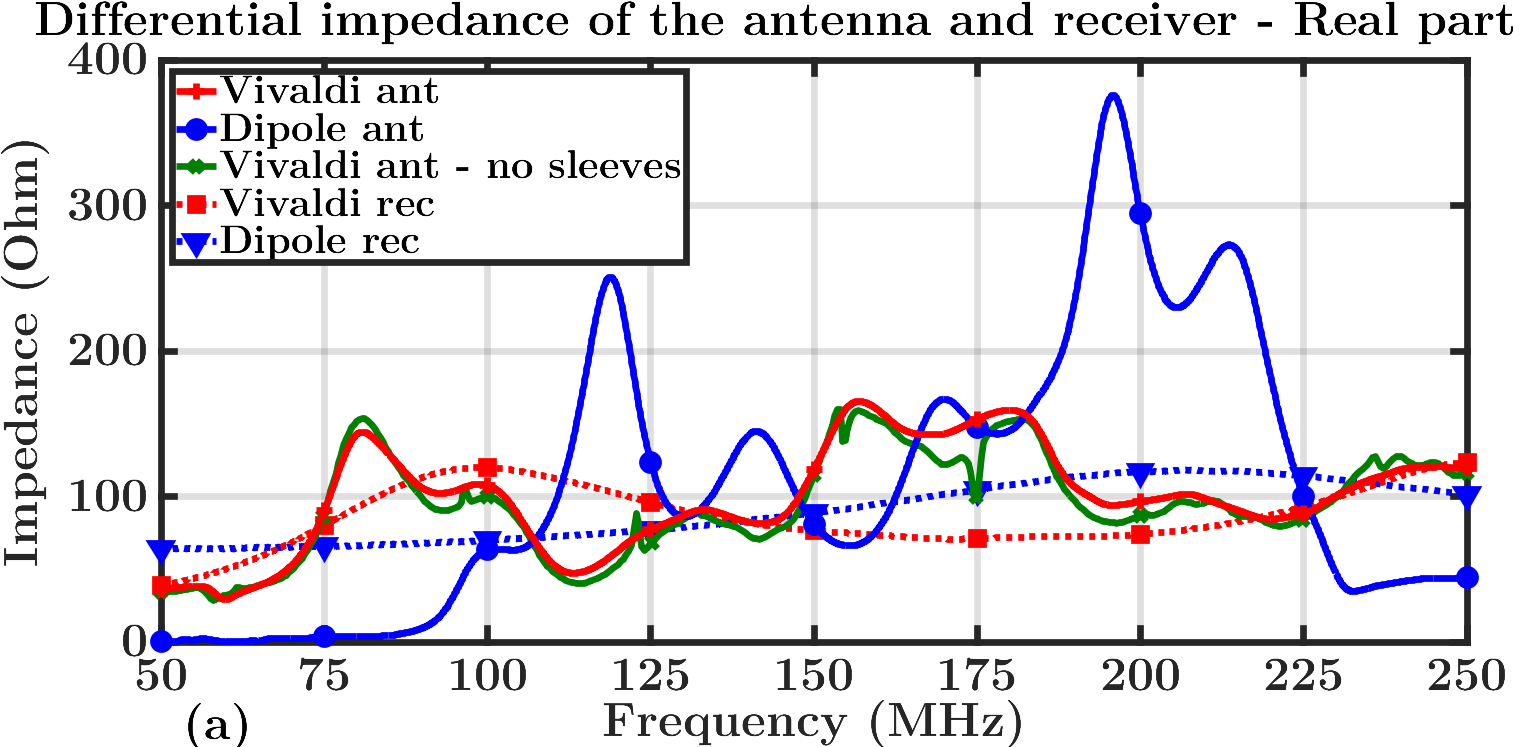}
    
    \vspace{2.7mm}
    
    \includegraphics[width=1.0\linewidth]{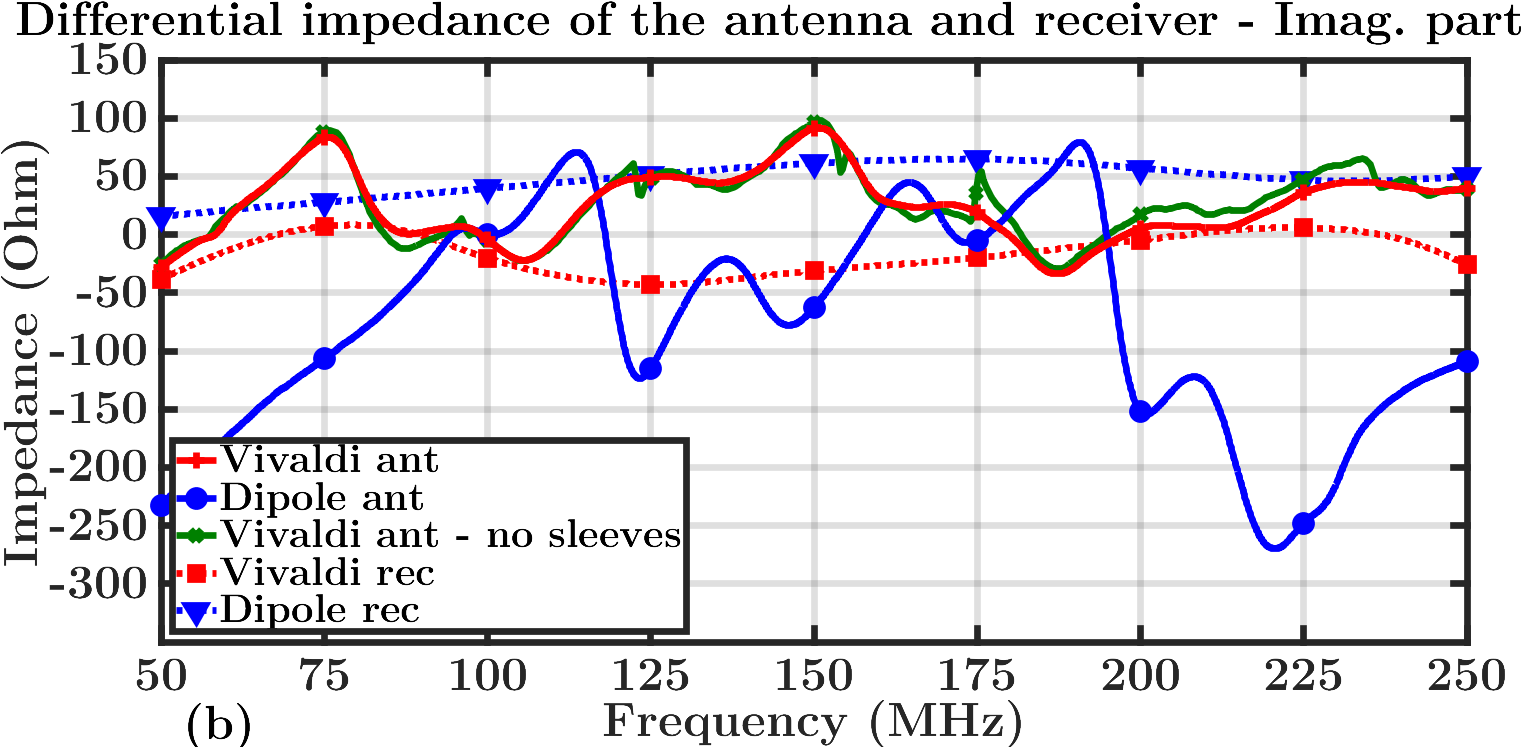}
    \caption{Measured antenna and receiver impedances. The measured impedance of the Vivaldi antenna does agree well with the simulation of the model \textit{(f)} in Fig. \ref{fig7:impEvol}. Note also the presence of spikes with the Vivaldi feed when the cables are not shielded by the sleeves.}
    \label{fig13:systImp}
\end{figure}

However, measurements performed with a vector network analyser (VNA) revealed the presence of periodic spikes in the reflectometry data. They are visible in the antenna impedance (cf. green curve in Fig. \ref{fig13:systImp}) and are the consequence of reflections coming from the cables below the feed. Converted into the time domain, these spectral features would jeopardise the EoR detection. The outer conductors of the two cables create a parallel transmission line along which the radiated E-field propagates. The signal is then reflected at the end of this line and standing waves form. The problem was recreated in simulations, and the propagation of the E-field between these cables can be visualised in Fig. \ref{fig15:cableSimu} \textit{(a)}. To avoid that, each coaxial cable is inserted into a long metal braided sleeve (cf. Fig. \ref{fig13:feedArea} and \ref{fig14:assembly}). The two metal sleeves are then tied together so that they remain in electrical contact from beginning to end. The two cables are shielded and the absence of gap between the sleeves prevents the E-field from propagating, as shown in Fig. \ref{fig15:cableSimu} \textit{(b)}. The spikes in the antenna impedance do disappear after the addition of the sleeves (cf. red curve in Fig. \ref{fig13:systImp}). Despite the proximity of the cables with the blades, this configuration does not impact too much the inter-port isolation between the X and Y-polarisations, which remains above 40 dB as long as the cables are accurately centred. Without cable, the isolation is above 50 dB (cf. Fig. \ref{fig30:Spara_10_1000MHz}).

%

%%%%%%%%%%%%%%%%% SECTION IV %%%%%%%%%%%%%%%%%%

\section{Figures of merit and performance}
\label{sec:4.Perf}

The performance of the system is assessed by simultaneously taking into account the electromagnetic properties of the antenna including the dish, along with the electrical properties of the receiver. The beams are simulated with CST, whereas the reflectometry data of the antenna and of the receiver are obtained with measurements performed with a VNA. The performance of the dipole system is also given for comparison.

\subsection{Radiation pattern}
\label{sec:4.1.RadPar}

\begin{figure}[H]
    \centering
    \includegraphics[width=1.0\linewidth]{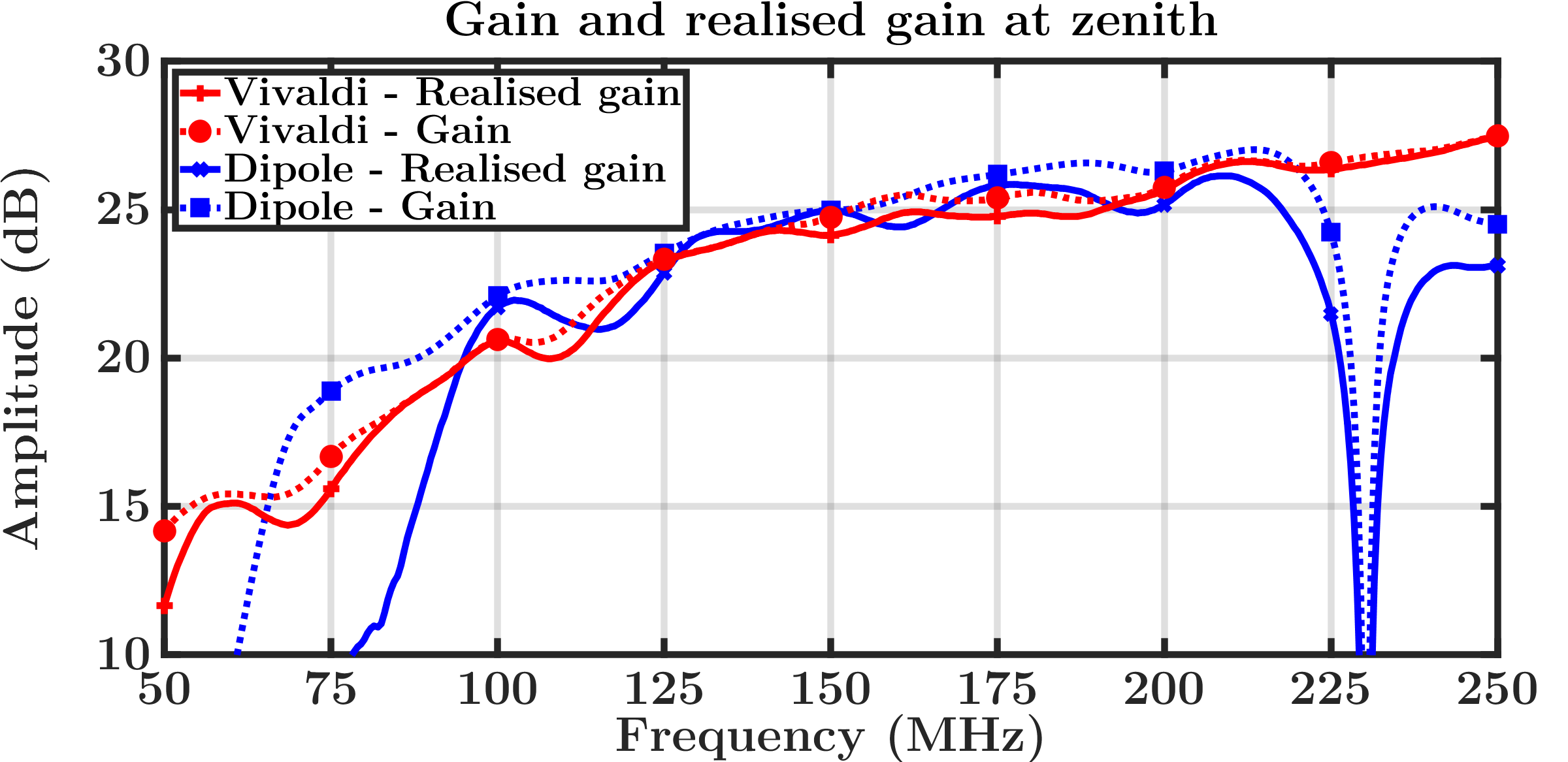}
    \caption{Gain and realised gains at zenith.}
    \label{fig17:realGain}
\end{figure}

\begin{figure}[H]
    \centering
    \includegraphics[width=1.0\linewidth]{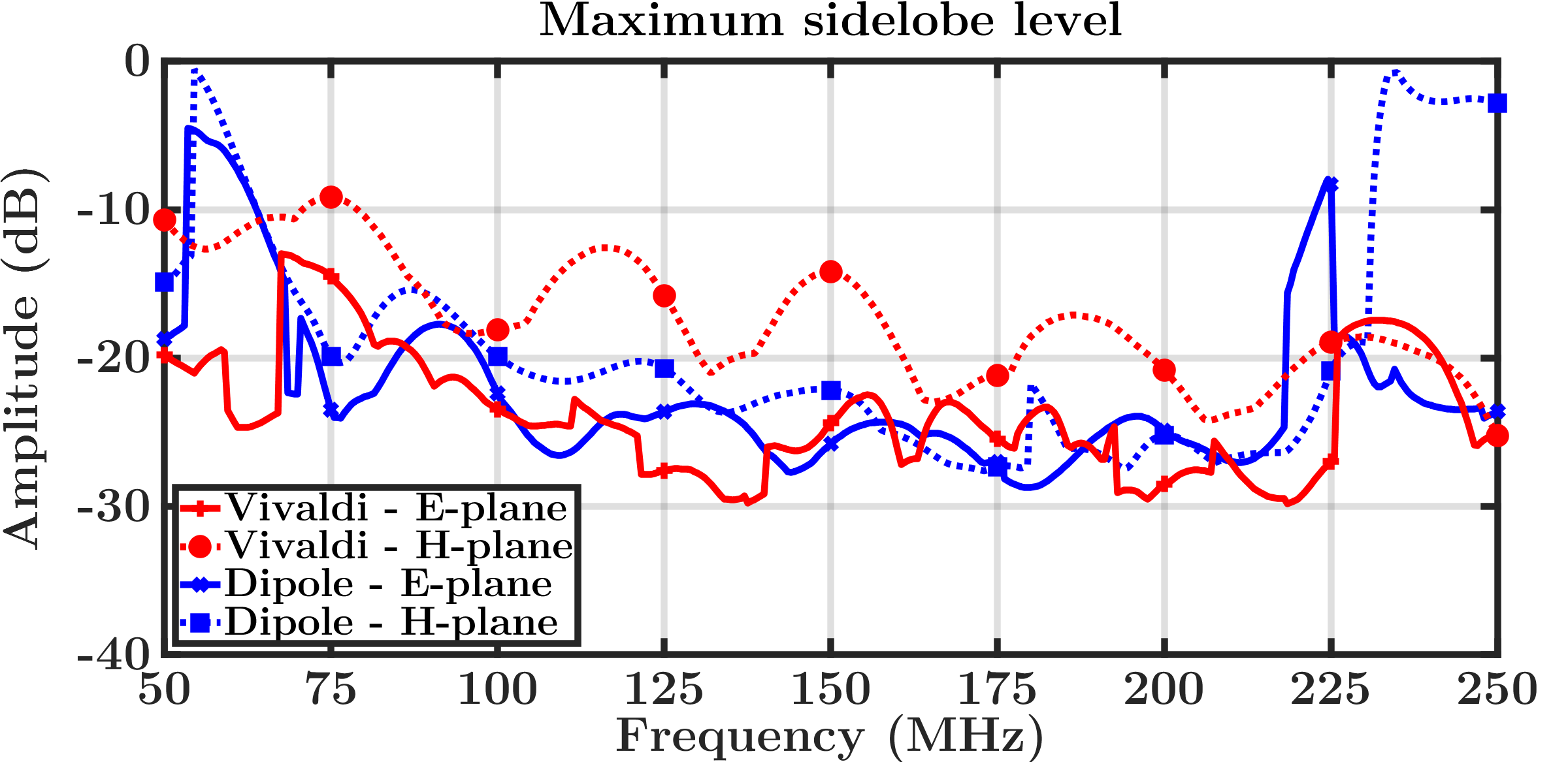}
    \caption{Maximum sidelobe level for an angle from zenith $\theta$ between -90$\degree$ and 90$\degree$, in the E and H-planes.} 
    \label{fig18:maxSLL}
\end{figure}

\begin{figure}[H]
    \centering
    \includegraphics[width=1.0\linewidth]{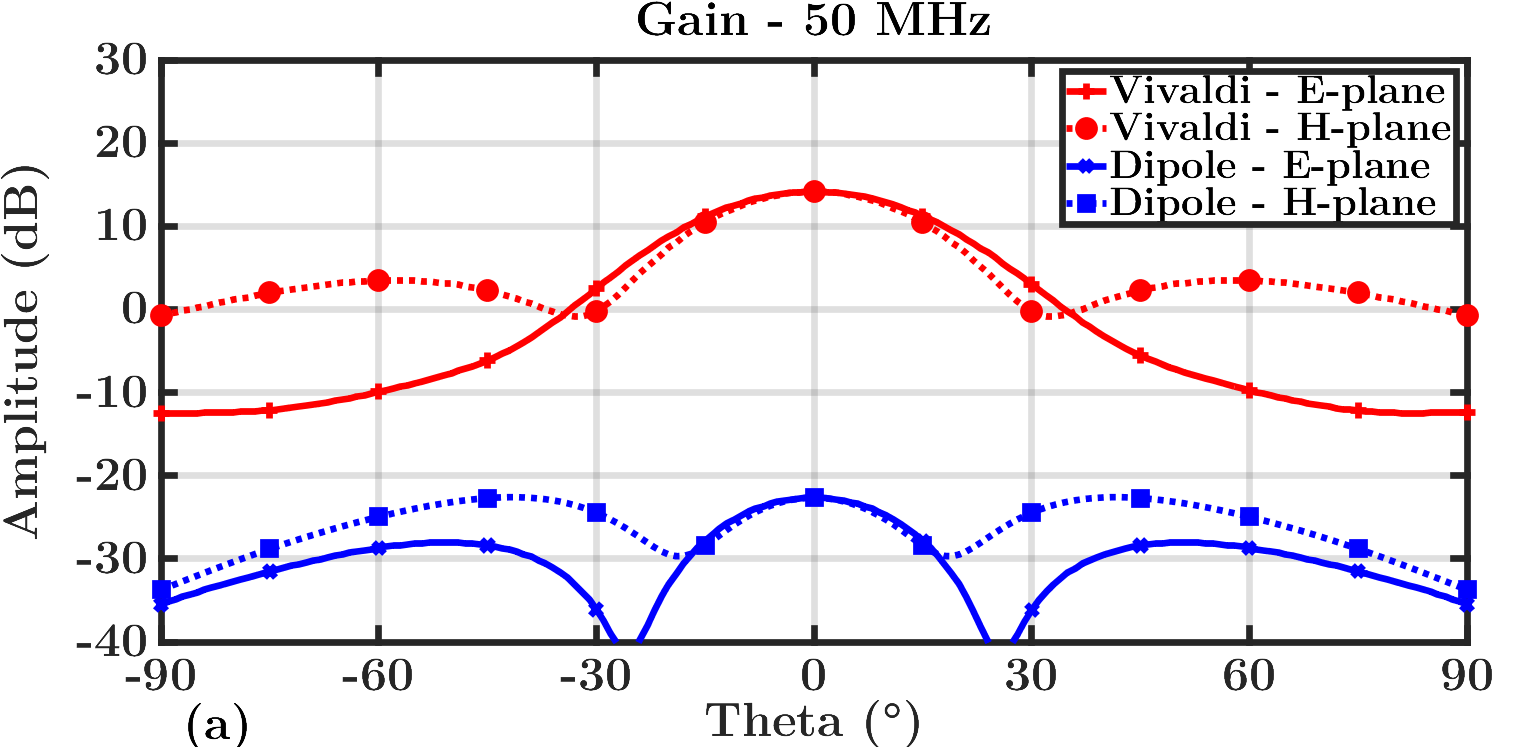}

    \vspace{3mm}
    
    \includegraphics[width=1.0\linewidth]{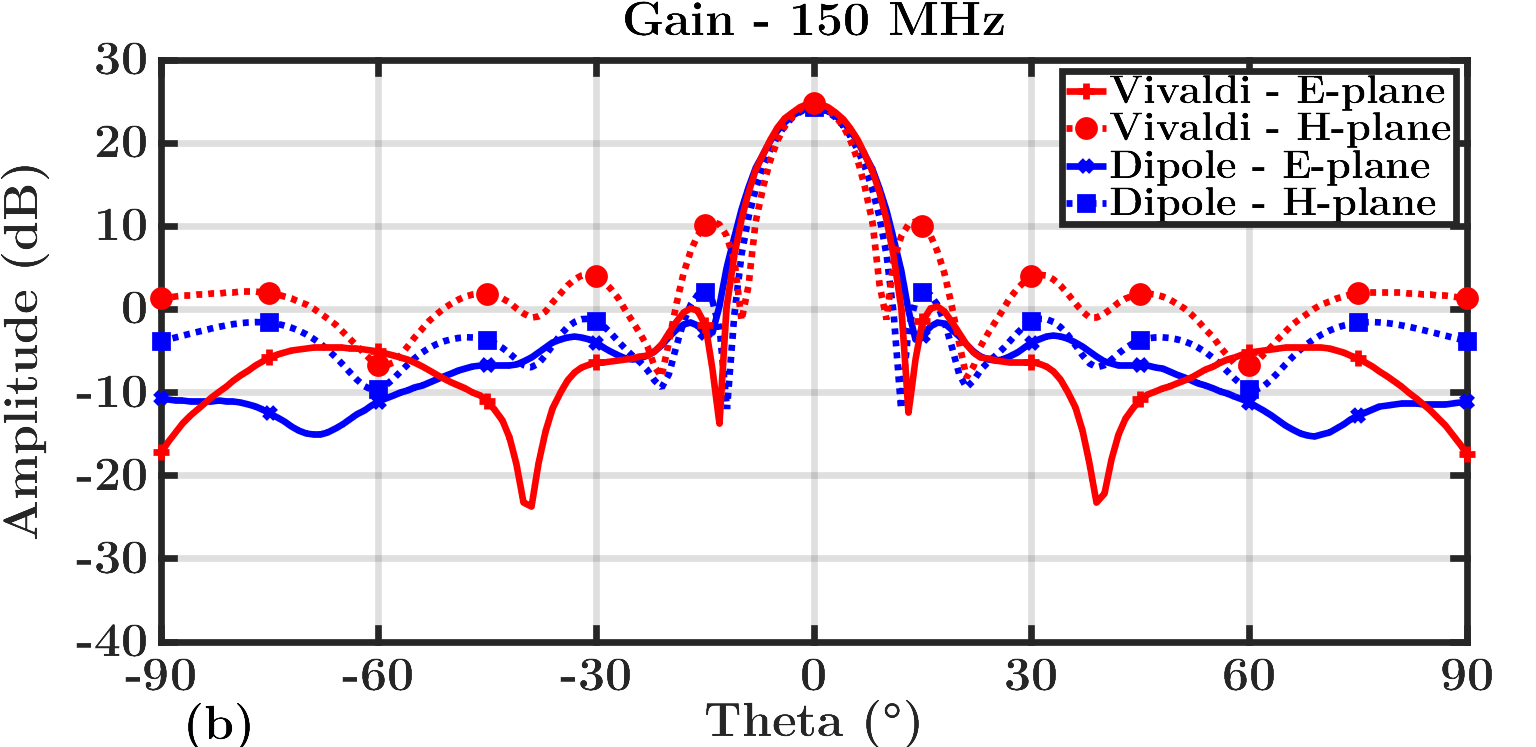}

    \vspace{3mm}

    \includegraphics[width=1.0\linewidth]{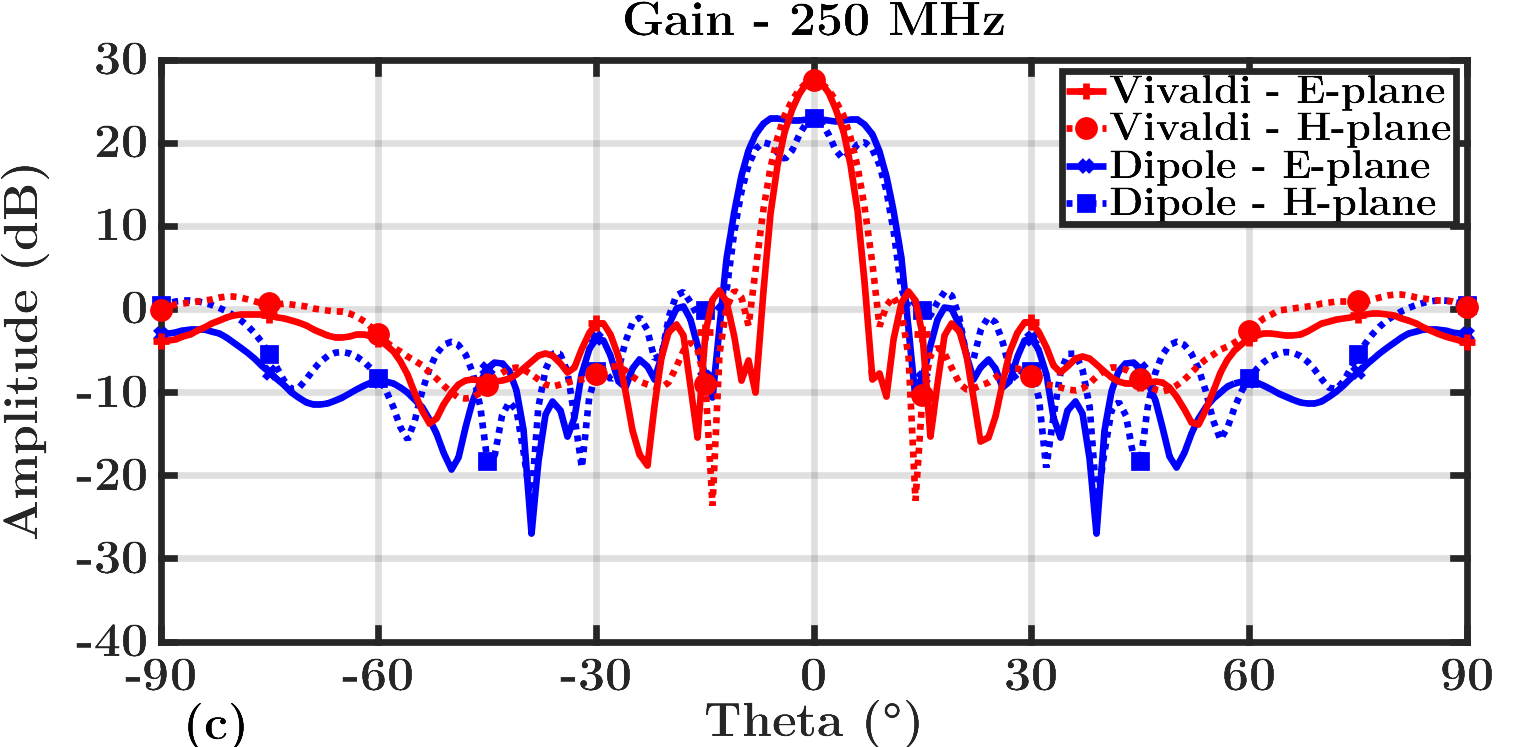}
    
    \vspace{3mm}
    
    \includegraphics[width=0.65\linewidth]{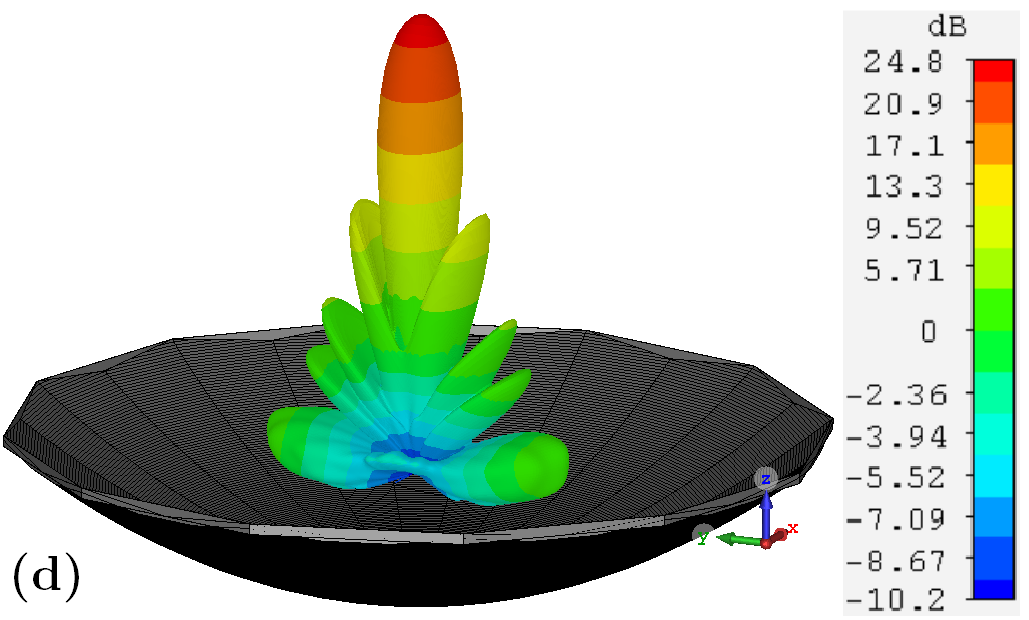}
    \caption{Cuts of the simulated antenna gain in the E and H-planes, at 50, 150, and 250 MHz, and 3D gain at 150 MHz for the X-polarisation.}
    \label{fig19:gainBeam}
\end{figure}

Fig. \ref{fig17:realGain}, \ref{fig18:maxSLL}, and \ref{fig19:gainBeam} present the characteristics of the simulated radiation pattern. Despite the cables, it is symmetric in the E and H-planes and smooth. This point is important since its spatial and frequency fluctuations directly affect the time response. The gain at zenith with the Vivaldi feed progressively increases from 14 dB at 50 MHz up to 28 dB at 250 MHz. The comparison between the gain and the realised gain shows that the losses caused by the impedance mismatch with the receiver are below 1 dB, except around 50 MHz. On average, the mismatch losses are even below 0.5 dB, which corresponds to a return loss above 9.6 dB (cf. Fig. \ref{fig30:Spara_10_1000MHz}). This is better than with the dipole and counterbalances the spillover losses. In the case of the dipole, the presence of a cage allows the dish illumination to be better controlled between 100 and 200 MHz, but causes the beam to fork above 230 MHz and complicates the impedance matching (cf. Fig. \ref{fig13:systImp}). Overall, with the new feed the realised gain is higher at low and high frequencies, and has a similar level at mid frequencies. 

Due to the planar configuration of the blades and the absence of a cage, the main beam is not perfectly circular. The maximum sidelobe level is around -17 dB in the H-plane, and -25 dB in the E-plane. Compared with the dipole, the level is similar in the E-plane, and 5 to 10 dB higher in the H-plane.


\subsection{Efficiencies}
\label{sec:4.2.efficiencies}

\begin{figure}[H]
    \centering
    \includegraphics[width=1.0\linewidth]{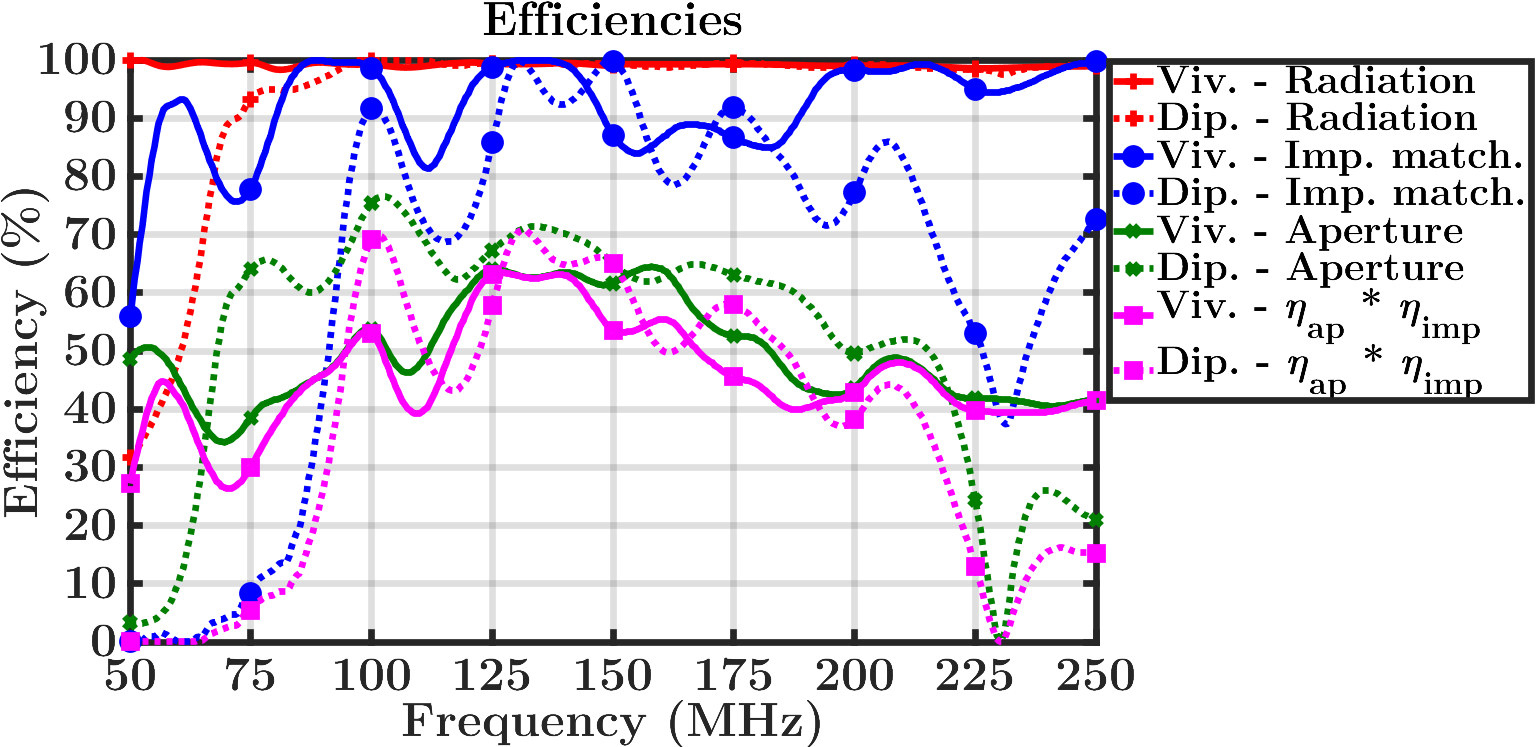}
    \caption{Radiation $\eta_{rad}$, impedance mismatch $\eta_{imp}$, aperture $\eta_{ap}$, and total $\eta_{ap}$*$\eta_{imp}$ efficiencies.}
    \label{fig20:efficiency}
\end{figure}

Fig. \ref{fig20:efficiency} presents different efficiency factors which describe how the power of the incident signal is lost at the antenna level. The ohmic and dielectric losses are quantified by the radiation efficiency $\eta_{rad}$. Despite the dielectric structure of the Vivaldi feed, these losses are negligible ($\eta_{rad}\approx 99\%$). This parameter also affects the antenna sensitivity (cf. Equation \eqref{eq6:sensitivity}). The impedance mismatch factor $\eta_{imp}$ quantifies the matching with the receiver in terms of power signal. It oscillates around $90\%$ for the Vivaldi feed. The effective collecting area is described by the aperture efficiency $\eta_{ap}$, which includes the effects of the dish illumination as well as $\eta_{rad}$. By combining these losses, we can see that about half of the incident power is lost, mainly because of the dish illumination and the impedance mismatch

\subsection{Frequency response of the system}
\label{sec:4.3.freqResp}

\begin{figure}[t]
    \centering
    \includegraphics[width=1.0\linewidth]{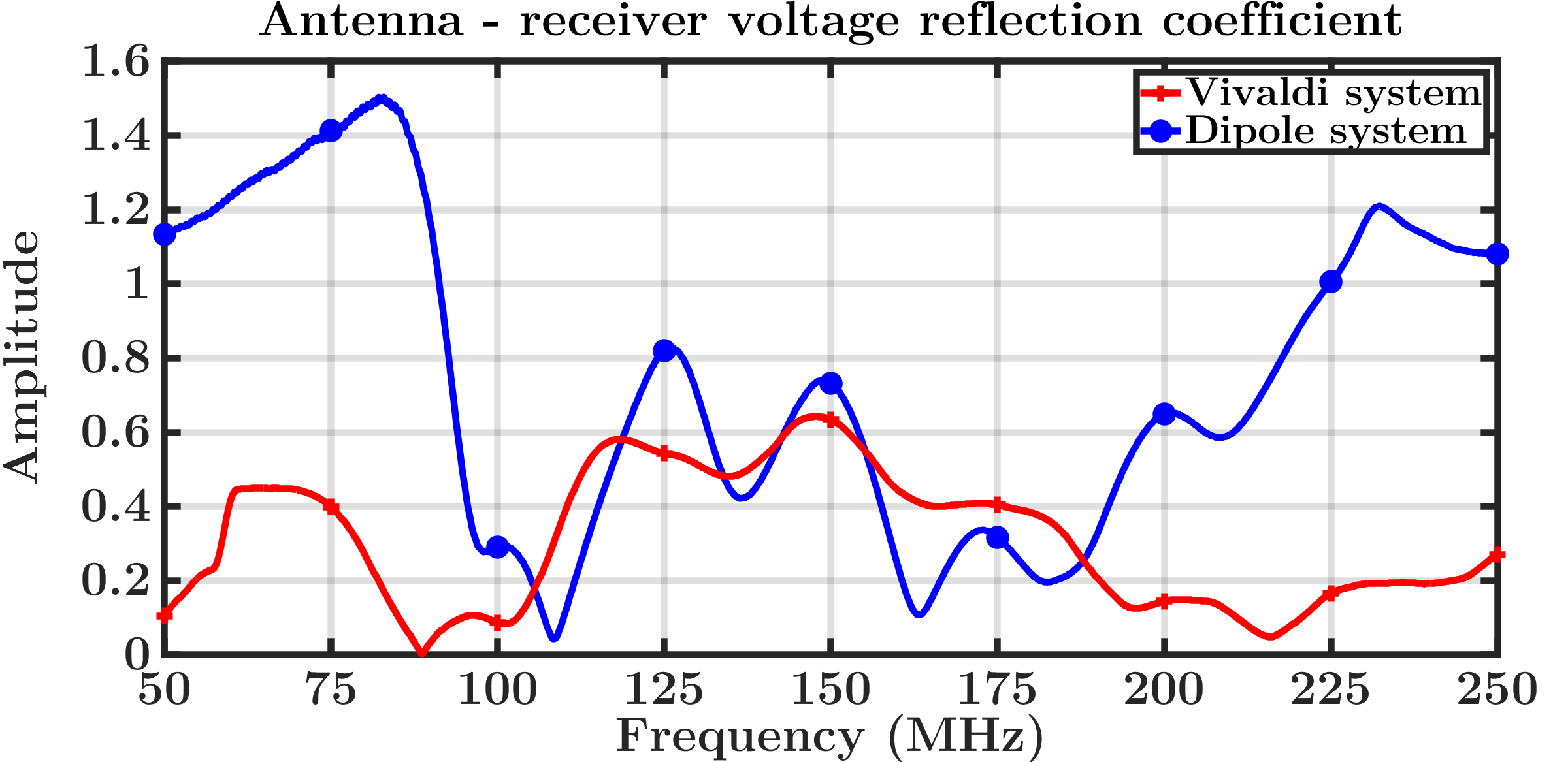}
    \caption{Magnitude of the measured voltage reflection coefficient $\Gamma$ at the interface between the antenna and the FEM input.}
    \label{fig21:voltRXcoef}
\end{figure}

The success of this experiment relies on the smoothness of the system voltage response. In particular, a fraction of the received signal is reflected multiple times between the feed and the dish, because of impedance mismatch with the receiver. The quality of the matching at this interface is assessed with the voltage reflection coefficient $\Gamma$ presented in Fig. \ref{fig21:voltRXcoef}. For each system, $\Gamma$ is calculated from the impedances of the antenna $Z_{\rm ant}$ and related receiver $Z_{\rm rec}$ given in Fig. \ref{fig13:systImp} \cite{Rahola2008}:
\begin{align}
    \Gamma  = \frac{{{Z_{\rm rec}} - {Z_{\rm ant}}}}{{{Z_{\rm rec}} + {Z_{\rm ant}}}}.
	\label{eq3:VolRxCoef}
\end{align}
Thanks to the wideband properties of the Vivaldi feed and the absence of a cage, the antenna impedance is much flatter. Therefore, the reflections at this interface can be mitigated in comparison with the dipole. One should not be surprised to see $\Gamma$ with a magnitude superior to 1. This can occur with complex impedances and on condition that ${Re[Z_{\rm rec}}]Re[{Z_{\rm ant}}] + Im[{Z_{\rm rec}}]Im[{Z_{\rm ant}}] < 0$ \cite{Vernon1969}. However, the reflected power is smaller than the incident power (cf. Fig. \ref{fig30:Spara_10_1000MHz}).

The propagation of the signal through the receiver can also be affected by reflections, in particular within the transmission cable. The method used to compute the system response is detailed in \cite{Fagnoni2021}. The antenna is excited by an electromagnetic plane wave $\boldsymbol{{E_{\rm in}}} \left( {\theta ,\;\varphi } \right)$, coming with an angle of arrival defined by the spherical angles $\theta$ and $\varphi$, and generates a voltage ${V_{\rm out}}\left( {\theta ,\;\varphi } \right)$ at the PAM output. Thus, the frequency voltage response of the system $\boldsymbol{H}\left( {\theta ,\;\varphi } \right)$ is defined in reception by:
\begin{align}
    {V_{\rm out}}\left( {\theta ,\;\varphi } \right) = \boldsymbol{H}\left( {\theta ,\;\varphi } \right)\boldsymbol{\cdot}\boldsymbol{{E_{\rm in}}} \left( {\theta ,\;\varphi } \right).
	\label{eq4:transFunc}
\end{align}
The system voltage response depends on the antenna E-farfield pattern ${\boldsymbol{{E_{\rm pat}}}} \left( {\theta ,\varphi } \right)$, and on the reflection and transmission properties of the receiver which can be described by a 2-port network and its 2x2 impedance matrix $Z_r$. Is is equal to:  
\begin{align}
    &\boldsymbol{H}\left( {\theta ,\;\varphi } \right) =  - \boldsymbol{{E_{\rm pat}}} \left( {\theta ,\varphi } \right)\left[ {\frac{{4\pi }}{{k{Z_{\rm fs}}}}\mathrm{j}} \right] \nonumber \\
    &\left[ {\frac{{{Z_{\rm ant}} + {Z_{\rm 0}}}}{{2{a_{\rm in}}\sqrt {{Z_{\rm 0}}} }}} \right]\left[ {\frac{{{Z_{\rm L}}{Z_{\rm r21}}}}{{\left( {{Z_{\rm L}} + {Z_{\rm r22}}} \right)\left( {{Z_{\rm ant}} + {Z_{\rm r11}}} \right) - {Z_{\rm r21}}{Z_{\rm r12}}}}} \right],
	\label{eq5:VolSystResp}
\end{align}
with $k$ the wavenumber, $Z_{\rm fs}$ the impedance of free space, $Z_{\rm L}$ the termination impedance of the receiver equal to 50 $\Omega$, and $Z_{\rm 0}$ the antenna termination impedance equal to 100 $\Omega$ and which is used in CST to excite the model with a incident travelling wave $a_{\rm in}$ and simulate the beam.

\begin{figure}[H]
    \centering
    \includegraphics[width=1.0\linewidth]{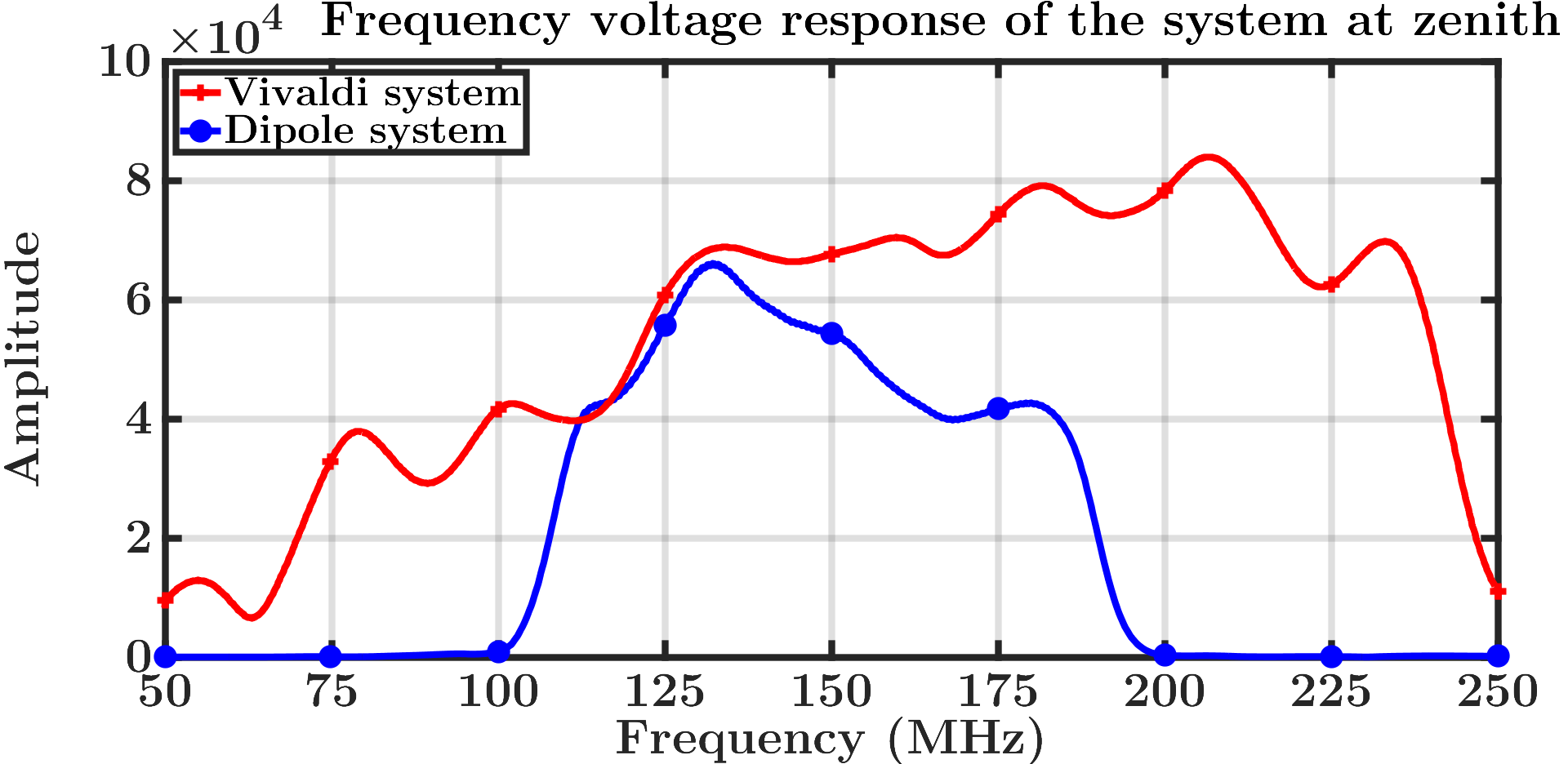}
    \caption{Magnitude of the frequency response of the antenna - receiver system, at zenith for the co-polarisation.}
    \label{fig22:freqRespAntRec}
\end{figure}

Fig. \ref{fig22:freqRespAntRec} presents the system frequency response at zenith for the co-polarisation. The input signal is amplified by the FEM and the PAM, and filtered. The periodic oscillations are associated with reflections occurring at a distance equivalent to the length between the feed and the vertex.

\subsection{Time response of the system}
\label{sec:4.4.timeResp}

\begin{figure}[t]
    \centering
    \includegraphics[width=1.0\linewidth]{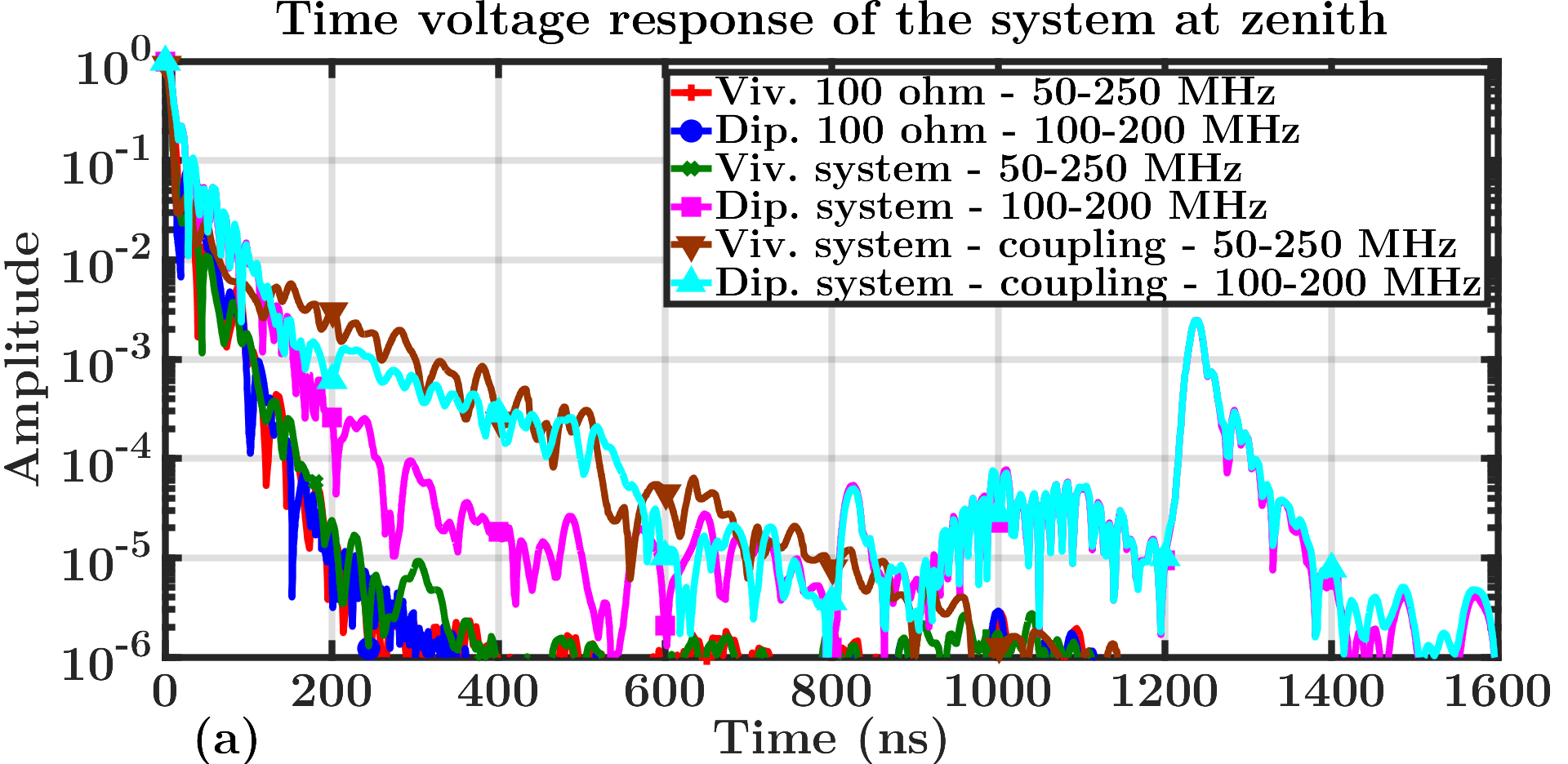}
    
    \vspace{3mm}
    
    \includegraphics[width=1.0\linewidth]{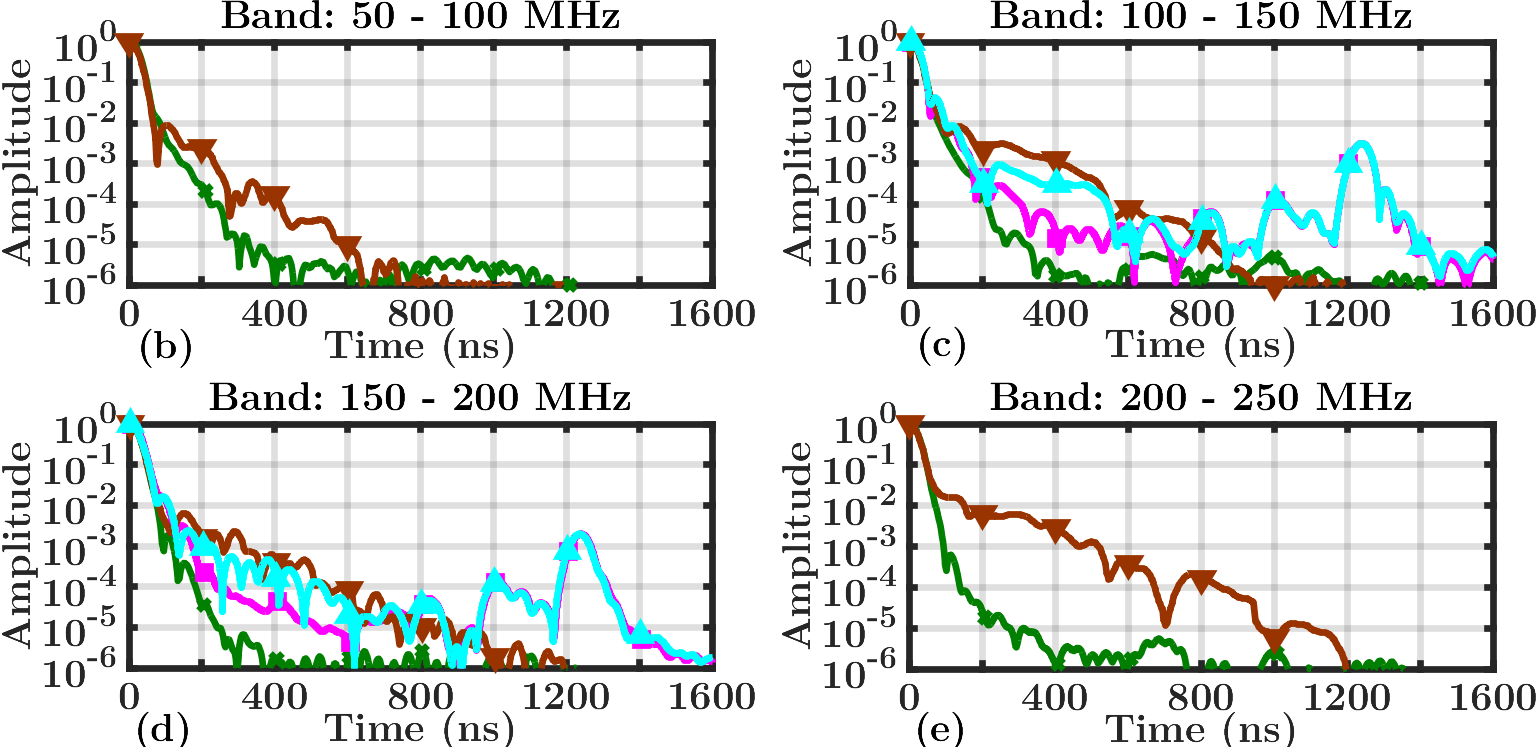}
    \caption{Time response of the antenna - receiver system, at zenith for the co-polarisation. It is windowed by a Blackman-Harris window. The response is also given when the antenna is terminated by 100 $\Omega$. The mutual coupling is simulated in a horizontal strip of 11 x 3 dishes. The top plot is obtained over the entire bandwidth, and the bottom plots over 50-MHz sub-bands.}
    \label{fig23:timeRespAntRec}
\end{figure}

The most important figure of merit is the time voltage response of the system, which is obtained by Fourier transform. This transformation reveals if the voltage reflection coefficient is low enough and if the spectral structures in the frequency response may affect the detection of the EoR. As detailed in \cite{Thyagarajan2016}, a Blackman-Harris window function is applied to the frequency signal before transformation. Compared with a square window, the signal leakage in the time domain caused by its transformation over a finite bandwidth is much more limited. The results are given in Fig. \ref{fig23:timeRespAntRec} when the antenna is terminated by the receiver or by a 100-$\Omega$ reference impedance, in order to differentiate the chromatic contributions from the antenna and from the receiver. As in \cite{Fagnoni2021}, the effects of mutual coupling are illustrated by simulating the response of an antenna at the edge of a horizontal strip of 11 x 3 dishes. This corresponds to 160 m, about half of the array width. This size is limited by our computational power. The time response is normalised and its maximum set at t = 0 to study its attenuation. The noise floor is around $10^{-6}$ and is limited by the precision of the frequency beams exported from CST.

The time responses obtained with the Vivaldi feed connected to the receiver (green curve) and to a 100-$\Omega$ impedance (red curve) are similar. The receiver is achromatic and the response is quickly attenuated by $10^5$ after 230 ns. As for the dipole system, its response is much more chromatic (pink curve). The intense spike at 1250 ns is caused by the reflection of the signal at the end of the 150-m cable connecting the FEM to the PAM. In addition, micro-reflections occur within the cable. This problem has been solved in HERA Phase II by using a radio-over-fibre system with a 500-m optical fibre which is less chromatic. Moreover, a reflection occurring at 500 m would not significantly limit the study of the EoR delay spectrum. 

When considering mutual coupling, the time response of both systems is similarly affected. \cite{Fagnoni2021} has studied in detail this aspect for the HERA Phase I system. The mutual coupling mainly comes from dish-dish interactions. This paper has shown how a received electromagnetic wave is able to propagate from one dish to another one, with little attenuation. The response slowly decays and eventually drops around 600 ns, once the reflected wave coming from the farthest antenna reaches the reference antenna. Therefore, the attenuation of the response depends on the antenna position, and the central antennas should be less affected. If the simulated antenna strip were longer, the extrapolation of the slope suggests that the response would be attenuated by a factor $10^{4}$ between 600 and 700 ns, and by a factor $10^{5}$ after about 1000 ns. By comparison, the core of the final array is about 300-m large along the X-axis and 250-m large along the Y-axis. In terms of propagation delay, this is respectively equivalent to 1000 ns and 833 ns. This indicates that any antenna could significantly interact with the dishes at the edges in the final array.

In practice, the received signal is analysed over small frequency bands associated with the cosmological evolution of the signal. In Fig. \ref{fig23:timeRespAntRec}, the time response is also computed over 50-MHz sub-bands to check if some sub-bands are more chromatic than others. With mutual coupling, the system is actually more chromatic in the higher band. The results for the dipole feed below 100 MHz and above 200 MHz are not particularly relevant because of the bandpass filter.  

\begin{figure}[t]
    \centering
    \includegraphics[width=1.0\linewidth]{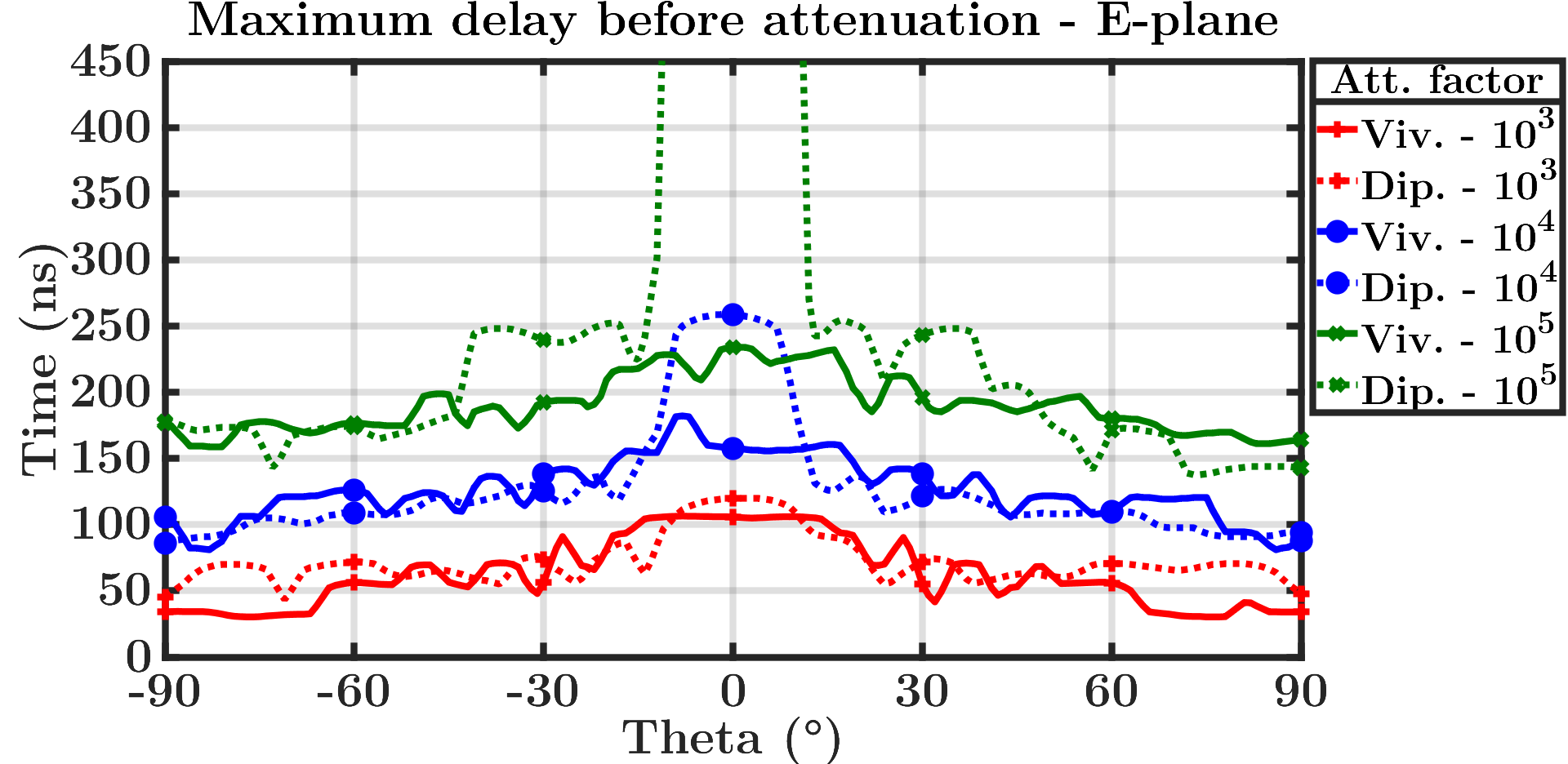}
    
    \vspace{3mm}
    
    \includegraphics[width=1.0\linewidth]{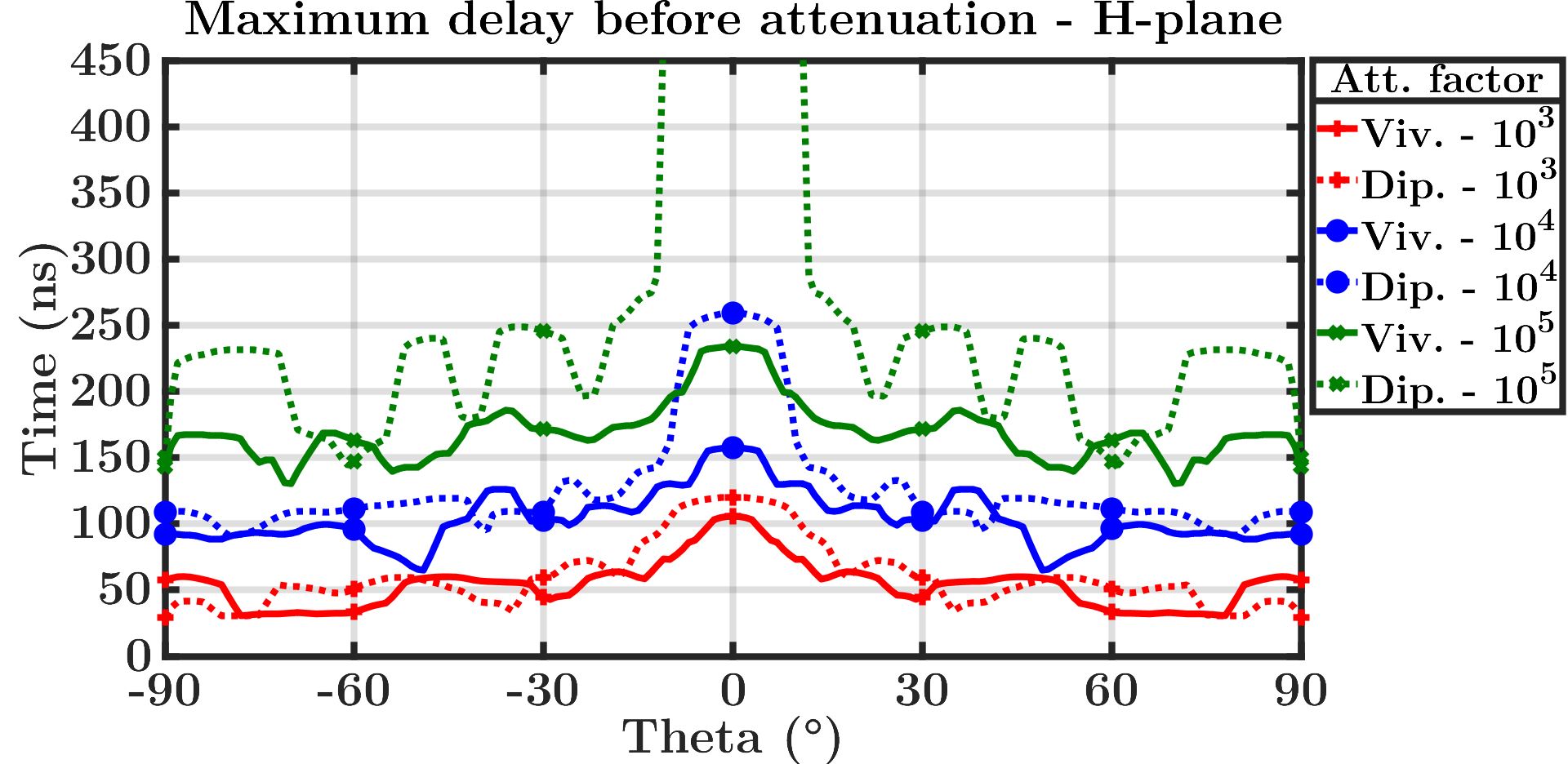}
    \caption{Maximum delays before the system response of a single antenna is attenuated by a certain factor, for the H and E-planes. Band: 50-250 MHz.}
    \label{fig24:maxDelEHplanes}
\end{figure}

As illustrated in Fig. \ref{fig19:gainBeam}, the antenna beams present some spatial and frequency variations which may translate into a high delay signal. In order to quantify the contribution from the sidelobes, we study the system time response as a function of the angle of incidence $\theta$. The results in the E and H-planes are presented for a single antenna. Fig. \ref{fig24:maxDelEHplanes} represents the maximum delay before the response is attenuated by a certain factor. Despite the fact that the sidelobe level is higher in the H-plane (cf. Fig. \ref{fig18:maxSLL}), the time response is actually slightly more chromatic in the E-plane where the variations in the radiation pattern are less smooth. The effect of the wider beamwidth in the E-plane, in particular at low frequencies, is also clearly visible. By considering similar excitation signals, the time response generated by the sidelobes is about 10 times lower than the response at zenith.

\begin{figure}[t]
    \centering
    \includegraphics[width=1.0\linewidth]{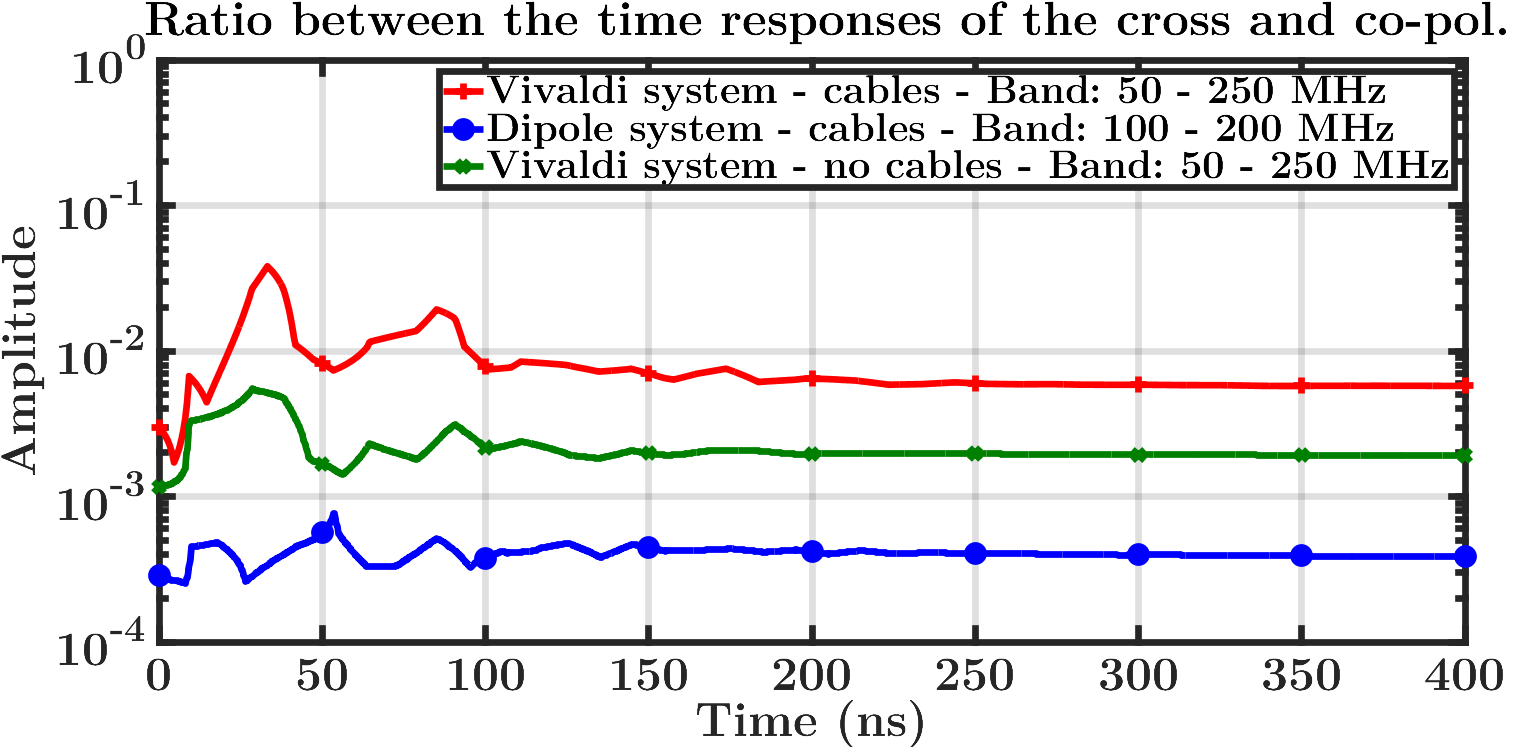}
    \caption{Ratio between the time responses of the cross and co-polarisations. The effect of the cables between the blades of the Vivaldi feed is studied.}
    \label{fig25:timeRespAntRecCrossPol}
\end{figure}

Lastly, polarisation leakage from polarised sources and synchrotron emission can also impact the detection of the EoR signal \cite{Nunhokee2017}. We study the case where the X-blade is excited by a plane wave with an orthogonal Y-polarisation coming from the zenith. The ratio between the time responses of the cross and co-polarisations is presented in Fig. \ref{fig25:timeRespAntRecCrossPol}. Due to the symmetry in the beams between the X and Y-polarisations of the antenna, this also corresponds to the ratio between the signals generated by the ports 1 and 2. With the Vivaldi feed, the time response of the cross-polarisation is about 200 times lower than the co-polarisation response. This is about one order of magnitude higher than with the dipole feed. This difference can be explained by the presence of the cables which increases the polarisation leakage by a factor 3, and the fact that signals can more easily be exchanged between the voluminous blades. These ratios do not significantly vary as a function of the incidence angle at high delays.


\subsection{Antenna sensitivity}
\label{sec:4.5.Sensi}

\begin{figure}[H]
    \centering
    \includegraphics[width=1.0\linewidth]{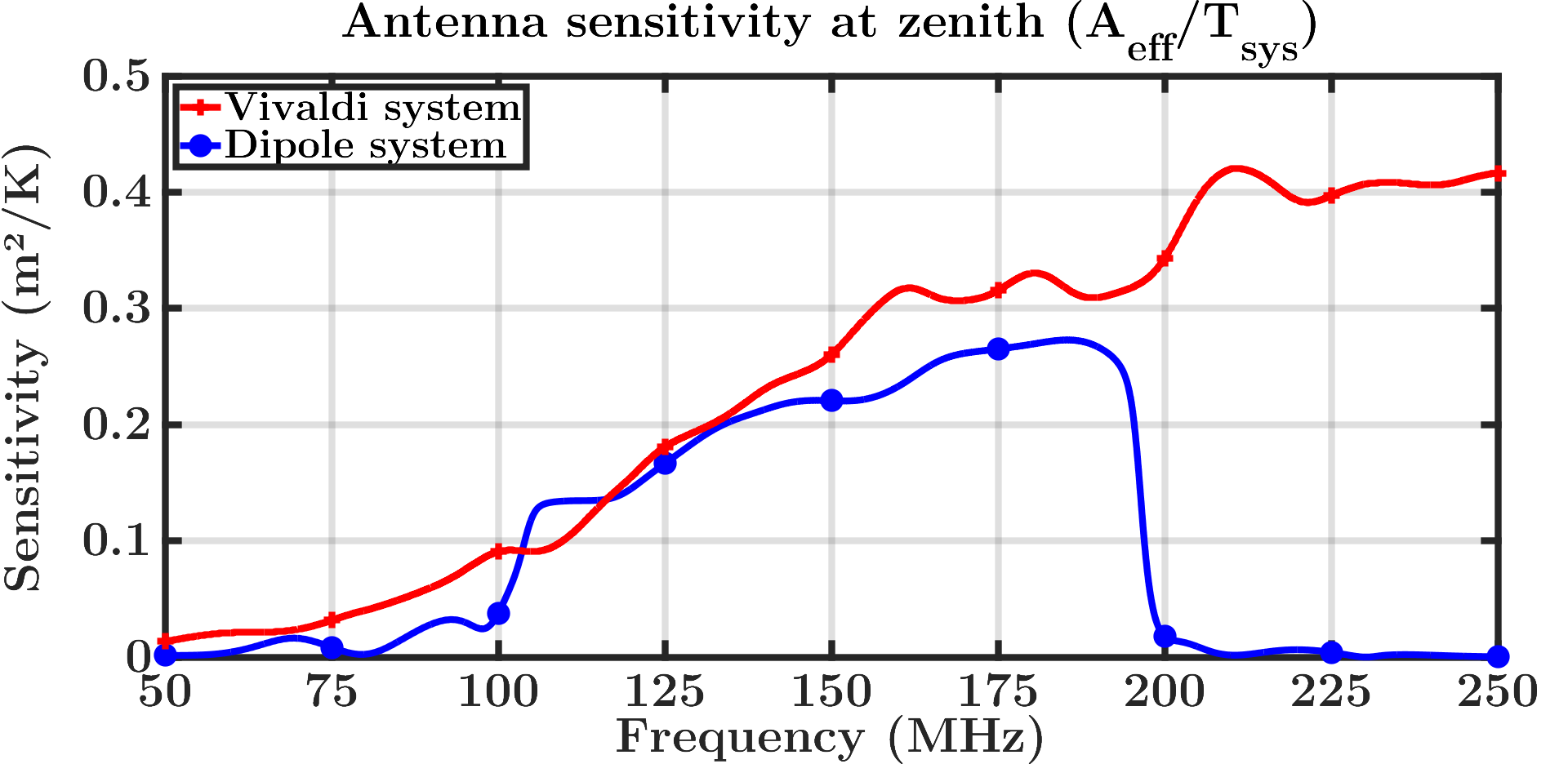}
    \caption{Antenna sensitivity at zenith ($A_{eff}/T_{sys}$).}
    \label{fig26:sensi}
\end{figure}

The sensitivity ${S_{\rm A}}$ at the feed point of the antenna, and for a given polarisation, is defined as the ratio between the effective antenna aperture ${A_{\rm eff}}$ and the system noise temperature ${T_{\rm sys}}$ \cite{Cortes-Medellin2007}. It is assumed that ${T_{\rm sys}}$ is equal to the receiver temperature ${T_{\rm rec}}$ (cf. Section \ref{sec:3.1.RFrec}), plus the antenna temperature corresponding to the noise received from the sky ${T_{\rm sky}}$, plus a term associated with the noise from the dissipative elements of the antenna at the temperature ${T_{\rm 0}}$ = 290 K and with the radiation efficiency ${\eta _{\rm rad}} \approx 99\%$. Thus, it is equal to:
\begin{align}
    {S_{\rm A}}\left( {\theta ,\;\varphi } \right) = \frac{{A_{\rm eff}\left( {\theta ,\;\varphi } \right)}}{{T_{\rm rec}+{\eta _{\rm rad}}T_{\rm sky}+\left(1-{\eta _{\rm rad}}\right)T_{\rm 0}}}.
	\label{eq6:sensitivity}
\end{align}
In order to simplify the comparison, first we consider that the sky contribution is uniform. The sky temperature was measured with EDGES \cite{Rogers2008} between 100 and 200 MHz, and for a cold patch it can be approximated by a power law:
\begin{align}
    T_{\rm sky}\left(\upsilon \right) = T_{\rm 150}{\left( {\frac{\upsilon }{150\;\rm{MHz}}} \right)^{-2.5}} + T_{\rm CMB},
	\label{eq7:Tsky}
\end{align}
with $T_{\rm 150}$ = 283.20 K the average sky temperature at 150 K, and $T_{\rm CMB}$ = 2.725 K the temperature of the cosmic microwave background. Except around 110 MHz, Fig. \ref{fig26:sensi} shows that the new system is more sensitive because the receiver temperature is lower thanks to the use of better amplifiers and a better noise matching with the antenna. The sharp drops in the dipole sensitivity at 105 and 195 MHz are caused by the bandpass filter of its receiver.

\begin{figure}[H]
    \centering
    \includegraphics[width=1.0\linewidth]{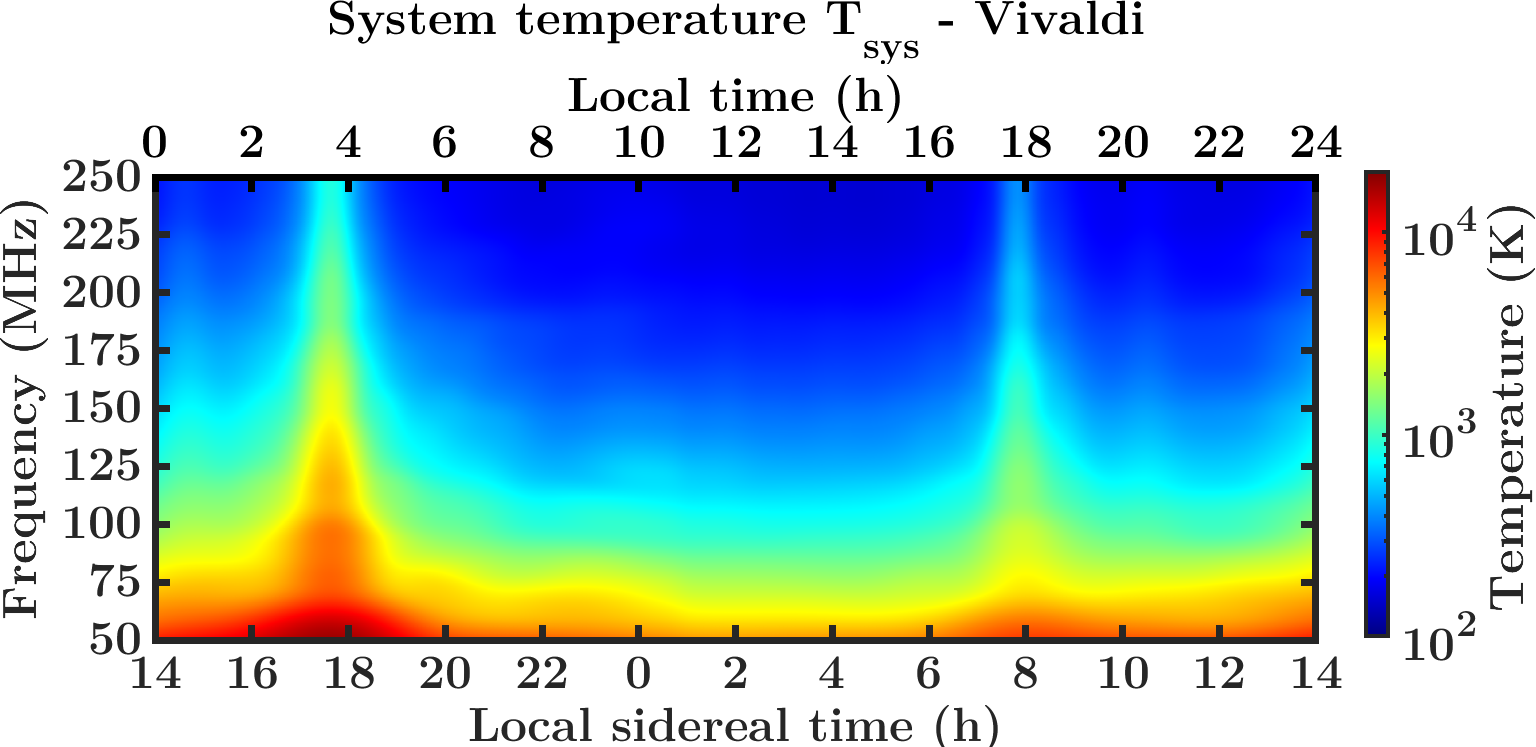}
    \caption{Simulated system temperature computed with the Haslam map and a spectral index of 2.5, on 1st May 2020. The frequency resolution is 1 MHz, and the time resolution is 5 minutes.}
    \label{fig27:Tsys}
\end{figure}

Lastly, the radiation pattern is combined with a map of the sky in order to simulate observations, which then can be used in sky-calibration techniques for the telescope \cite{Kern2020b}. Fig. \ref{fig27:Tsys} presents the simulated system temperature obtained with the reprocessed all-sky 408 MHz map of Haslam including the strong extra-galactic radio sources \cite{Remazeilles2015}. The spikes at 18 h and 8 h local sidereal time correspond to the transit of the galactic plane above the beam. The flux density measured at 408 MHz has been scaled to 50 -- 250 MHz by applying a spectral index of 2.5. In practice, this index varies between 2.1 and 2.7 depending on the region of the sky observed. However, this remains a good first approximation for regions far from the galactic centre, typically observed by HERA \cite{Rogers2008}.

\subsection{Effects of the cable misalignment}
\label{sec:4.6.cableMisalign}

\begin{figure}[H]
    \centering
    \includegraphics[width=1.0\linewidth]{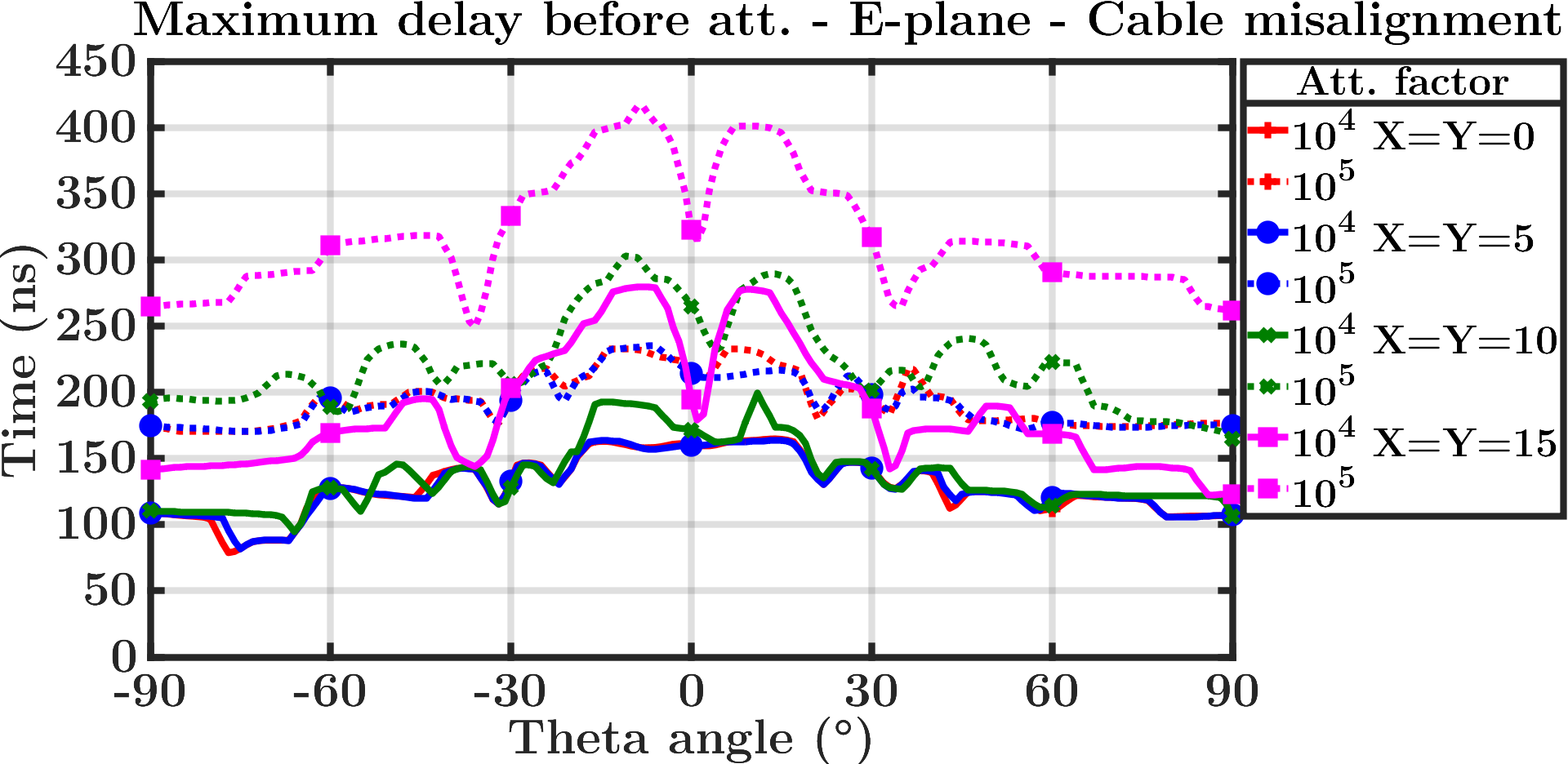}
    \caption{Maximum delays before the system response is attenuated by a certain factor, for the E-plane, when the cables are diagonally misaligned by 0, 5, 10, and 15 cm with respect to the vertex. Band: 50-250 MHz.}
    \label{fig28:timeRespAntRecCable}
\end{figure}

One of the specificity of the design is the presence of cables passing through the aperture. We study the effects of the cable misalignment on the time response. The cables are inserted into a PVC pipe when passing between the blades. This ensures that they are straight, centred, and not in contact with the blades. They can only move by 4 mm inside. The pipe ends at half the length of the slot. Below the aperture, the cables are held by ropes which are attached to the frame (cf. Fig. \ref{fig11:CSTmecha}). They are tied to the vertex as well to keep them tight. In this analysis, we assume that the cables are shifted by 4 mm along the X and Y-axes inside the pipe. Then, their position at the vertex is shifted along the X and Y axes. In simulations, we found that the performance is the most degraded when the cables are diagonally moved. Fig. \ref{fig28:timeRespAntRecCable} shows the attenuation delays in the E-plane, which is the most affected, when the cables are shifted by 5, 10, and 15 cm. Table \ref{tab3:beamVar} presents the variations in the gain at zenith, maximum sidelobe level, and beam pointing. Variations in the beams due to the cable misalignment, but also due to the feed positioning and mutual coupling, may cause baseline non-redundancy which affects the calibration \cite{Orosz2019}. In the case of the cable misalignment, the variations are very limited if the shift is below 5 cm, which is achievable under moderate wind condition.

\begin{table}[H]
\caption{Beam variations caused by cable misalignment}
\begin{tabular}{|c|c|c|c|c|}
\hline
\textbf{Shift at vertex (cm)}                                                           & X:0 Y:0   & X:5 Y:5   & X:10 Y:10 & X:15 Y:15 \\ \hline
\textbf{Gain at zen. var. (dB)}                                                         & 0.1       & 0.1       & 0.2       & 0.3       \\ \hline
\textbf{\begin{tabular}[c]{@{}c@{}}Max. SLL var. (dB)\\ E-plane / H-plane\end{tabular}} & 1.3 / 0.1 & 1.5 / 0.1 & 2 / 0.2   & 2.3 / 0.5 \\ \hline
\textbf{Beam pointing var. (°)}                                                         & 0.2       & 0.2       & 0.3       & 0.6       \\ \hline
\end{tabular}
\label{tab3:beamVar}
\end{table}

%%%%%%%%%%%%%%%%%%%% CONCLUSION %%%%%%%%%%%%%%%%%%

\section{Conclusion}
\label{sec:5.Conclu}

We have presented the design of the new Vivaldi feed for the HERA radio-telescope. Illuminating a 14-m diameter dish, this feed has been developed in conjunction with a new receiver using radio-over-fibre. The bandwidth has been increased from 100 -- 200 MHz to 50 -- 250 MHz in order to study the Cosmic Dawn and the whole EoR. The goal was also to create a system with a smooth voltage response. This aspect is essential to limit the leakage of the foreground into the EoR delay spectrum.

The initial design has been trimmed to reduce its weight and the wind load, while maintaining good performance. The FEM is placed inside the cavity with minimal effects on the electromagnetic properties of the feed. The two control and power cables pass through the aperture, exactly in the middle of the slot, to keep the beam symmetric. They are inserted into metal braided sleeves which are kept in contact to prevent the formation of a parallel transmission line between the outer conductors. A cable misalignment within 5 cm with respect to the vertex has very little effect on the antenna performance.

Compared with the dipole system, the bandwidth has been extended and the sensitivity improved, while mitigating the reflections in particular in the transmission line. However, the absence of elements around the feed to taper the radiation pattern, such as a metal cage, makes the dish illumination less efficient and the sidelobe level a bit higher, at low frequencies. The cross-polarisation response has also been degraded, but still remains 200 times lower than the co-polarisation response. 

Lastly, the initial goal was to attenuate the system time response by $10^{5}$, ideally after 150 ns. The response of a single element reaches this target after 230 ns, which is not too far. However, this radically changes when considering mutual coupling. Because the antennas are spaced 60 cm apart, an incident wave scattered by a dish is able to propagate through the whole array. Thus, the attenuation of the response depends on the array size and on the antenna position. Mutual coupling is mainly caused by dish-dish interactions and both feeds show similar behaviour. This implies that the largest spatial scales with the highest signal to noise measurements will not be characterisable with a strict foreground avoidance method, which reduces the sensitivity of the telescope to the EoR power spectrum \cite{Ewall-Wice2016}. Alternative methods, based on absolute calibration strategies for instance \cite{Kern2020b}, are being developed to try to mitigate the foreground contribution. The mutual coupling in the array is a complex problem and is still being investigated. More accurate models and solutions to limit the propagation of the signal from one dish to another one are also being studied.

%%%%%%%%%%%%%%%%%%%% APPENDICES %%%%%%%%%%%%%%%%%%

\appendix

The antenna has been optimised to work between 50 and 250 MHz, and the bandwidth is limited by a bandpass filter. However, due to the wideband nature of the Vivaldi feed, the antenna could receive signals at lower and much higher frequencies. In this appendix, the performance between 10 and 1000 MHz are presented, as it may be of interest to some readers and open future research opportunities, in particular in the mapping of the neutral hydrogen for more recent periods. With respect to a 100-$\Omega$ termination, S$_{11}$ is below -7 dB up to 700 MHz, thanks to the stability of the antenna impedance. As for the low frequencies, the cut-off frequency occurs at 40 MHz. The beam pattern remains also stable. Lastly, as the frequency increases the performance will get much more sensitive to cable misalignment and reflector surface errors.

\begin{figure}[H]
    \centering
    \includegraphics[width=1.0\linewidth]{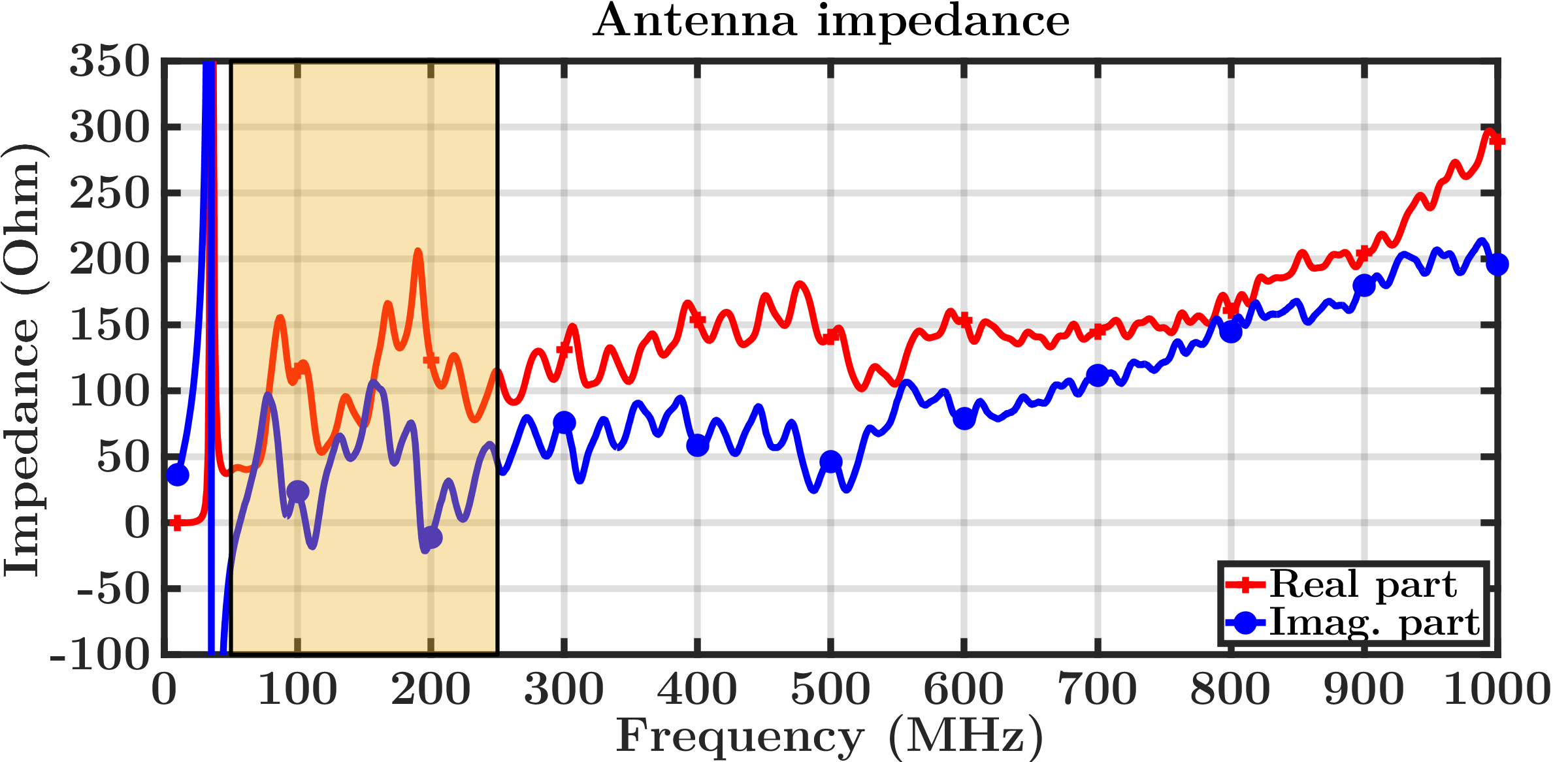}
    \caption{Differential antenna impedance.}
    \label{fig29:antImp_10_1000MHz}
\end{figure}
\begin{figure}[H]
    \centering
    \includegraphics[width=1.0\linewidth]{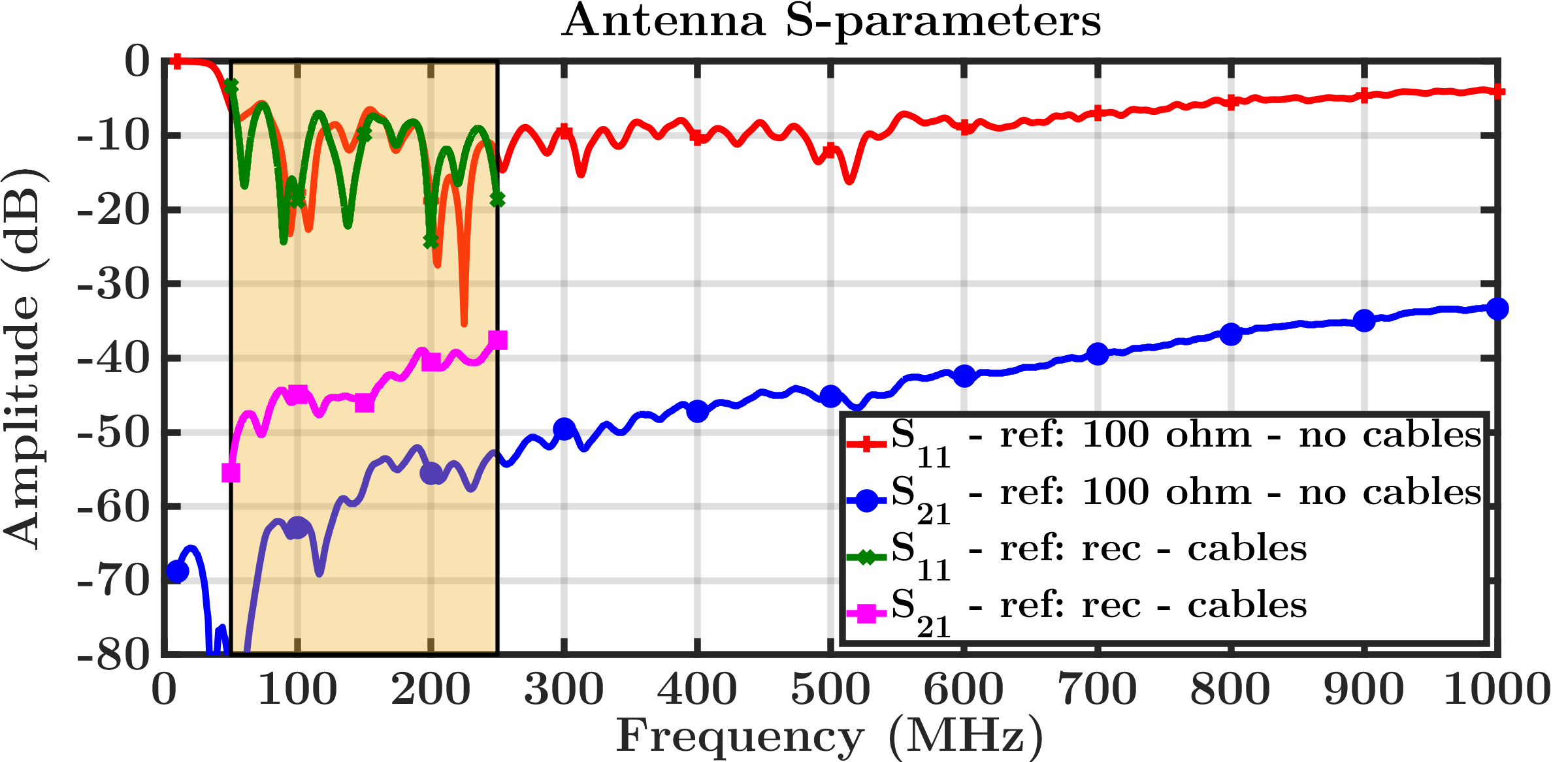}
    \caption{Differential antenna S-parameters. Termination impedance: 100 $\Omega$ and receiver impedance. The effect of the cables are also illustrated.}
    \label{fig30:Spara_10_1000MHz}
\end{figure}
\begin{figure}[H]
    \centering
    \includegraphics[width=1.0\linewidth]{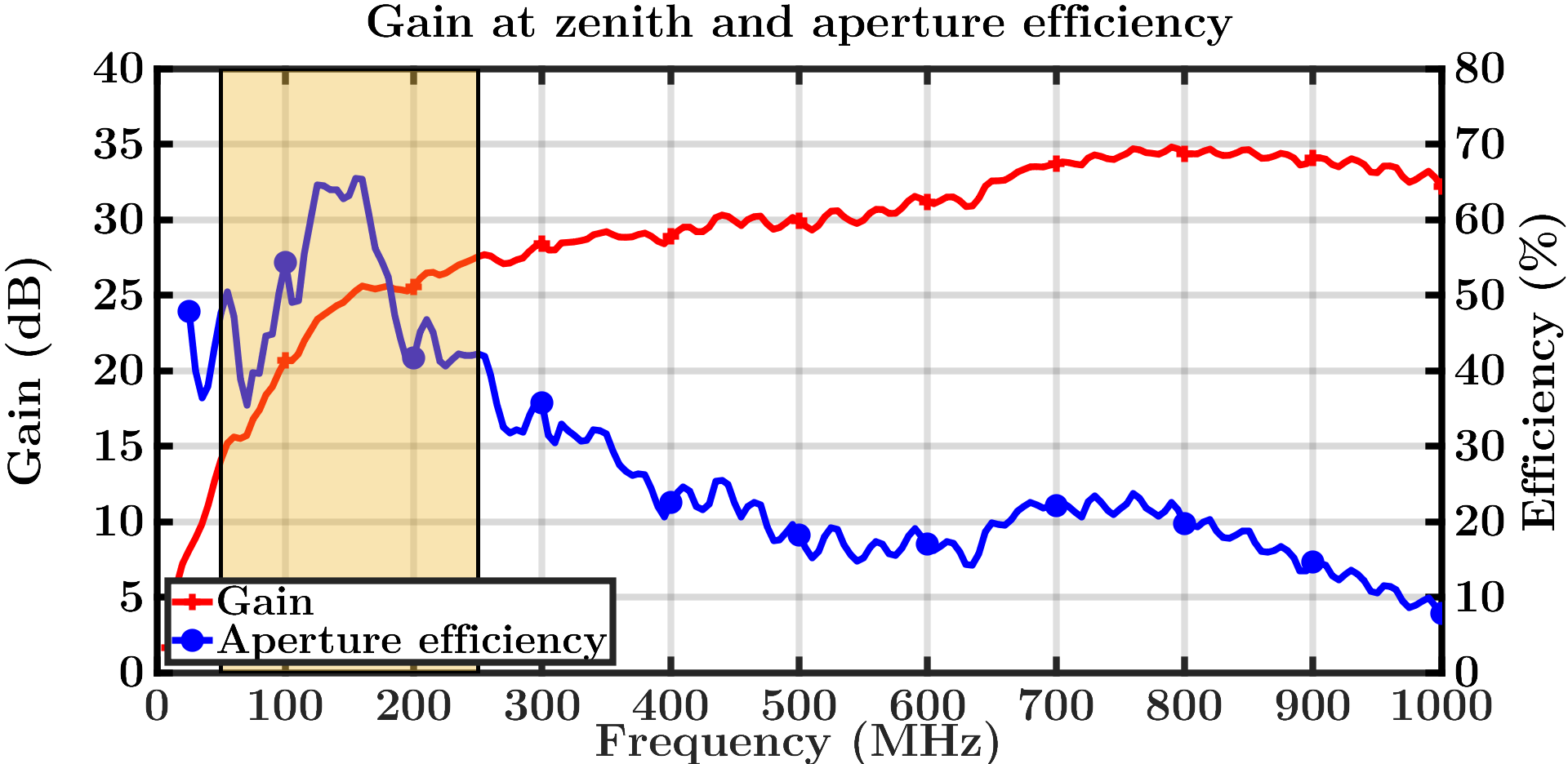}
    \caption{Antenna gain at zenith and aperture efficiency.}
    \label{fig31:antGain_10_1000MHz }
\end{figure}
\begin{figure}[H]
    \centering
    \includegraphics[width=1.0\linewidth]{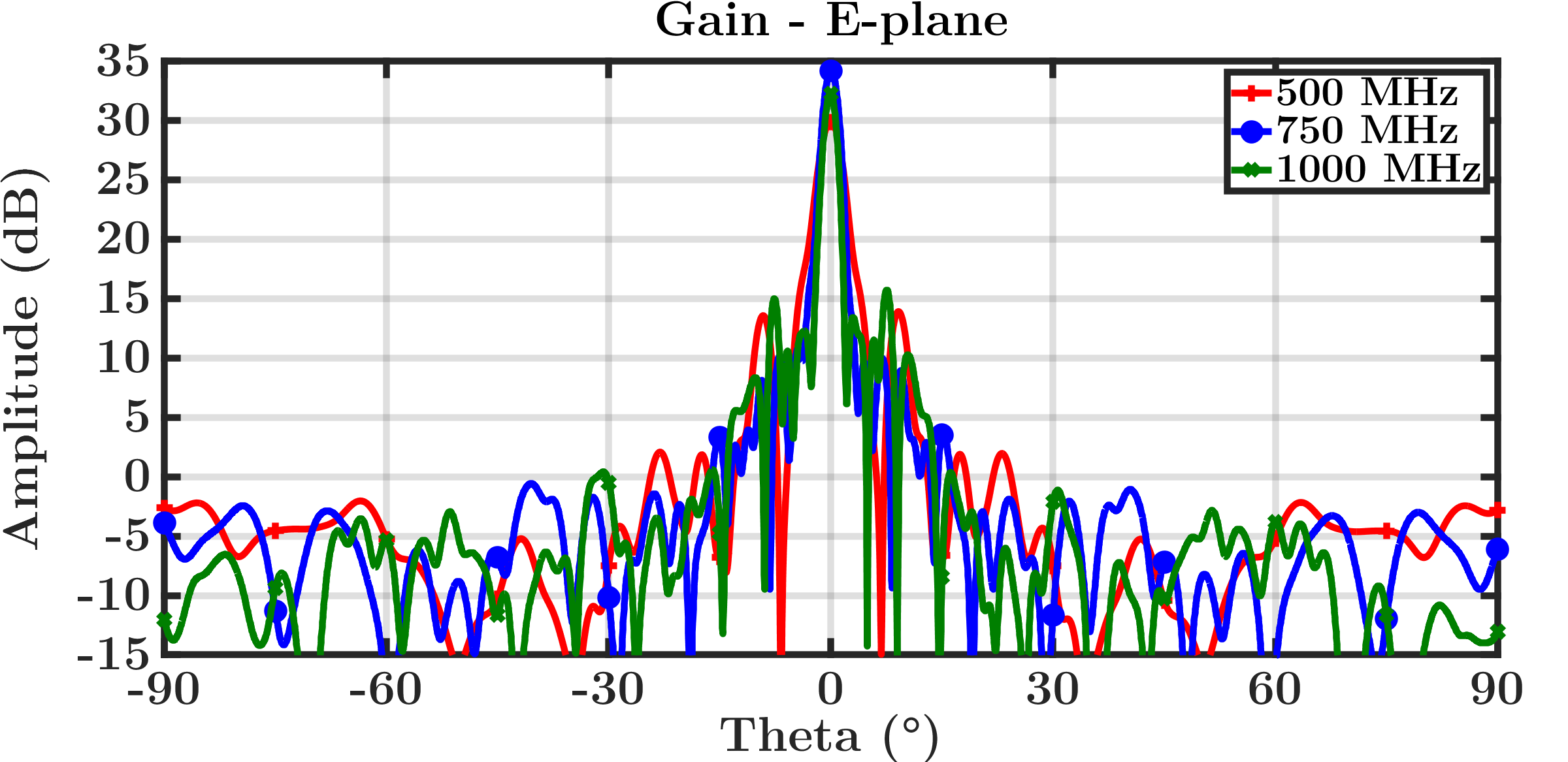}
    
    \vspace{3mm}
    
    \includegraphics[width=1.0\linewidth]{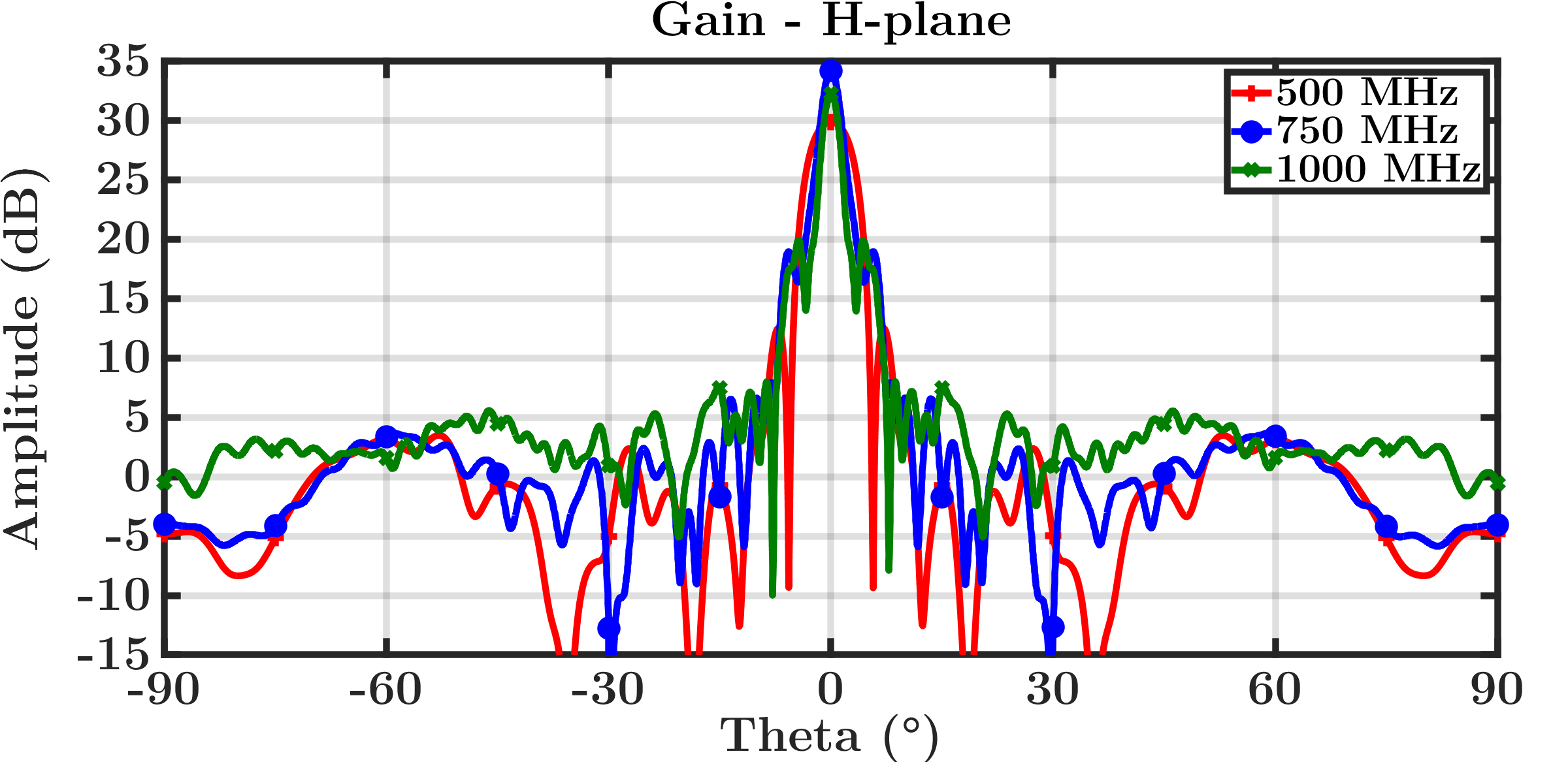}        
        
    \caption{Cuts of the antenna gain at 500, 750, and 1000 MHz.}
    \label{fig32:antGain_cut_10_1000MHz}
\end{figure}
%

%%%%%%%%%%%%%%%%%%%% ACKNOWLEDGMENT %%%%%%%%%%%%%%%%%%

\section*{Acknowledgment}

The authors acknowledge the UK Science and Technology Facilities Council (STFC), as well as the Institute of Astronomy and the Physics Department of the University of Cambridge for their financial support via the Isaac Newton Studentship.

%%%%%%%%%%%%%%%%%%%% REFERENCES %%%%%%%%%%%%%%%%%%

\bibliographystyle{IEEEtran.bst}
\bibliography{biblio} %Biblio.bib
\nocite{*}

%%%%%%%%%%%%%%%%%%%% BIOGRAPHIES %%%%%%%%%%%%%%%%%%

% Nicolas Fagnoni

\begin{IEEEbiography}
[{\includegraphics[width=1in,height=1.25in,clip,keepaspectratio]{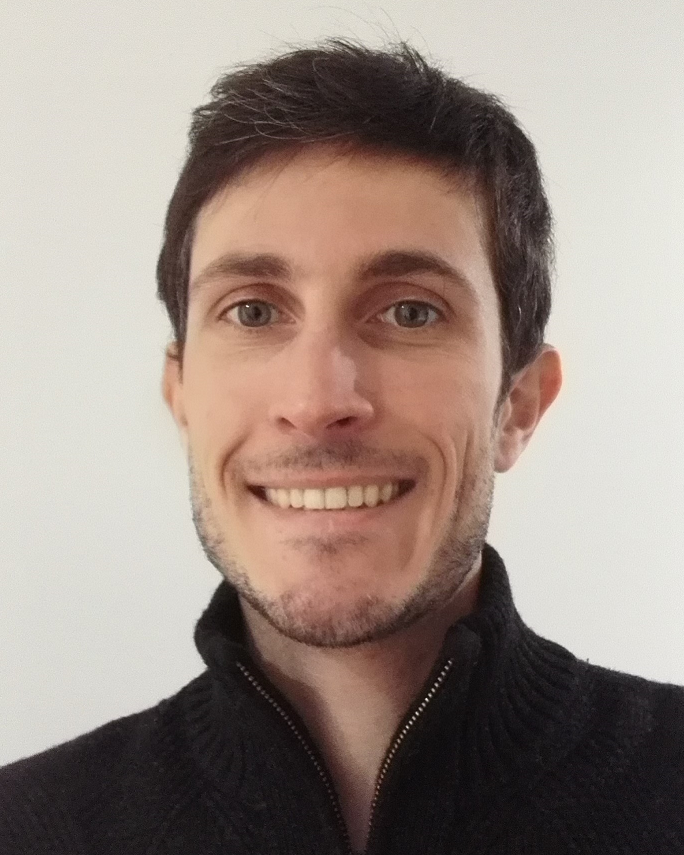}}]{Nicolas Fagnoni} has an Engineer's degree in Telecommunication from Telecom SudParis as well as a Master of Science in "Spacecraft Technology and Satellite Telecommunication" with Distinction, from the University College London. 

After graduating, Nicolas worked for 4 years for Eutelsat as a satellite service engineer in a ground station near Paris.
In 2015, Nicolas started a Ph.D. in Astrophysics at the University of Cambridge, supervised by Dr Eloy de Lera Acedo. He worked on the development of new receivers for radio-telescopes for HERA, REACH, and the SKA, in order to study the formation and evolution of the first galaxies during the Cosmic Dawn and the Epoch of Reionization. In 2019, he joined Airbus Defence \& Space UK where he designs antenna systems for telecom satellites and ESA science missions.
\end{IEEEbiography}

% Eloy de Lera Acedo

\begin{IEEEbiography}
[{\includegraphics[width=1in,height=1.25in,clip,keepaspectratio]{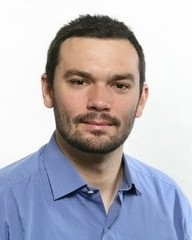}}]{Eloy de Lera Acedo} received his Ph.D. from the University Carlos III of Madrid in 2010.

He is a Principal Investigator at the Cavendish Laboratory of the University of Cambridge. His research interests include 21-cm cosmology, technology development for radio-astronomy and ultra fast digital communications, and electromagnetic modelling. He led the design of the array antennas for the SKA1-LOW project, and he is currently the Principal Investigator of the REACH telescope. 
\end{IEEEbiography}

% Nick Drought

\begin{IEEEbiography}
[{\includegraphics[width=1in,height=1.25in,clip,keepaspectratio]{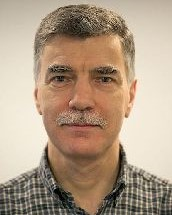}}]{Nick Drought} has a B.Sc. in Mechanical Engineering from the University of Bristol, which was followed by a 1-year postgraduate training course at the University of Cambridge. Then, he completed a Teaching Company Scheme with the University of Lancaster and Leyland Vehicles Ltd, where his interest in product design engineering was formed.

Next, he worked for 3-years at Black \& Decker, gaining experience in design for low cost manufacture and assembly. In 1988, Nick joined Cambridge Consultants, and over the following 32 years worked on many client projects in a variety of technical fields. These include the low and mid frequency antenna structures for the SKA, HERA, various robust enclosures and novel antenna structures for radio, radar and telecommunication systems, and a novel flow reversing valve for full-scale hurricane wind load testing in the "3 Little Pigs" facility at the University of Western Ontario. 
\end{IEEEbiography}

% David DeBoer

\begin{IEEEbiography}
[{\includegraphics[width=1in,height=1.25in,clip,keepaspectratio]{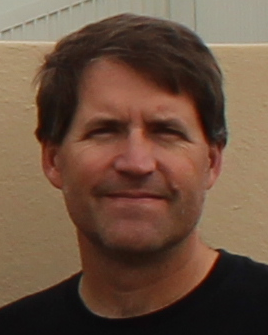}}]{David DeBoer} graduated cum laude from Harvard University in 1989, with a Bachelor's degree in Astronomy and Astrophysics. In 1995, he graduated from the Georgia Institute of Technology with a Ph.D. in Electrical and Computer Engineering. His thesis work investigated the constituency of the atmospheres of the outer planets. 

After a brief stint working out of NASA's Goddard Space Flight Center, Dr. DeBoer returned to Georgia Tech to oversee the development of a 100-ft radio telescope for use in the SETI Institute's Project Phoenix. He then went on to become an Assistant Professor of Electrical and Computer Engineering at Georgia Tech. He left Georgia Tech to become the Project Engineer/Project Manager for the Allen Telescope Array project before joining CSIRO in Australia. Dr. DeBoer served as the Chair of the US National Research Council Committee on Radio Frequencies (CORF) and Chair of Commission J of the US National Committee of the International Union of Radio Science (URSI). He served on the governing board of the SKA in the US and Australia. He is now the Project Manager of HERA, the Breakthrough Listen program, and a member of the research staff at the University of California at Berkeley.
\end{IEEEbiography}

% Daniel Riley

\begin{IEEEbiography}
[{\includegraphics[width=1in,height=1.25in,clip,keepaspectratio]{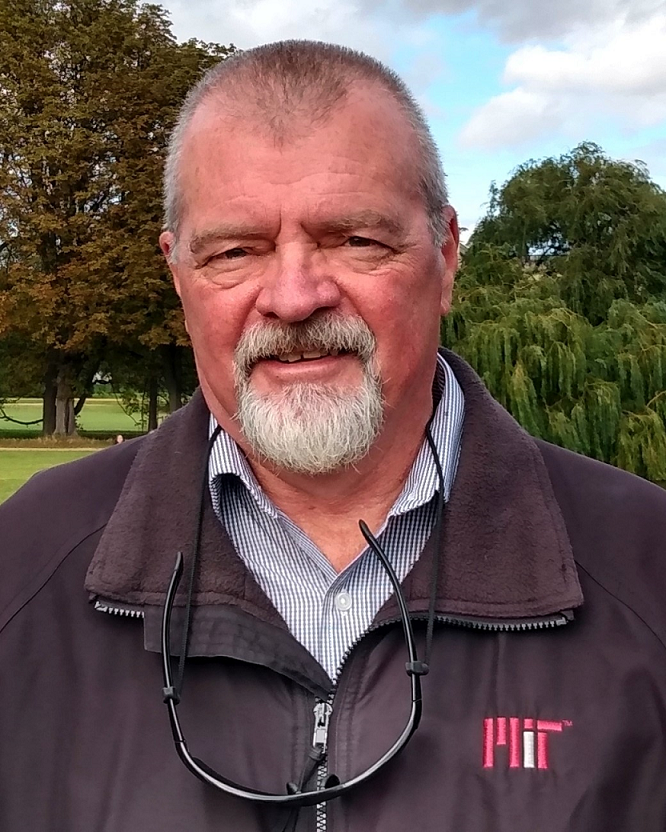}}]{Daniel Riley} is a 1980 graduate apprentice of General Dynamic’s machining program. 

In 1984, he left General Dynamics and for the next 17 years worked in a high end prototype machine shop. The next 10 years, he worked in the medical device industry making prototypes. In 2012, he went to work as a Prototype Machinist at the Harvard Smithsonian Center for Astrophysics. While there, he worked on “SWEAP” (Solar Wind Electrons Alphas and Protons) and “HI-C” (NASA's High Resolution Coronal Imager) missions. In 2015, he joined the Kavli Institute for Astrophysics and Space Research at the MIT as an Instrument Maker/Prototype Machinist, where he worked on “TESS” (NASA’s - Transiting Exoplanet Survey Satellite) mission. After a successful launch in 2018, he went on to fabricate various antenna feed designs for the HERA project. Once the design was finalized, he set up manufacturing in South Africa and instructed the crew on assembly of the feeds.
\end{IEEEbiography}

% Nima Razavi-Ghods

\begin{IEEEbiography}
[{\includegraphics[width=1in,height=1.25in,clip,keepaspectratio]{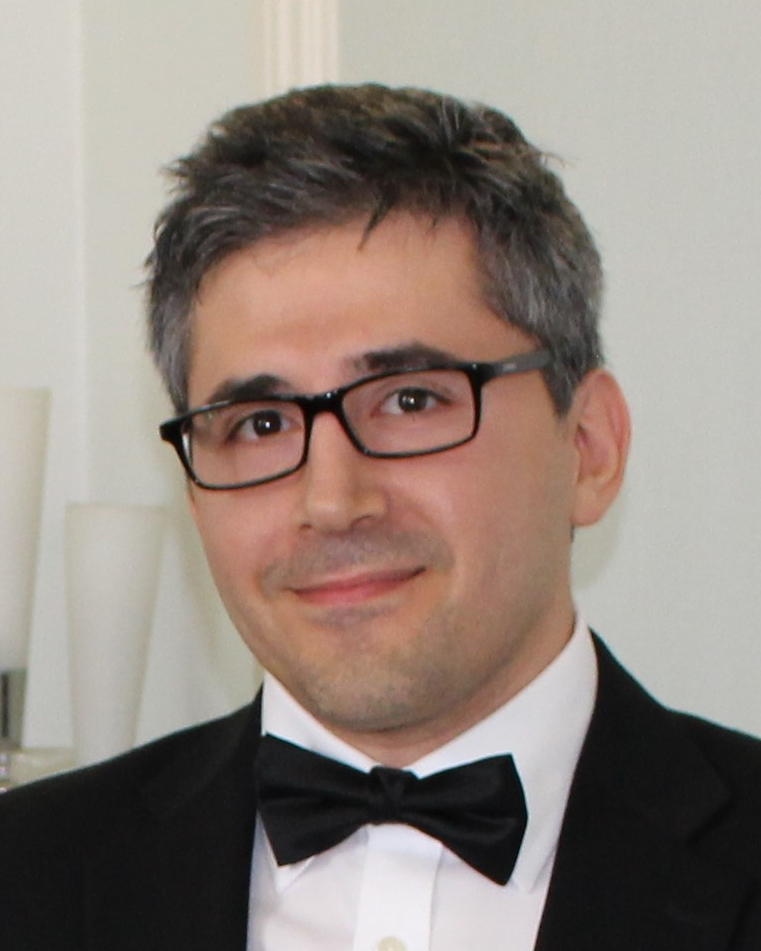}}]{Nima Razavi-Ghods} received his Ph.D. in 2007 (CASE award) from the University of Durham. 

He is a Principle Research Associate at the University of Cambridge, where his primary research interests are in instrumentation and calibration of 21-cm cosmology experiments. He has led the RF design and instrumentation effort for a number of radio telescopes including HERA. 
\end{IEEEbiography}

% Steven Carey

\begin{IEEEbiography}
[{\includegraphics[width=1in,height=1.25in,clip,keepaspectratio]{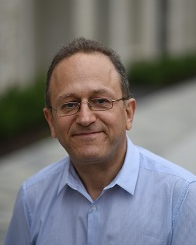}}]{Steven Carey} graduated from the London South Bank University in 2004 with a first class Bachelor of Engineering degree in Telecommunications and Computer Network Engineering. 

This was obtained by part time study while working as a Higher Technologist for the Metropolitan Police Service where he was responsible for the specification and design of RF sections of custom electronic equipment for police use. He graduated from the University of Surrey in 2012 with a Master of Science degree in Microwave Engineering and Wireless Subsystems Design. He then contributed to the design and engineering of analogue sections of the HERA radio-telescope arrays, and is now involved in the design of the REACH receiver.
\end{IEEEbiography}

% Aaron Parsons

\begin{IEEEbiography}
[{\includegraphics[width=1in,height=1.25in,clip,keepaspectratio]{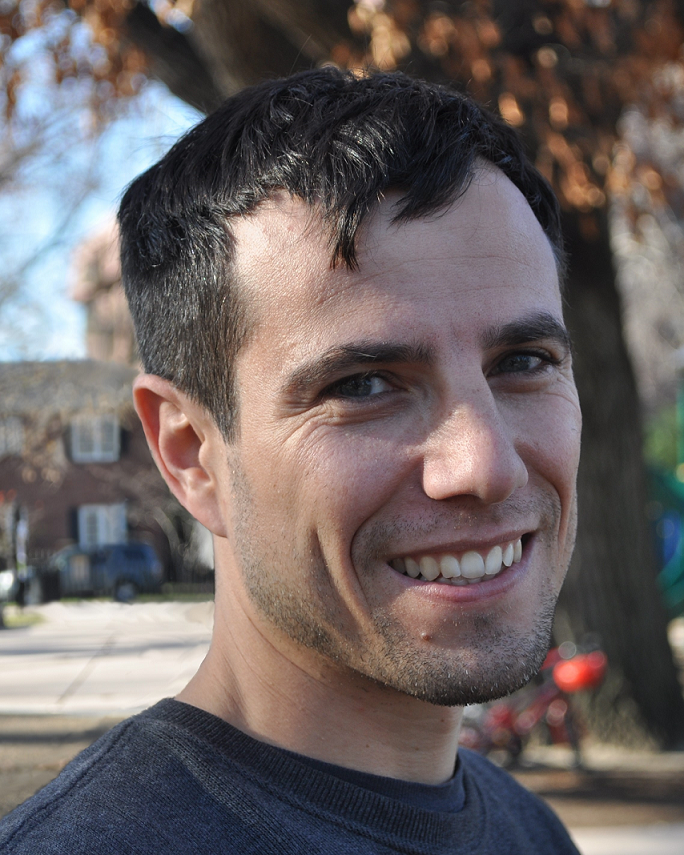}}]{Aaron Parsons} received his Ph.D. in Astronomy in 2009 from the University of California, Berkeley. 

He is currently a Professor of Astrophysics and Director of the Radio Astronomy Laboratory at the University of California, Berkeley. He is the Principal Investigator of the HERA telescope and in 2019 was awarded the Presidential Early Career Award for Scientists and Engineers.
\end{IEEEbiography}

%%%%%%%%%%%%%%%%%%%%%%%%%%%%%%%%%%%%%%%%%%%%%%%%%%

\end{document}